\documentclass[a4paper,longbibliography]{styleFiles/jpconf}
\usepackage{graphicx}
\usepackage{xspace}
\usepackage{sidecap}
\usepackage{amssymb}
\usepackage{url}

\newcommand{\nova}{{NO$\nu$A}\xspace}

\def\Cer {\v{C}erenkov\xspace}

\newcommand{\numutonue}{\ensuremath{\nu_{\mu}\rightarrow\nu_{e} }\xspace}
\newcommand{\numubartonuebar}{\ensuremath{\overline{\nu}_{\mu}\rightarrow\overline{\nu}_e}\xspace}

\newcommand{\numu}{\ensuremath{\nu_{\mu}}\xspace}
\newcommand{\numubar}{\ensuremath{\overline{\nu}_{\mu}}\xspace}

\newcommand{\nue}{\nu_{e}}
\newcommand{\nuebar}{\overline{\nu}_{e}}

\newcommand{\dm}{{\Delta}m^2}

\newcommand{\thetatwothree}{\theta_{23}}
\newcommand{\thetaonethree}{\theta_{13}}
\newcommand{\thetaonetwo}{\theta_{12}}
\newcommand{\snthetaonethree}{\sin^2(2\theta_{13})}
\newcommand{\snthetaonethreeNOBrackets}{\sin^22\theta_{13}}

\newcommand{\snthetatwothree}{\sin^2\theta_{23}}
\newcommand{\sintttwothree}{\ensuremath{\sin^2(\theta_{23})}\xspace}
\newcommand{\sintttwotwothree}{\ensuremath{\sin^2(2\theta_{23})}\xspace}
\newcommand{\sintttwoonethree}{\ensuremath{\sin^2(2\theta_{13})}\xspace}

\newcommand{\nuthree}{\nu_{3}}

\newcommand{\nust}{\nu_{s}}
\newcommand{\nutau}{\ensuremath{\nu_{\tau}}\xspace}
\newcommand{\dmsqtwo}{|\Delta m^{2}_{32}|}

\newcommand{\dmbaratm}{\ensuremath{\Delta \overline{m}^{2}_{{\rm atm}}}\xspace}
\newcommand{\dmatm}{\ensuremath{\Delta m^{2}_{{\rm atm}}}\xspace}
\newcommand{\snthetabar}{\ensuremath{\sin^{2}(2\overline{\theta})}\xspace}
\newcommand{\sntheta}{\ensuremath{\sin^{2}(2\theta)}\xspace}

\newcommand{\musec}{\ensuremath{\mathrm{\mu s}}\xspace}

\begin{document}
\title{Long-baseline Neutrino Oscillation Experiments}

\author{G. J. Feldman} \address{Department of Physics, Harvard
  University, Cambridge, Massachusetts 02138, USA.\@}
\ead{gfeldman@fas.harvard.edu}
 
\author{J. Hartnell} \address{Department of Physics and Astronomy,
  University of Sussex, Brighton. BN1 9QH\@. United Kingdom.}
\ead{j.j.hartnell@sussex.ac.uk}

\author{T. Kobayashi} \address{Institute for Particle and Nuclear
  Studies, High Energy Accelerator Research Organization (KEK), 1-1,
  Oho, Tsukuba, 305-0801, Japan.} \ead{takashi.kobayashi@kek.jp}

\begin{abstract}
A review of accelerator long-baseline neutrino oscillation experiments
is provided, including all experiments performed to date and the
projected sensitivity of those currently in progress. Accelerator
experiments have played a crucial role in the confirmation of the
neutrino oscillation phenomenon and in precision measurements of the
parameters. With a fixed baseline and detectors providing good energy
resolution, precise measurements of the ratio of distance/energy
($L/E$) on the scale of individual events have been made and the
expected oscillatory pattern resolved. Evidence for electron neutrino
appearance has recently been obtained, opening a door for determining
the CP violating phase as well as resolving the mass hierarchy and the
octant of $\theta_{23}$: some of the last unknown parameters of the
standard model extended to include neutrino mass.
\end{abstract}

\section{Introduction}

Neutrino oscillation experiments are normally categorized into
short-baseline and long-baseline experiments. For experiments using
accelerator neutrinos as the source, the long-baseline means that $E/L
\simeq\Delta m^2 \sim2.5\times 10^{-3}$~eV$^2$, where $E$ and $L$ are
the neutrino energy and flight distance respectively. In this article,
accelerator long-baseline (LBL) neutrino oscillation experiments are
reviewed. The recent reactor neutrino experiments
to look for non-zero $\theta_{13}$ at $\Delta m^2 \sim2.5\times
10^{-3}$~ eV$^2$ and atmospheric neutrino experiments are covered
elsewhere in this Special Issue on Neutrino Physics.

Neutrino beams for the LBL experiments are produced in the
``conventional'' method where a high energy proton beam hits a target
and the pions that are produced then decay in flight to give muon
neutrinos. The typical neutrino energy thus produced is 0.5--10~GeV
and that sets the necessary distance to a neutrino detector to be
several hundreds of kilometers such that the neutrino oscillation
driven by $\Delta m^2 \sim2.5 \times 10^{-3}$~eV$^2$ can be
investigated. This review describes KEK~\cite{kek:1998},
NuMI~\cite{Anderson:1998zza}, CNGS~\cite{Acquistapace:1998rv} and
J-PARC~\cite{Abe:2011ks} neutrino beams and their associated
experiments.

The goals of the first LBL experiments proposed in 1990s,
K2K~\cite{Ahn:2006zza}, MINOS~\cite{Ambats:1998aa} and CERN to Gran
Sasso (CNGS) experiments OPERA~\cite{Acquafredda:2009zz} and
ICARUS~\cite{Arneodo:2001tx} were to clarify the origin of the anomaly
observed in the atmospheric neutrino measurements of
Kamiokande~\cite{Hirata:1988uy} and IMB~\cite{Haines:1986yf} and later
to confirm the discovery of neutrino oscillations by Super-Kamiokande
(SK) in 1998~\cite{Fukuda:1998mi}. Kamiokande observed a deficit of
muon neutrinos coming through the earth, which could have been
interpreted as muon to tau neutrino oscillation and/or to electron
neutrino oscillation. Soon afterwards, the CHOOZ
experiment~\cite{Apollonio:2002gd} excluded the possibility that muon
to electron neutrino oscillation is the dominant mode. Therefore, the
goal of the first generation LBL experiments was focused on confirming
muon to tau neutrino oscillation. The K2K and MINOS experiments, which
used beams with neutrino energies of a few-GeV, focused on detecting
muon neutrino disappearance because the energy of the neutrinos was
rarely high enough to make $\nutau$ charged current interactions
(threshold energy is about 3.5~GeV). In contrast, the CNGS experiments
make use of a higher energy ($\sim$20~GeV) neutrino beam and OPERA is
optimized for the detection of tau neutrino appearance.

Soon after the discovery of neutrino oscillation by SK, the importance
of the sub-leading electron neutrino appearance channel was pointed
out.  In the three flavor mixing picture, the probability of electron
neutrino appearance gives a measure of the mixing angle
$\theta_{13}$. The existence of electron neutrino appearance at the
atmospheric oscillation length means non-zero $\thetaonethree$. Only
an upper bound of $\snthetaonethree=0.14$~(90\%~C.L.) from the CHOOZ
experiment was known until very recently. Because the CP violating
observable, the phase $\delta$, appears always in the product with
$\sin(2\theta_{12}) \sin(2\theta_{23}) \sin(2\theta_{13})$ and
$\thetatwothree$ and $\thetaonetwo$ are known to be large, the size of
$\theta_{13}$ is a major factor in the feasibility of the future CP
violation search.

With the goal to discover electron neutrino appearance and determine
$\theta_{13}$, the T2K experiment~\cite{Abe:2011ks} in Japan started
taking data in 2010 and the \nova
experiment~\cite{Ayres:2002ws,Ayres:2004js,Ayres:2007tu} in the USA is
now under construction and will start measurements in 2013. The design
of these experiments was optimized for detection of electron neutrino
appearance. Both T2K and \nova adopted a novel ``off-axis'' beam
technique that provides a narrow peak in the energy spectrum, tuned to
be at the expected oscillation maximum, while at the same time
reducing the unwanted high energy tail. The $\numu\rightarrow\nue$
transition is a sub-dominant effect and the oscillation probability to
be probed is small. To have enough sensitivity, beam powers of order
1~MW and detector masses of order 10~kilotons are required and as such
these experiments are sometimes called ``superbeam'' experiments.

With evidence of $\nue$ appearance from early T2K results and the
recent measurement of $\nuebar$ disappearance by the reactor
experiments~\cite{Abe:2011fz,An:2012eh,Ahn:2012nd}, the major focus
for the future will be to determine the mass hierarchy and search for
evidence of CP violation. \nova will have the longest baseline of all
second-generation experiments at 810~km, which will give enhanced sensitivity to the
neutrino mass hierarchy due to the neutrino-matter interaction in the
Earth as the neutrinos propagate. Information on the mass
hierarchy and the expected precision measurement of $\thetaonethree$
from the reactor experiments will be crucial to resolve degeneracies
in the grand combination of T2K, \nova and reactor experiments to
reveal information on what nature has chosen for leptonic CP
violation.

Beyond oscillations, the provision of intense and relatively well
understood neutrino beams along with the large detectors in these
experiments has opened up whole new avenues to look for new
physics. This review provides a concise overview of searches for
sterile neutrinos, velocity measurements of neutrinos and searches for
violation of Lorentz symmetry. In the future, the MINOS+
experiment~\cite{Tzanankos:2011zz} will focus on searches for new
physics through high-precision, high-statistics measurements with the
NuMI beam operating at a peak on-axis energy of 7~GeV.

This review paper is structured as follows. Section~\ref{sec:beams}
describes the beams and section~\ref{sec:detectors} gives an overview
of the detectors. The results from long-baseline neutrino oscillation
experiments are presented here in three parts:
section~\ref{sec:resultsDom} describes the measurements made using the
dominant $\numu \rightarrow \nutau$ oscillation mode;
section~\ref{sec:resultsSubdom} details the recent detection of
sub-dominant $\numu \rightarrow \nue$ oscillations; and
section~\ref{sec:resultsNewPhysics} describes the results from
searches for new physics such as sterile neutrinos. Future
sensitivities are described in section~\ref{sec:future} and a
conclusion is given in section~\ref{sec:conclusion}.

\section{Neutrino Beams}
\label{sec:beams}
The accelerator neutrino beams used by the experiments covered in this
review article are described in this section. As in other areas of
particle physics, the experiments' detectors exist in a strongly
coupled relationship with the beam and it is important to consider
both beam and detector to understand the design and performance of the
experiments.

An interesting feature of neutrino beams is that multiple detectors
can be simultaneously exposed to the same individual beam spills with
no noticeable effect on the beam itself. This is true for Near and Far
detectors but also, for example, where there are multiple experiments
in the same underground laboratory.

An advantage of accelerator beams is the ability to exploit the pulsed
nature of the beams to reject backgrounds from cosmic rays and
atmospheric neutrinos. With beam pulses lasting tens of microseconds
and accelerator cycle times measured in seconds, a background
rejection factor of $10^5$ is typical.

The beams used in long-baseline experiments are described here in the
following order: section~\ref{sec:beamKEK} describes the beam used by
K2K; section~\ref{sec:beamNuMI} describes the NuMI beam used by MINOS
and in future \nova and MINOS+; section~\ref{sec:beamCNGS} describes
the CNGS beam used by OPERA and ICARUS; and
section~\ref{sec:beamJPARC} describes the J-PARC beam used by T2K.


\subsection{KEK Beam}
\label{sec:beamKEK}
\begin{figure}
\centering
\includegraphics[width=0.9\columnwidth]{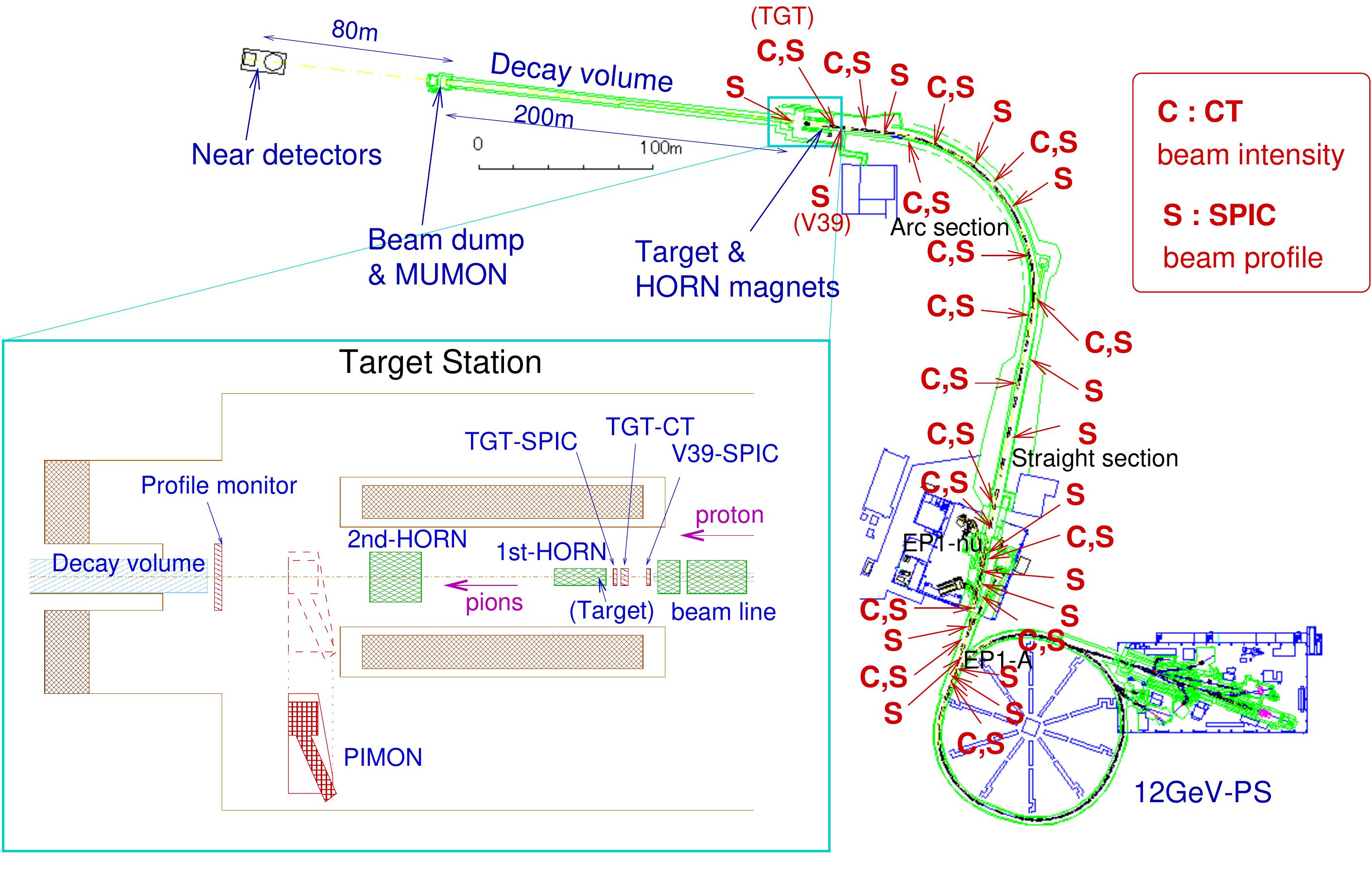}
\caption{A schematic of the K2K beamline that includes the primary
  proton beamline.}
\label{fig:beamK2K} 
\end{figure}

In this section, the beam for the first LBL experiment K2K in Japan
which was in operation from 1999 to 2004 is
described~\cite{Ahn:2006zza}. A schematic layout of the K2K beam line
is shown in Figure~\ref{fig:beamK2K}.
The beam of muon neutrinos was produced with the KEK 12~GeV proton
synchrotron~(PS) and was sent towards Super-Kamiokande, which is
located 250~km from KEK\@. The central axis of the neutrino beam was
aligned to aim at the center of Super-Kamiokande giving an on-axis
wideband beam.

The proton beam was extracted from the PS in a single turn with a 2.2~s
cycle time. The spill was 1.1~$\mu$s long and consisted of nine
bunches. The proton beam intensity reached about $6\times 10^{12}$
protons/pulse, corresponding to a beam power of about 5~kW\@.

\begin{figure}
\centering
\includegraphics[width=0.9\columnwidth]{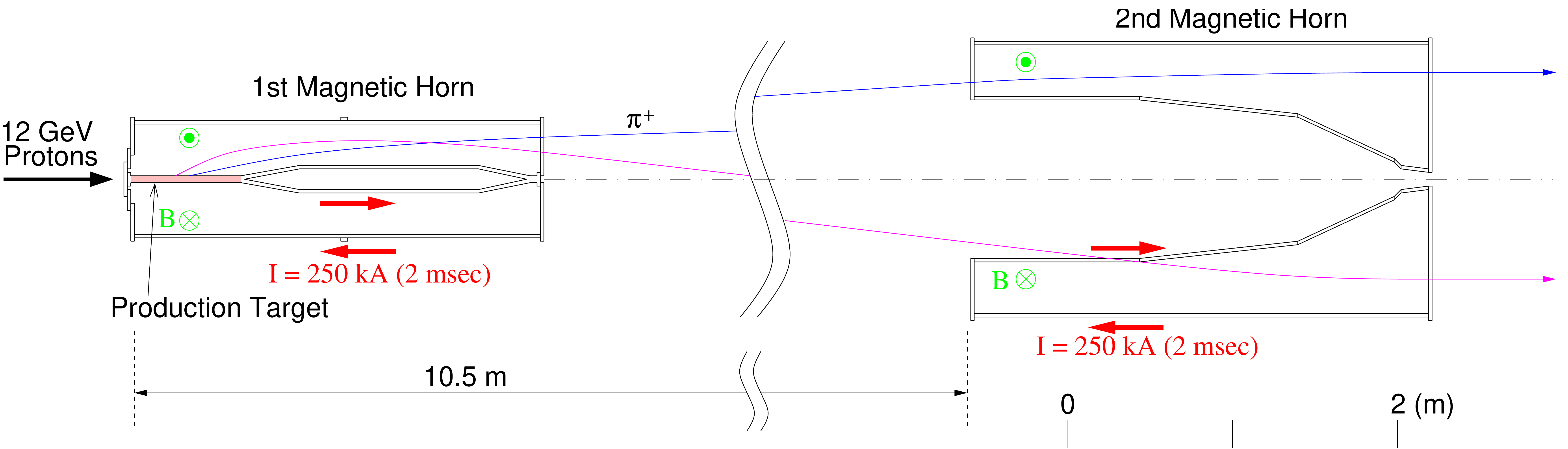}
\caption{A schematic showing the layout and operation of the K2K
  beamline target and horns.}
\label{fig:hornK2K} 
\end{figure}

Initially the target was a 66~cm long, 2~cm diameter Al rod but this
was replaced with a wider, 3~cm diameter rod in November
1999. Secondary positive pions were focused by two electromagnetic
horns~\cite{Yamanoi:2000hg}. Both horns had a pulsed current about
1~ms long with a 200~kA peak for the June 1999 run, and that was
increased to a 250~kA peak for runs after November 1999. The target
was embedded in the first horn and played a role as an inner conductor
as shown in Figure~\ref{fig:hornK2K}.

Measurements of the momenta and angular distribution of secondary
pions, $N(p_\pi, \theta_\pi)$, were made using the pion monitor. This
detector was a gas \Cer detector occasionally placed just downstream of
the second horn in the target station. The results of the pion monitor
measurements were used in calculations of the ratio of the flux at SK
to the flux at the near detector (ND), $R_\Phi (E_\nu) \equiv
\Phi_{\rm SK}(E_\nu)/\Phi_{\rm ND}(E_\nu)$.

The target region was followed by a 200~m long decay pipe where pions
decayed in flight to muon neutrinos and muons. At the downstream end
of the decay pipe, there was a beam dump made of iron 3~m thick and
followed by 2~m thick concrete. Muons above 5~GeV could penetrate the
beam dump and be detected by the muon monitors installed just behind
the beam dump. The muon monitors consisted of $2~{\rm m}\times 2~{\rm
  m}$ segmented ionization chambers along with an array of silicon pad
detectors and provided spill-by-spill monitoring of the beam profile
and intensity.

Beam line components were aligned with Global Positioning
System~(GPS)~\cite{Noumi:2004es}. The alignment uncertainty from the
GPS survey was $\lesssim 0.01$~mrad while that of the civil
construction was $\lesssim 0.1$~mrad, both of which were much better
than physics requirement of 1~mrad.

\begin{figure}
\centering
\includegraphics[width=0.4\columnwidth]{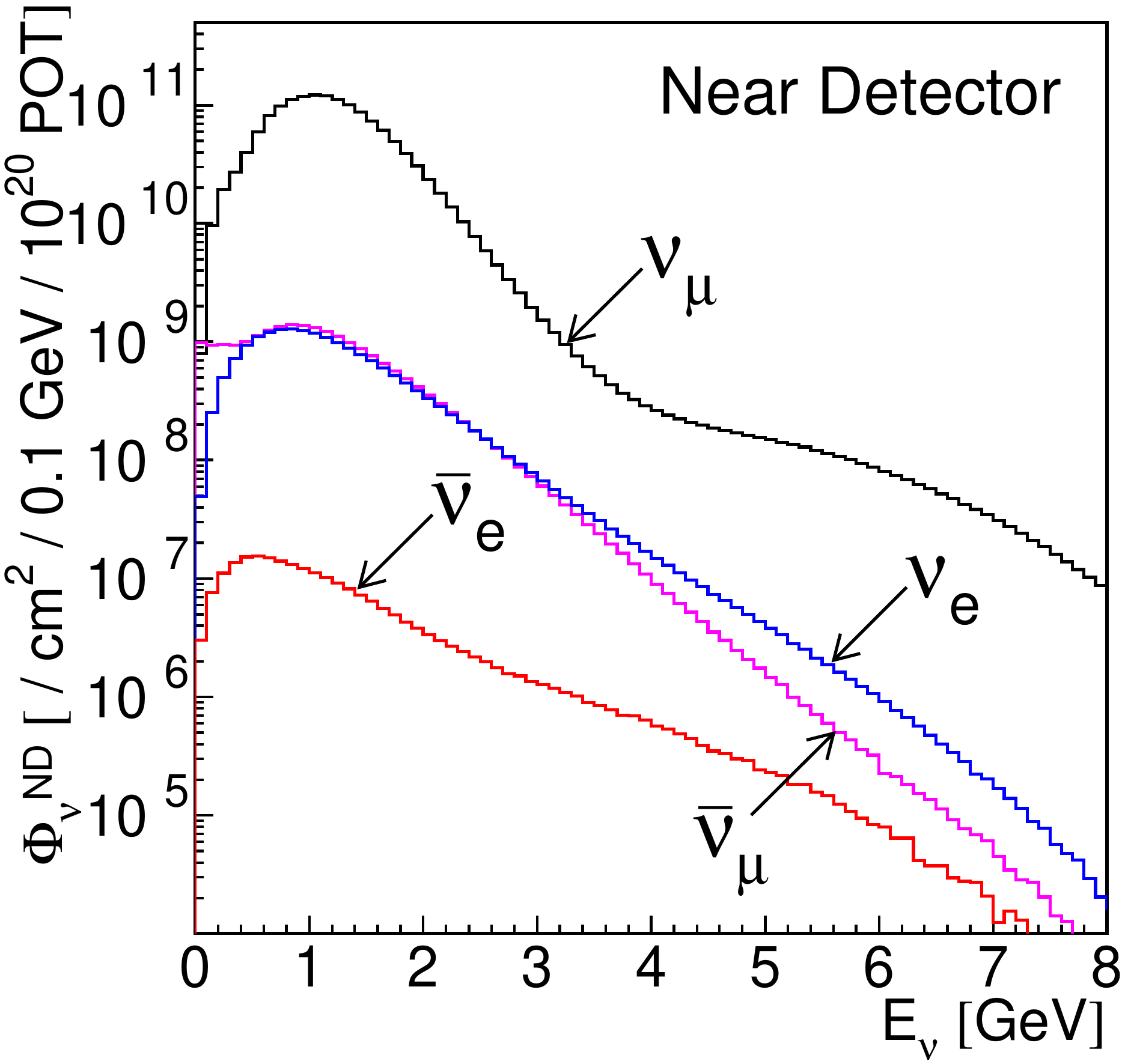}
\includegraphics[width=0.4\columnwidth]{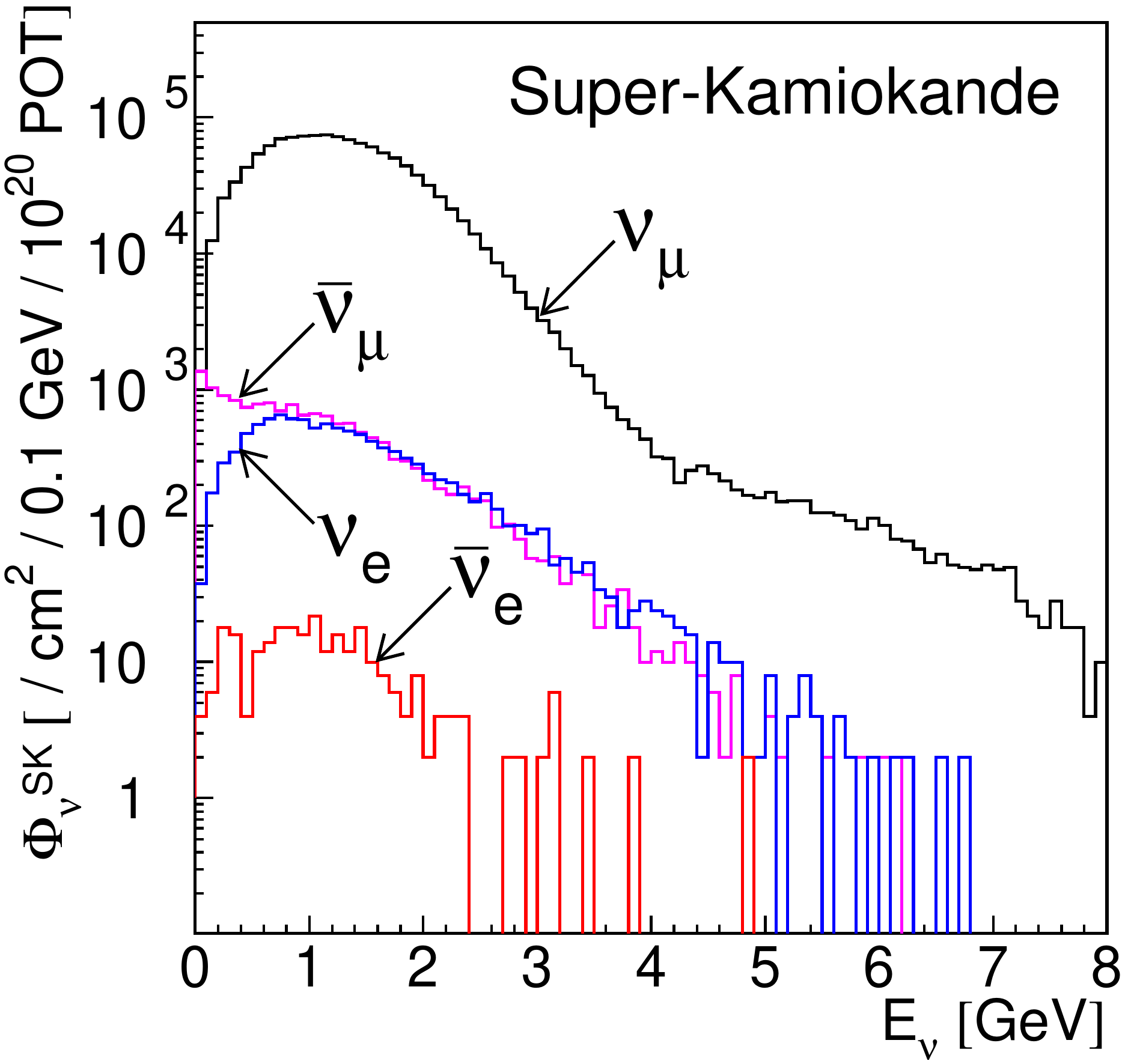}
\caption{Simulated neutrino flux of K2K beam.}
\label{fig:K2Kflux} 
\end{figure}

The expected neutrino spectra at SK are plotted in
Figure~\ref{fig:K2Kflux}. The average neutrino energy was 1.3~GeV and
the purity of $\numu$ in the beam was estimated by Monte Carlo~(MC)
simulation to be 98.2\% and $\nue$ contamination to be 1.3\%.

The K2K experiment started physics data taking in June 1999 and
finished in November 2004. The total number of protons on target~(POT)
delivered was $1.049\times 10^{20}$, of which $0.922\times
10^{20}$~POT were used in the final physics analysis.


\subsection{NuMI Beam}
\label{sec:beamNuMI}
The NuMI beam~\cite{Anderson:1998zza} is located at the Fermi National
Accelerator Laboratory in Illinois, USA, and it was initially
constructed primarily for the MINOS experiment. In this section a
description of NuMI as it was operated for the last 7 years is given
first, before going on to discuss the upgrades for the \nova
experiment that are underway at the time of writing. MINOS measured
the NuMI flux at distances of 1~km and 735~km from the target and
\nova will have the longest baseline of all such experiments at
810~km.

Protons from the Main Injector~(MI) accelerator with a momentum of
120~GeV/c are used for the production of neutrinos and antineutrinos
in the NuMI beamline. Typically, either 9 or 11 slip-stacked batches
of protons from the MI are extracted in a single-shot onto the NuMI
target giving neutrino pulses either 8~or~10~$\mu$s long. Filling the
MI with 8~GeV/c protons from the Booster accelerator takes about 0.7~s
and then acceleration to 120~GeV/c takes a further 1.5~s, giving a
total cycle time of about 2.2~s. A single-shot extraction from the MI
contains around $3\times10^{13}$ protons and the beam operated at a
power of 300--350~kW over the last few years. By the time of the
long-shutdown that started on 1st May 2012, NuMI had received nearly
$16\times10^{20}$ protons on target.

\begin{figure}
\centering
\includegraphics[trim = 0.1mm 0.1mm 0.1mm 0.1mm, clip, width=0.9\columnwidth]{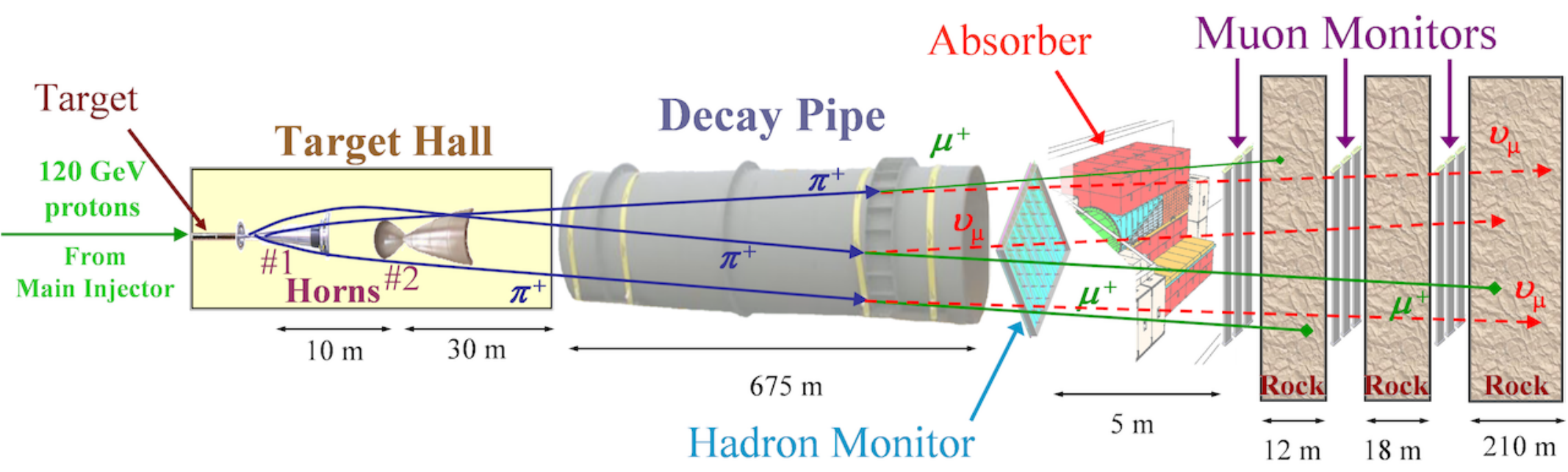}
\caption{A schematic of the NuMI beamline. Protons from the Main
  Injector strike a graphite target, shown at the far left, and the
  resulting negatively or positively charged hadrons are focused by
  two magnetic horns. A 675~m long decay pipe gives the short-lived
  hadrons and muons time to decay. All hadrons remaining at the end of
  the decay volume are stopped by the absorber leaving just muons and
  neutrinos. The remaining muons are stopped by nearly 250~m of
  rock. Figure from~\cite{Zwaska:2005be}.}
\label{fig:beamNuMI} 
\end{figure}

Figure~\ref{fig:beamNuMI} shows a schematic of the NuMI beamline and
the components are described in sequence, starting on the far left
with the protons coming from the MI\@. A water-cooled, segmented
graphite target 2.0~interaction lengths long is used to produce the
short-lived hadrons that give rise to the neutrinos. Two magnetic
horns focus either positively or negatively charged particles towards
a 675~m long decay volume, previously evacuated but now filled with
helium. At the end of the decay volume a hadron absorber stops any
remaining hadrons leaving just neutrinos and muons. Beyond that nearly
250~m of rock attenuates the muons leaving just the neutrinos.

The NuMI beamline was designed to be flexible in its operation with a
number of parameters that could be adjusted to optimise the
sensitivity to the physics topics of interest. The position of the
target with respect to the first horn, the position of the second
horn, the horn current and polarity could all be adjusted. The vast
majority of data were taken in a ``low energy'' configuration that
optimized the sensitivity to the atmospheric mass squared splitting by
providing as large a flux as possible at the oscillation maximum for
MINOS (around 1.4~GeV). This was achieved by inserting the target as
far into the first horn as safely possible and having the second horn
close to the first. A horn current of 185~kA was routinely
used. Approximately 80\% (20\%) of the data were taken with the horn
current polarity set to focus positively (negatively) charged hadrons
enhancing the production of neutrinos (antineutrinos). The energy
spectrum measured by MINOS is shown in the results section in
Figure~\ref{fig:enSpect}.

The neutrino flavor composition of the on-axis NuMI beam is as
follows: firstly, with the magnetic horn polarity set to focus
positive hadrons a neutrino-enhanced beam is produced, giving rise to
interactions in the (on-axis) MINOS near detector that are
91.7\%~$\numu$, 7.0\%~$\numubar$ and 1.3\%~$\nue+\nuebar$; secondly,
with the opposite polarity an antineutrino-enhanced beam is produced,
giving near detector interactions that are 40\%~$\numubar$,
58\%~$\numu$, 2\%~$\nue+\nuebar$~\cite{HimmelThesis}. However, it
should be noted that in the antineutrino-enhanced beam the $\numubar$
component comprises about 80\% of the interactions below 6~GeV in the
region where the oscillation effect is largest.

On a number of occasions and for relatively short periods the NuMI
beamline was operated in non-standard configurations. These special
runs were used to constrain uncertainties in analyses and better
understand the beam. Examples include: runs with the horn current at
170~kA, 200~kA and 0~kA; and runs with the target pulled back out of
the first horn by up to 2.5~m.

At the time of writing, the long accelerator shutdown to upgrade the
NuMI beam for \nova is underway. With the shutdown of the Tevatron,
two relatively straightforward changes will allow the NuMI beam power
to be doubled to 700~kW\@. Previously the Recycler, a fixed field ring
in the MI tunnel, was used to store antiprotons but now for \nova it
will accumulate protons from the Booster while the MI is ramping. By
parallelizing the accumulation and acceleration of protons for NuMI, and with
a small increase in the MI ramp rate,
the cycle time will be reduced from 2.2~s to 1.33~s. The second change
is that the number of batches in the MI ring will be increased from 11
to 12 and the two that were previously used to produce antiprotons
will now be used for NuMI.

\begin{figure}
\centering \includegraphics[trim = 0.1mm 0.1mm 0.1mm 0.1mm, clip,
  width=0.6\columnwidth]{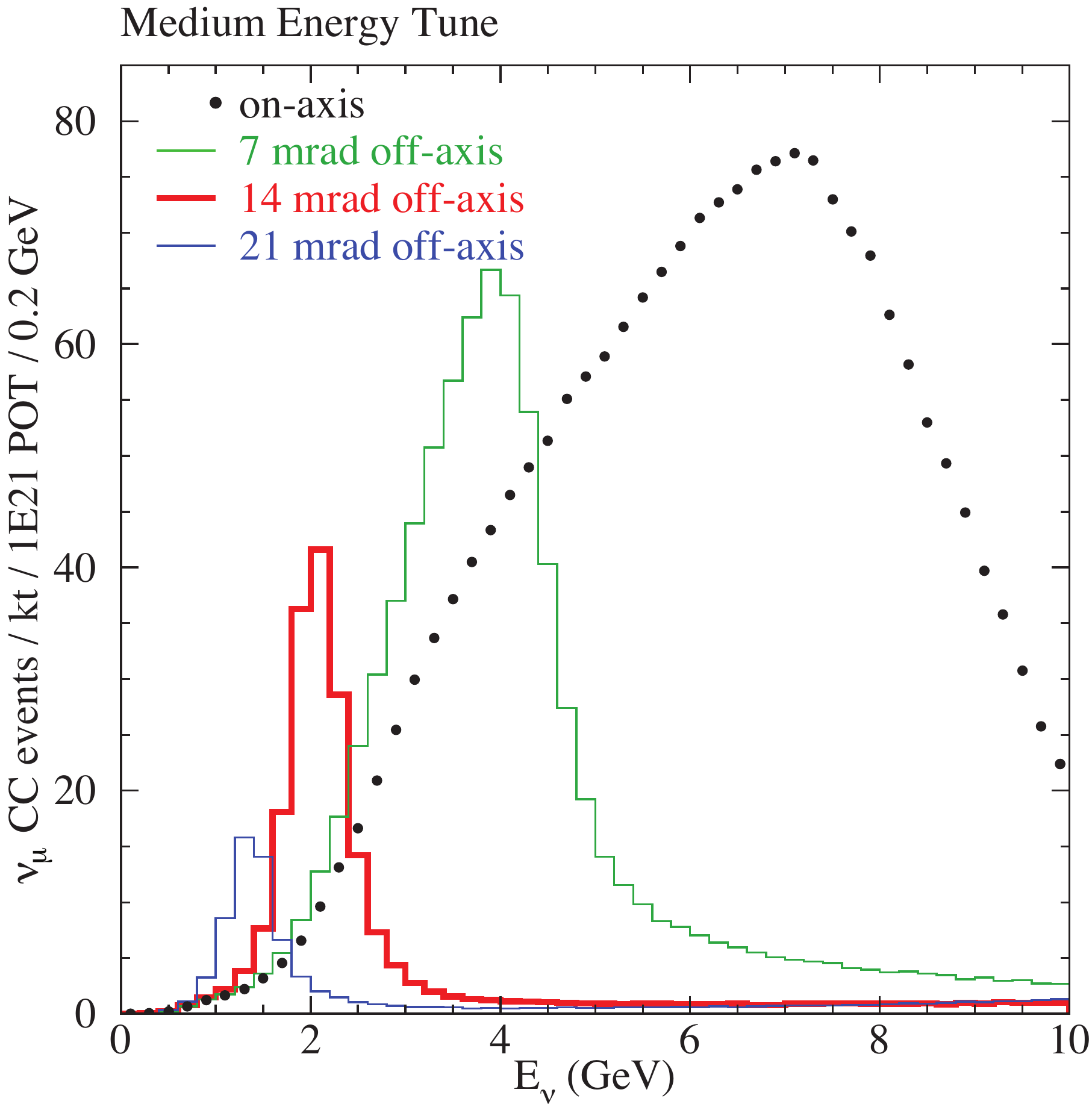}
\caption{The simulated NuMI energy spectrum as it will be in the
  \nova-era with the beamline in the ``medium energy''
  configuration. The \nova detectors will sit 14~mrad off-axis and the
  energy spectrum at that angle is shown by the red histogram. In
  contrast, the MINOS+ experiment will sit on-axis and so collect
  thousands of neutrino interactions per year that will be measured
  with an $L/E$ resolution at the 10\% level: the on-axis spectrum is
  shown by the black dots. The green and blue histograms further
  illustrate how the spectrum changes with the off-axis angle.}
\label{fig:spectraNuMI} 
\end{figure}

In addition to the upgrades to the accelerator for \nova,
modifications will also be made to the NuMI beamline. For the \nova
detectors the position of the peak in the energy spectrum will be
determined by the off-axis angle and so the flux will be optimized by
focusing the maximum number of pions into the decay pipe with energies
that allow a substantial fraction of them to decay within the 675~m
long decay volume. The optimal configuration of the NuMI beamline for
\nova will be to operate in a so-called ``medium energy'' configuration
with the target sitting a meter or so back from the first horn and
with the second horn positioned further downstream. This medium energy
beam will have a peak energy of around 7~GeV for the on-axis
experiments (e.g. MINOS+) compared to 1.9~GeV for \nova. The simulated
energy spectrum is shown in Figure~\ref{fig:spectraNuMI}. The \nova
detectors, sitting 14~mrad off-axis, will see a beam flux with
significantly higher purity than is obtained on-axis, having only
about 1\% $\numubar$ contamination of the $\numu$-enhanced beam and
about 5\% $\numu$ contamination of the $\numubar$-enhanced beam.

The target for the \nova era has been redesigned since there is no
longer the constraint that it should be placed inside the first horn
and increased reliability is expected. Beyond the upgrades underway
for \nova, there is the possibility of increasing the beam power
further; for example, the first phase of a proton driver could deliver
1.1~MW.


\subsection{CNGS Beam}
\label{sec:beamCNGS}
The CNGS beam~\cite{Acquistapace:1998rv,Meddahi:1051376} is located at
CERN on the border of Switzerland and France and the neutrinos are
measured by experiments at the Gran Sasso Laboratory in Italy, 730~km
away. CNGS uses 400~GeV/c protons from CERN's SPS accelerator that are
fast extracted in two 10.5~$\mu$s spills 50~ms apart every 6~s. Each
spill contains typically $2\times10^{13}$ protons to give an average
power of around 300~kW. The CNGS beam was commissioned in 2006 and the
total exposure is expected to reach $1.9\times10^{20}$ protons on
target by the end of the 2012 run.

The CNGS target assembly consists of a magazine containing 5 separate
targets, of which one is used at a time and the others are in-situ
spares. Each target consists of a series of thirteen graphite rods
10~cm long, the first two are 5~cm in diameter and the remainder are
4~cm. The magnetic focusing system consists of a horn and a reflector
that are pulsed at 150~kA and 180~kA respectively. An evacuated decay
volume 1000~m long and 2.5~m in diameter allows the short-lived
hadrons time to decay. At the end of the decay volume there is a
graphite and iron hadron stop. Beyond that, two detector stations
measure the remaining muons, which are used to derive the intensity and
profile of the neutrino beam.

The CNGS beam is operated in a neutrino-enhanced mode and provides a
high purity $\numu$ source with $\numubar$-contamination of 2\% and
$\nue$+$\nuebar$-contamination of less than 1\%. The number of prompt
$\nutau$ in the beam is negligible~\cite{Agafonova:2010dc}.

At the time of writing, no formal proposal for running the CNGS beam
beyond the long LHC-shutdown in 2013 has been made by OPERA or other
Gran Sasso experiments.


\subsection{T2K Beam}
\label{sec:beamJPARC}
The neutrino beam for the Tokai-to-Kamioka (T2K) experiment is
produced at the Japan Proton Accelerator Research Complex (J-PARC) and
measured by both near detectors locally and by Super-Kamiokande,
295~km from J-PARC\@. The T2K beam is an off-axis narrow band
beam. Details of the experimental apparatus for T2K including the
beamline are described in~\cite{Abe:2011ks}.

J-PARC is a high intensity proton accelerator complex located in Tokai
village, Japan, whose construction was completed in 2009. The accelerator
chain consists of a 181~MeV LINAC, 3~GeV
Rapid Cycling Synchrotron and a 30~GeV Main Ring~(MR). The design beam
power of the MR is 750~kW. The proton beam used to produce the
neutrino beam is extracted from MR in a single turn (fast extraction)
with repetition cycle of 3.52~s at the beginning of operation in 2010
and 2.56~s now in 2012. The beam pulse of the single extraction
consist of 8~bunches, 580~ns apart, making the pulse about 5~$\mu$s
long. The beam power achieved for stable operation as of summer 2012
was 200~kW which corresponds to $1.1\times 10^{14}$~protons/pulse~(ppp)
or $1.3\times 10^{13}$~protons/bunch~(ppb).

\begin{figure}
\centering
\includegraphics[width=0.6\columnwidth]{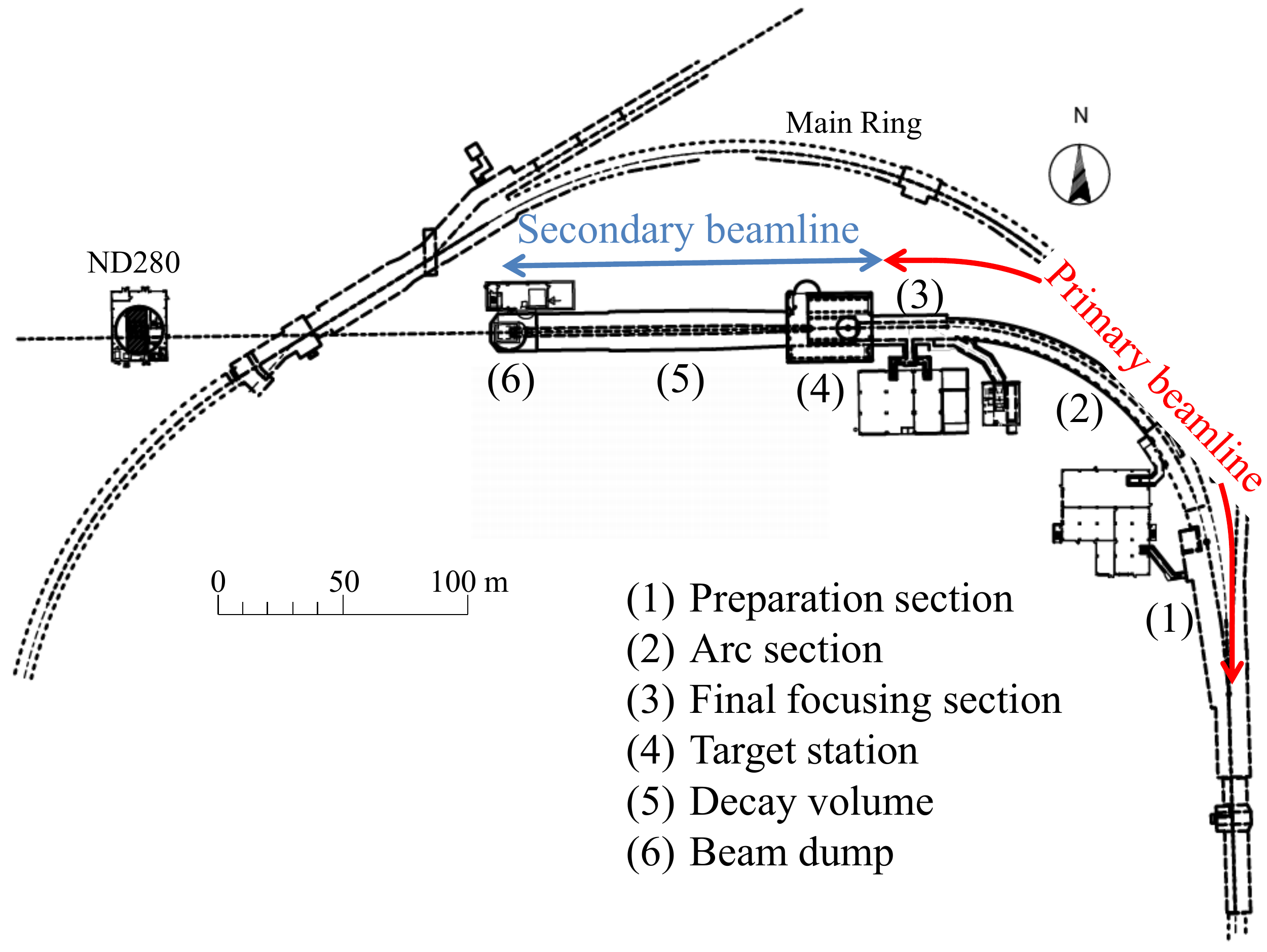}
\caption{A schematic of the overall T2K beam facility, showing the
  primary and secondary beamlines plus the location of the ND280
  detector complex.}
\label{fig:T2K-beamline} 
\end{figure}

The layout of the neutrino beam facility at J-PARC is illustrated in
Figure~\ref{fig:T2K-beamline}.
The extracted beam from MR is bent  by about 90$^\circ$ to point
in the Kamioka direction using 28 superconducting combined function
magnets~\cite{Ogitsu:2004ad,Nakamoto:2004nj,Ogitsu:2010zz} and
delivered to the production target.

\begin{figure}
\centering
\includegraphics[width=0.6\columnwidth]{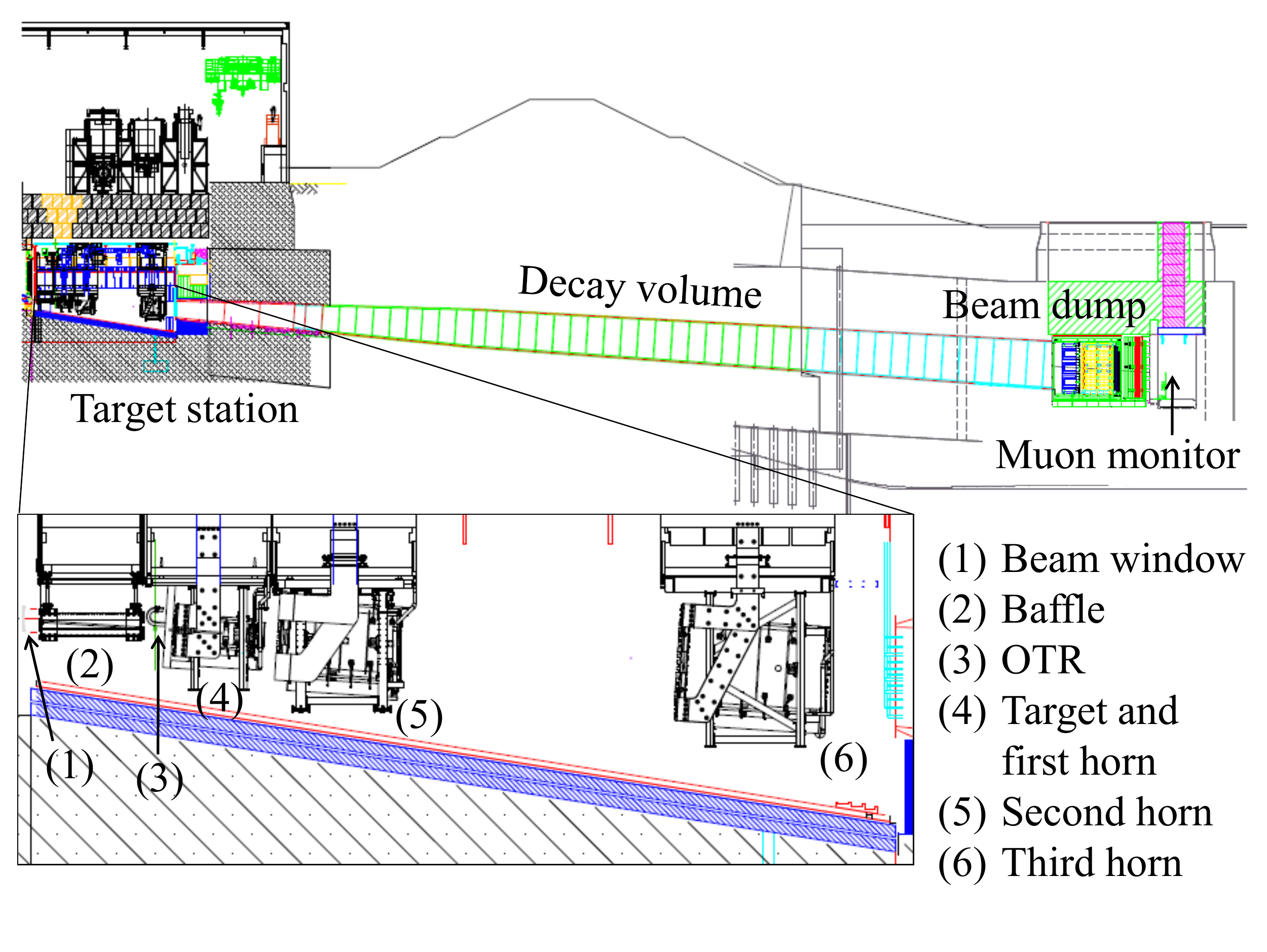}
\caption{Schematic of T2K secondary beam line.}
\label{fig:T2K-secondarybeam} 
\end{figure}

The secondary beamline where the neutrinos are produced is shown in
Figure~\ref{fig:T2K-secondarybeam}.
The production target is a 26~mm diameter and 90~cm long graphite rod,
corresponding to 2~interaction lengths, in which about 80\% of
incoming protons interact. The secondary positive pions (and kaons)
from the target are focused by three electromagnetic horns operated at
a 250~kA pulsed current.

The target region is followed by a 110~m long decay volume filled with
helium gas in which pions and kaons decay in flight into
neutrinos. The beam dump, which consists of graphite blocks about
3.15~m thick followed by iron plates 2.5~m thick in total, is placed
at the downstream end of the decay volume.

Muon monitors (MUMON) are placed just behind the beam dump to monitor
the intensity and the profile of muons which pass through the beam
dump on a spill-by-spill basis. High energy muons of $>5$~GeV can
penetrate the beam dump and reach the MUMONs.

The design principle of the J-PARC neutrino facility is that all parts
which can never be replaced later, for example, the decay volume shielding
and cooling pipes, beam dump cooling capacity, etc, are built such
that they can be operated with up to 3~MW of beam power from the
beginning. Parts that can be replaced are designed to be operated with
a beam power up to 750~kW and have a safety factor of 2 to 3.

\begin{figure}
\centering
\includegraphics[width=0.6\columnwidth]{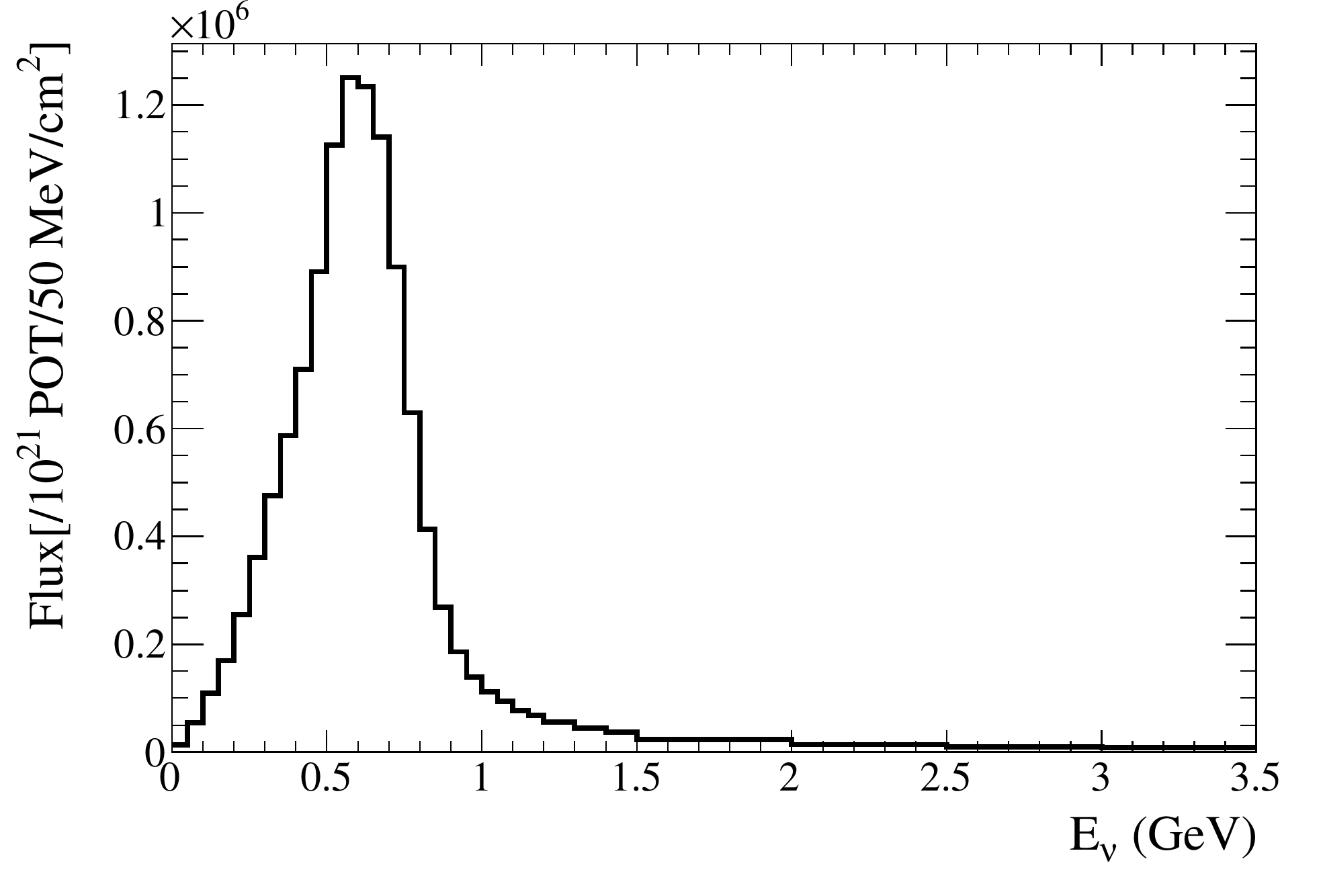}
\caption{The unoscillated $\numu$ flux at Super-Kamiokande with an
  off-axis angle of 2.5$^{\circ}$ and operation of the electromagnetic
  horns at 250~kA.}
\label{fig:T2K-flux}
\end{figure}

The neutrino beamline is designed so that the neutrino energy spectrum
at Super-Kamiokande can be tuned by changing the off-axis angle down
to a minimum of $2.0^{\circ}$ from the current (maximum) angle of
2.5$^{\circ}$. The unoscillated $\numu$ energy spectrum at
Super-Kamiokande with a 2.5$^\circ$ off-axis angle is shown in
Figure~\ref{fig:T2K-flux}.

The construction of the neutrino facility started in 2004 and was
completed in 2009. Stable beam production for physics measurements
started in January 2010 after careful commissioning.  The Great East
Japan Earthquake on March 11, 2011 damaged J-PARC and stopped the
operation of the accelerators. After recovery work, the accelerator
restarted operation in December 2011 and stable beam for T2K data
taking was achieved in March 2012.

The J-PARC neutrino facility will provide an integrated number of
protons on target of $7.5\times 10^{21}$ (equivalent to $750~{\rm
  kW}\times 5\times 10^7$s), which is the approved exposure for
T2K. With the present power upgrade scenario, this will take about 10
years.


\section{Detectors}
\label{sec:detectors}
In this section the detectors used by the experiments to achieve their
diverse physics goals are described. Design of these detectors took
into account multiple factors such as target mass, cost-effectiveness,
particle flavor identification purity and efficiency, the beam energy
spectrum and required baseline. The subsections below are time ordered
and include K2K Near detectors and Super-Kamiokande (SK), MINOS, OPERA,
ICARUS, the T2K ND280 complex and \nova.


\subsection{K2K Near Detectors}
The K2K Near detector complex was located at the KEK laboratory in
Japan. The detectors were about 300~m from the beam-target, about 70~m
of which was taken up with earth shielding. The detectors were
designed to measure the flux and energy spectrum of the beam as it
leaves KEK\@. Their mass composition was chosen to be primarily water
so as to largely cancel common systematic uncertainties with
Super-Kamiokande. These goals were achieved using a 1~kiloton water
\Cer detector (the ``1~kt'') and fine-grained detectors~(FGD). A
scintillating fiber detector~(SciFi)~\cite{Suzuki:2000nj},
scintillating counters, a lead glass array~(LG) and a muon range
detector~(MRD)~\cite{Ishii:2001sj} comprised the FGDs. For the second
phase of K2K, the LG was replaced by the fully active scintillator-bar
detector~(SciBar)~\cite{Nitta:2004nt}.

The 1~kt used the same technology as the Super-Kamiokande far detector
with the same arrangement of photomultiplier tubes and the same
40\%~coverage. In total, 680 50 cm photomultiplier tubes were used to
line an 8.6~m diameter, 8.6~m high cylinder.

The SciFi tracking detector used 20 layers of scintillating fibers, closely packed
together in $2.6~{\rm m}\times2.6$~m sheets that were separated by
9~cm. These layers were interleaved with 19 layers of water target
contained in extruded aluminum boxes and read
out using image-intensifier tubes and CCD cameras. The energy and
angle of the muons produced in $\numu$~CC interactions were measured
using the MRD\@. This detector was designed to be big enough ($7.6~{\rm
  m} \times 7.6$~m in the plane transverse to the beam) to measure
both the flux and the profile of the beam. The MRD consisted of
12~layers of iron absorber with vertical and horizontal drift tubes
in between. The first 4~(upstream) layers were 10~cm thick and the
remaining 8~layers were 20~cm thick. With 2.00~m of iron in total, up
to 2.8~GeV/c muons could be stopped and their total energy measured.

The SciBar detector was an upgrade to the near detectors designed with
the aim of improving the measurement of CC quasi-elastic interactions
and was installed in 2003. It was designed with the requirement of
high purity and efficiency, with the suppression of inelastic CC
interactions involving pions in the final state one of the main
goals. The detector was ``totally active'' and could measure $dE/dx$ for
individual particles such as protons and pions. The SciBar detector consisted of
14,848 extruded scintillator strips (of dimension
$1.3\times2.5\times300$~cm$^3$) packed tightly together to make up the
tracker part of the detector.  On the downstream side of the tracker
was an electromagnetic calorimeter, 11~radiation lengths thick and made
of scintillating fibres \& lead foils, called the Electron
Catcher. This calorimeter was used to aid the measurement of electron
showers and $\pi^0$ produced by neutrino interactions.


\subsection{Super-Kamiokande Detector}
The Super-Kamiokande detector~\cite{Fukuda:2002uc} is the world's
largest land-based water \Cer detector with a total mass of
50~kilotonnes. SK is a 39~m diameter and 41~m high stainless steel
cylindrical tank filled with ultra pure water that is located 1~km
underneath Mt. Ikenoyama in Japan.  The water tank is optically
separated into a 33.8~m diameter and 36.2~m high cylindrically-shaped
inner detector (ID) and outer detector (OD) by opaque black sheets and
Tyvek sheets attached to a supporting structure.  There are 11,129
inward-facing 50~cm diameter photomultiplier tubes (PMTs) lining the
ID giving 40\% coverage, and 1885 outward facing 20~cm diameter PMTs
on the inner wall of the OD\@. The ID and OD are optically separated
to allow interactions produced within the ID to be distinguished from
those entering from outside (e.g. cosmic rays).

A key feature of SK is the ability to separate $\numu$~CC events from
$\nue$~CC by identifying the electron or muon. The muons, being
heavier, produce sharper \Cer cones whereas electrons scatter more
easily and the resulting ``fuzzy'' \Cer cone is effectively the sum of
multiple overlapping cones all pointing in slightly different
directions. The vertex for each interaction is reconstructed using the
timing from all the hit PMTs and used to define the fiducial volume of
22.5~kilotonnes.


\subsection{MINOS Detectors}
\begin{figure*}[htpb]
\begin{center}
\includegraphics[trim = 0.1mm 0.1mm 0.1mm 0.1mm, clip,width=0.45\textwidth,keepaspectratio=true]{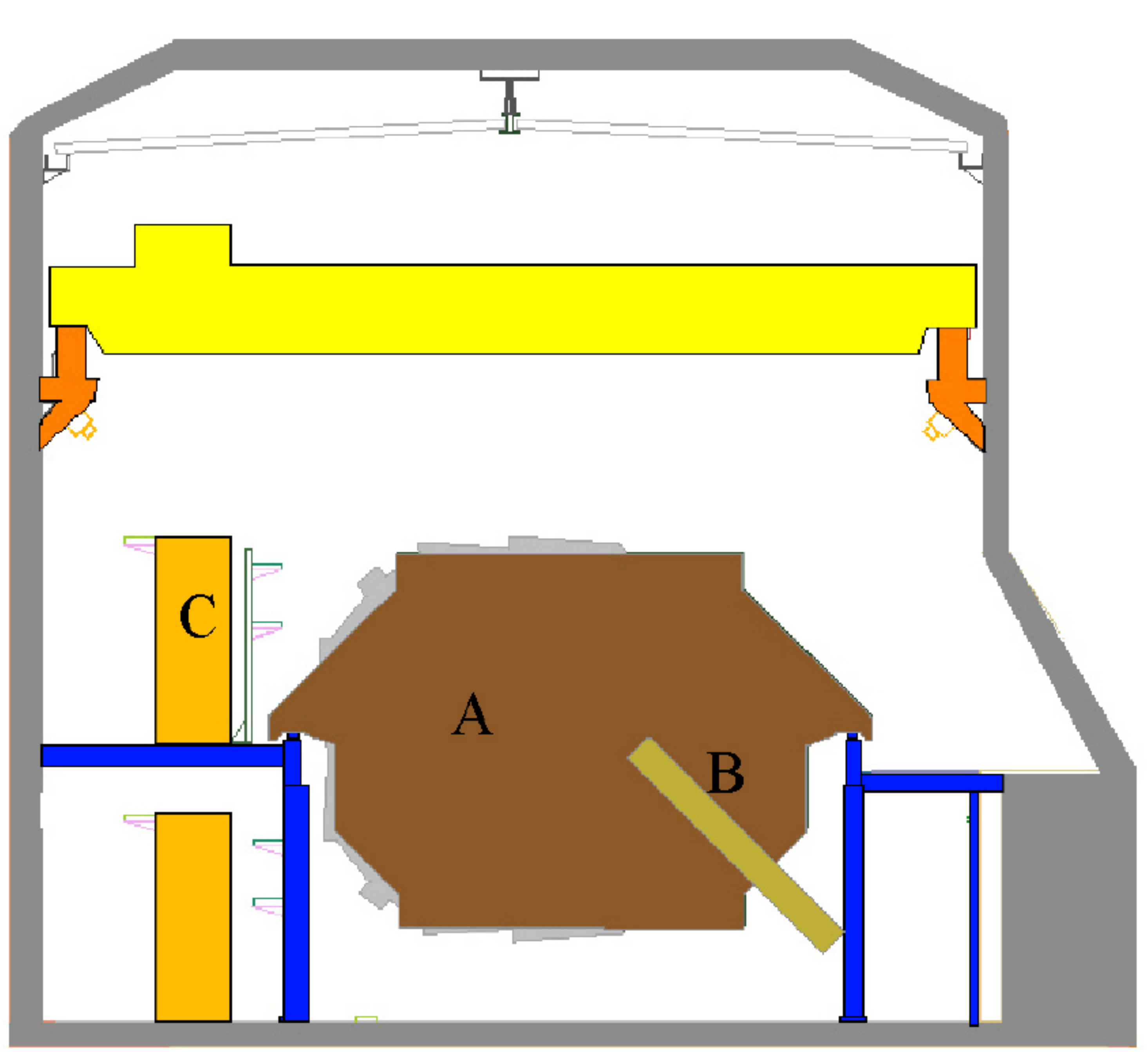}
\hspace{0.05\textwidth}
\includegraphics[trim = 0.1mm 0.1mm 0.1mm 0.1mm, clip,width=0.45\textwidth,keepaspectratio=true]{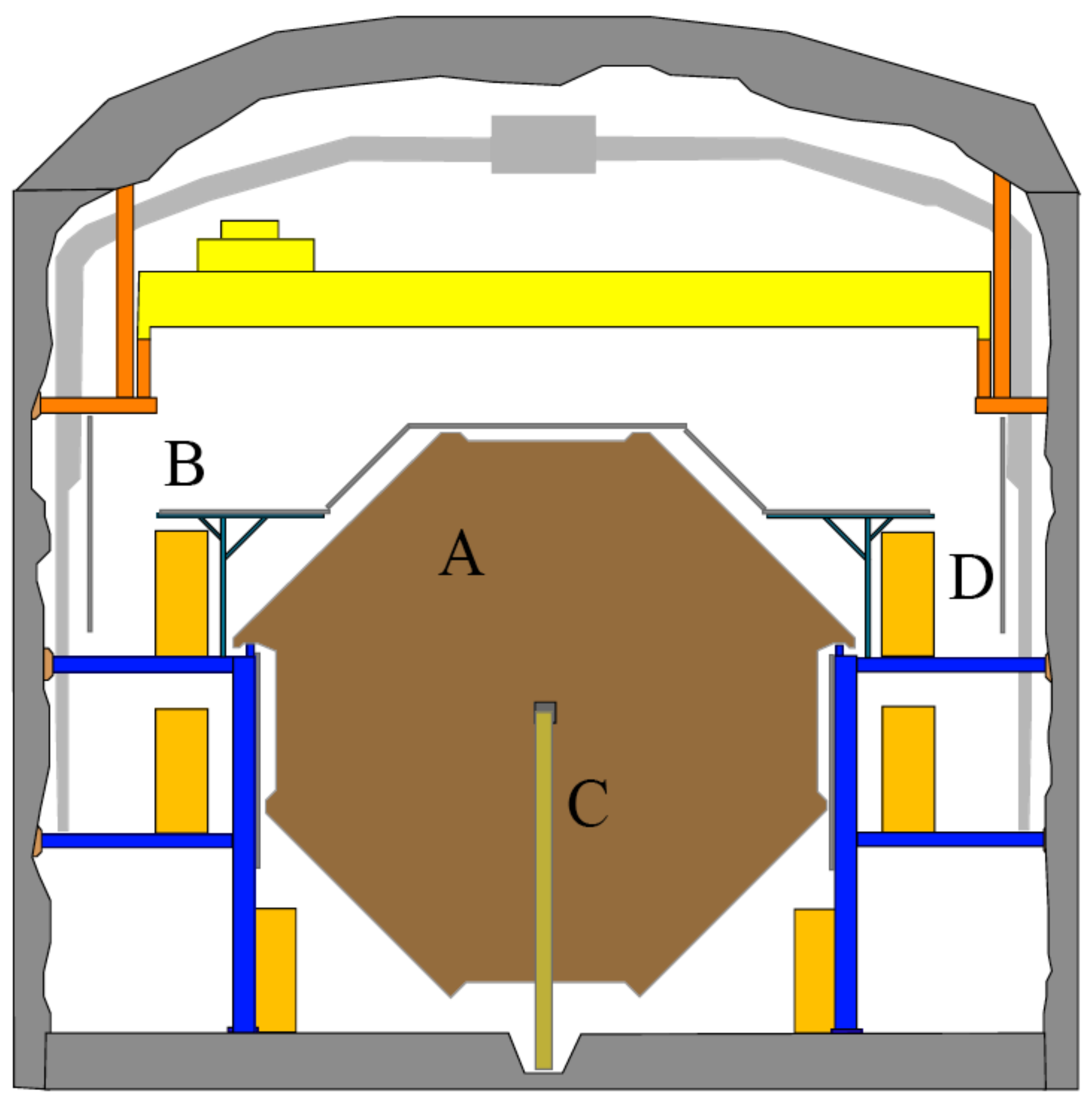}
\caption{Schematics showing the end views of the MINOS Near (left) and
  Far (right) detectors. For the Near detector the label `A'
  identifies the upstream steel plate, `B' is the magnet coil and `C'
  is an electronic rack. For the Far detector, `A' identifies the
  steel plane at the end of the second supermodule, the furthest
  downstream in the beam, `B' is the cosmic ray veto shield, `C' is a
  magnet coil and `D' is an electronics rack. The detectors are shown
  with different scales: the Near detector is 3~m wide compared to 8~m
  for the Far detector. Figure from~\cite{Michael:2008bc}.}
\label{fig:MINOSdetectors}
\end{center}
\end{figure*}

The MINOS detectors~\cite{Michael:2008bc} are magnetized tracking
calorimeters made of steel and plastic scintillator optimized for
measurements of muon neutrinos and antineutrinos with energies of a
few-GeV\@. The Near Detector at Fermilab has a mass of 0.98~kilotonnes
and the Far Detector at the Soudan Underground Laboratory in
Minnesota, USA has a mass of 5.4~kilotonnes. The detectors have a
planar geometry with the active medium comprised of solid plastic
scintillator strips with neighboring planes having their strips
orientated in perpendicular directions to give three dimensional
tracking capability. The planes are hung vertically so as to be
approximately perpendicular to the path of the beam neutrinos. In the
detectors' fiducial volumes 80\% of the target mass is provided by
steel planes and they are magnetized to provide average fields of
1.28~T and 1.42~T for the Near and Far detectors respectively.\ The
steel planes are 2.54~cm thick (1.45 radiation lengths) and mounted on
each one is, at most, a single 1.0~cm thick scintillator plane. Each
scintillator plane comprises of up to 192 strips that are 4.1~cm wide
and up to 8~m in length. There is an air gap between each plane of
2.4~cm in which the magnetic field is substantially smaller. A
schematic of the Near and Far detectors is shown in
Figure~\ref{fig:MINOSdetectors}. The Far Detector planes are an 8~m
wide octagonal shape and grouped together into two separately
magnetized supermodules that are about 15~m in length. The Near
Detector planes have a squashed octagon shape that is about 3~m wide
and 2~m high. The Near Detector has two main parts: a fully
instrumented region used for calorimetry and a muon spectrometer that
is located downstream in the neutrino beam.

MINOS scintillator is made of polystyrene, doped with the fluors PPO
(1\%) and POPOP (0.03\%), which is co-extruded with a thin 0.25~mm
TiO$_2$ layer. A groove runs along the length of each strip into which
a 1.2~mm wavelength-shifting (WLS) fibre optic cable is glued. On exiting
the ends of the strips, the WLS fibers run together in a manifold to
terminate in a connector. Clear fibre optic cables, with a longer 12~m
attenuation length, are used to route the light to multi-anode
photomultiplier tubes.

The Near and Far detectors were designed to be as similar as possible,
although due to their different environments it was necessary to use
different front-end electronics. On average, several neutrino
interactions occur in the Near detector in every beam spill, whereas in
the Far detector only a handful of neutrino interactions occur per
day. The Near detector electronics digitizes the signal from each PMT
pixel continuously during each beam spill at the frequency of the beam
RF structure of 53.103~MHz. In contrast, the Far detector electronics
has a dead time of at least 5~$\mu$s after each PMT dynode
trigger. The Far detector self-triggers with high efficiency on
neutrino interactions. In addition, the beam spill time is
sent over the internet and used to record all detector activity in a
100~$\mu$s window around the beam spill. Both Near and Far detectors
also record cosmic ray events, and at the Far detector, atmospheric
neutrino events can be selected.

Neutrino energy reconstruction in MINOS involved both calorimetry of
showers (although later analyses also used topological information to
improve shower energy resolution) and either range or curvature of
muon tracks. The calorimetric energy resolution of the MINOS detectors
was determined to be 21.4\%$/\sqrt{E} \oplus 4\%/E$ for
electromagnetic showers and $56\%/\sqrt{E} \oplus 2\%$ for hadronic
showers. The accuracy of the simulation of protons, pions, electrons
and muons was determined using a specially constructed calibration detector
that was exposed to CERN test-beams~\cite{Adamson:2006xv}. The
test-beam data was also used to demonstrate that differences in the
Near and Far detector readout systems could be corrected for by the
calibration and the detector simulation~\cite{Cabrera:2009fi} down to
the 1\% level.

In the Far detector the optimal fiducial volume of 4.2~kilotonnes
included as many events as possible to reduce the statistical
uncertainty on the oscillation parameters. Whereas in the Near
detector, with millions of events, the fiducial volume was optimized to
make the best possible measurement of the neutrino energy spectrum and
had a mass of 23.7~tonnes.


\subsection{OPERA Detector}
The OPERA detector is located 1400~m underground in Hall~C at the Gran
Sasso Laboratory, Italy and is optimized to enable a high purity
selection of tau neutrino interactions on an individual event basis. A
key signature of a $\nutau$ event is the topology of the tau
decay. Substantial energy is carried away by the $\nutau$ produced in
tau decay and due to the large tau mass the effect of missing
transverse momentum often gives rise to a substantial change in
direction (or ``kink'') at the point along a track where the tau
decays. With a mean lifetime of 0.29 picoseconds, corresponding to
87~$\mu$m at the speed of light, directly observing the tau in a
necessarily massive detector is an experimental challenge.

The detector used by the OPERA collaboration is a hybrid consisting of
a target constructed of fine grained emulsion and electronic
detectors. Neutrino events are localized in the target using the
scintillator target tracker (TT) detector and a spectrometer is used
to measure the momentum and charge of muons. The target is divided
into two supermodules with veto planes upstream. Each target region
contains 75\,000 emulsion cloud chambers (ECC), or ``bricks'', which
are constructed from 56 lead plates 1~mm thick that are interleaved
with 57 nuclear emulsion films. Each ECC weighs 8.3~kg for a total
target mass of around 1.25~kilotonnes. An automated system is used to
extract the bricks identified by the TT from the detector. Scanning of
the emulsion films is performed by automated microscopes located on
the surface in Europe and Japan.

\subsection{ICARUS Detector}
The ICARUS T600 detector~\cite{Rubbia:2011ft} is located in Hall~B of
the Gran Sasso Laboratory, Italy and consists of 760~tonnes of
ultra-pure liquid argon (LAr) held at 89~K\@. The argon provides the
target mass and the ionization medium for four time projection
chambers (TPCs). These four TPCs come in two pairs, with each pair
occupying a volume of $3.6\times3.9\times19.6$~m$^3$. A shared cathode
plane runs down the centre of each volume separating the two TPCs, 
giving a maximum drift path of 1.5~m. This detector provides exquisite
electronic imaging of neutrino interactions in three dimensions with a
position resolution of around 1~mm$^3$ over the whole detector active
volume of about 170~m$^3$.

An electric field of 500~V/cm is used to drift ionization electrons
towards three parallel planes of wires arranged at 0$^\circ$,
+60$^\circ$ and -60$^\circ$ to the horizontal. These planes are situated
along one side of each TPC and are separated by 3~mm. In total there
are 53248 wires that have a pitch of 3~mm and lengths up to 9~m
long. The first two planes (Induction-1 and Induction-2) provide
signals in a non-destructive way before the charge is finally
integrated on the Collection plane. Position information along the
drift direction is provided by combining measurement of the absolute
time of the ionising event with knowledge of the drift velocity (about
1.6~mm/$\mu$s at the nominal electric field
strength). VUV~scintillation light from the liquid argon, measured by
PMTs operating at cryogenic temperatures, provides the absolute timing
information.

Electronegative impurities such as O$_2$, CO$_2$ and H$_2$O were
initially reduced by evacuating the detector for 3~months before
filling and are generally maintained at below the 0.1~ppb level by
recirculating the LAr through purification systems. Full volume
recirculation can be accomplished in 6 days. A free electron lifetime
of 1~ms corresponds to a 1.5~m drift distance and this has been
successfully maintained for the vast majority of the time since the
detector started operation in mid-2010.


\subsection{T2K ND280 Detectors}
\label{sec:detectorsT2KND280}
The ND280 detector complex is located on the site of the J-PARC
accelerator complex about 280~m downstream of the production
target. The T2K experiment is formed of the ND280 detectors, the
beamline and Super-Kamiokande. The ND280 detectors measure the
neutrino energy spectrum and flavor content of the beam before it
oscillates. Since the far detector is located $2.5^\circ$ off-axis,
the primary near detector is also located off-axis at the same
angle. An on-axis near detector, INGRID, measures the neutrino beam
profile and intensity.

\begin{figure}
\centering
\includegraphics[trim = 0.1mm 0.1mm 0.1mm 0.1mm, clip, width=0.6\columnwidth]{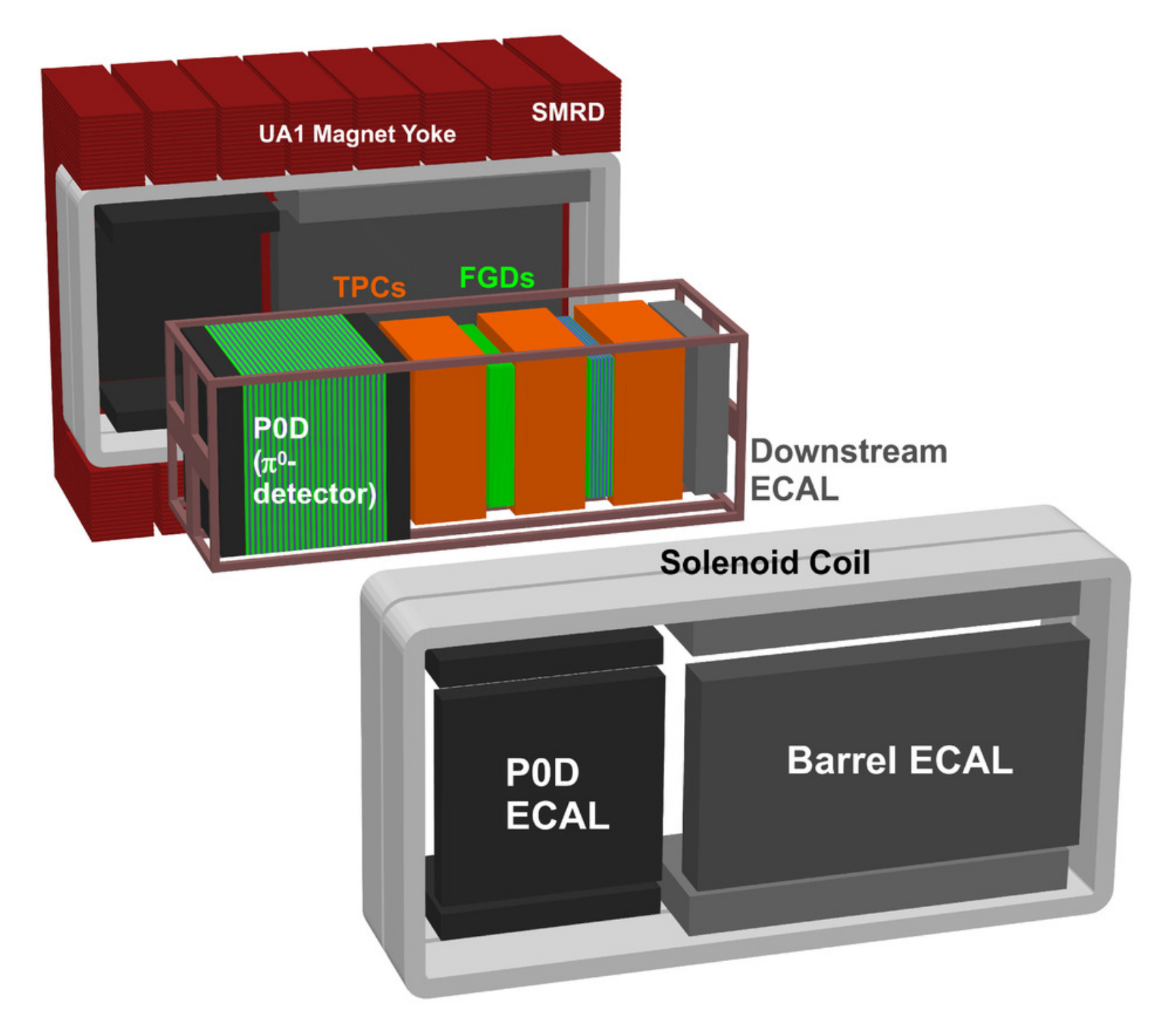}
\caption{An exploded view of the ND280 off-axis near detector for the
  T2K experiment. The ND280 is a magnetized tracking detector
  comprising of several sub-detectors located inside the UA1 magnet
  (see the main body of text for detailed descriptions). Figure
  from~\cite{Abe:2011ks}.}
\label{fig:nd280} 
\end{figure}

The off-axis near detector is a magnetized tracking detector
comprising of several sub-detectors located within the magnet recycled
from the UA1 experiment at CERN\@. Figure~\ref{fig:nd280} shows an
exploded view of the off-axis ND280 detector displaying the  $\pi^0$
detector~(P0D), the tracker comprising of fine-grained
detectors~(FGDs) and time projection chambers~(TPCs), the
electromagnetic calorimeter~(ECal), and side muon range
detector~(SMRD).
The P0D consists of scintillating bars alternating with either a water
target or brass or lead foil (to limit the range of any $\pi^0$s). The
FGDs consist of layers of finely segmented scintillator bars used to
measure charged current interactions. These inner detectors are all surrounded by
the ECal to catch any $\gamma$-rays that do not convert in the inner
detectors. Finally, the SMRD sits in the return yoke of the magnet and
measures the range of muons that exit the sides of the detector.

The on-axis INGRID detector consists of 14 identical modules arranged
in a cross pattern with two groups: extending 10~m along the
horizontal and vertical axes. A further two modules are located at
off-axis positions a few meters above the horizontal and to each side
of the vertical part of the cross. Each module is constructed from 9
steel plates 6.5~cm thick interleaved with 11 tracking scintillator
planes. The planes consist of two sets of 24 scintillator bars
measuring $1.0\times5.0\times120.3$~cm$^3$, one set arranged to run
vertically and the other horizontally. INGRID measures the center of
the beam to a precision of 10~cm, equivalent to 0.4~mrad.


\subsection{\nova Detectors}
\label{sec:detectorsNOvA}
The \nova~\cite{Ayres:2007tu} far detector will be located 14 mr off
the NuMI beam axis, 810 km from the NuMI target, off the Ash River
Trail in northern Minnesota, USA. The Ash River Trail is the most
northern road in the United States near the NuMI beam line.  The \nova
near detector will be located on the Fermilab site about 1 km from the
NuMI target, also at an angle of 14 mr to NuMI beam.

The \nova detectors can be described as totally active, tracking,
liquid scintillator calorimeters.  The basic cell of the far detector
is a column or row of liquid scintillator with approximate transverse
dimensions 4 cm by 15.6 m and longitudinal dimension 6 cm encased in a
highly-reflective polyvinyl chloride (PVC) container. A module of 32
cells is constructed from two 16-cell PVC extrusions glued together
and fitted with appropriate end pieces. Twelve modules make up a
plane, and the planes alternate in having their long dimension
horizontal and vertical.  The far detector will consist of a minimum
of 928 planes, corresponding to a mass of approximately 14 kt.
Additional planes are possible depending on available funds at the end
of the project. Each plane corresponds to 0.15~radiation lengths.

\begin{figure}[htb]
\centering
\includegraphics[height=8cm]{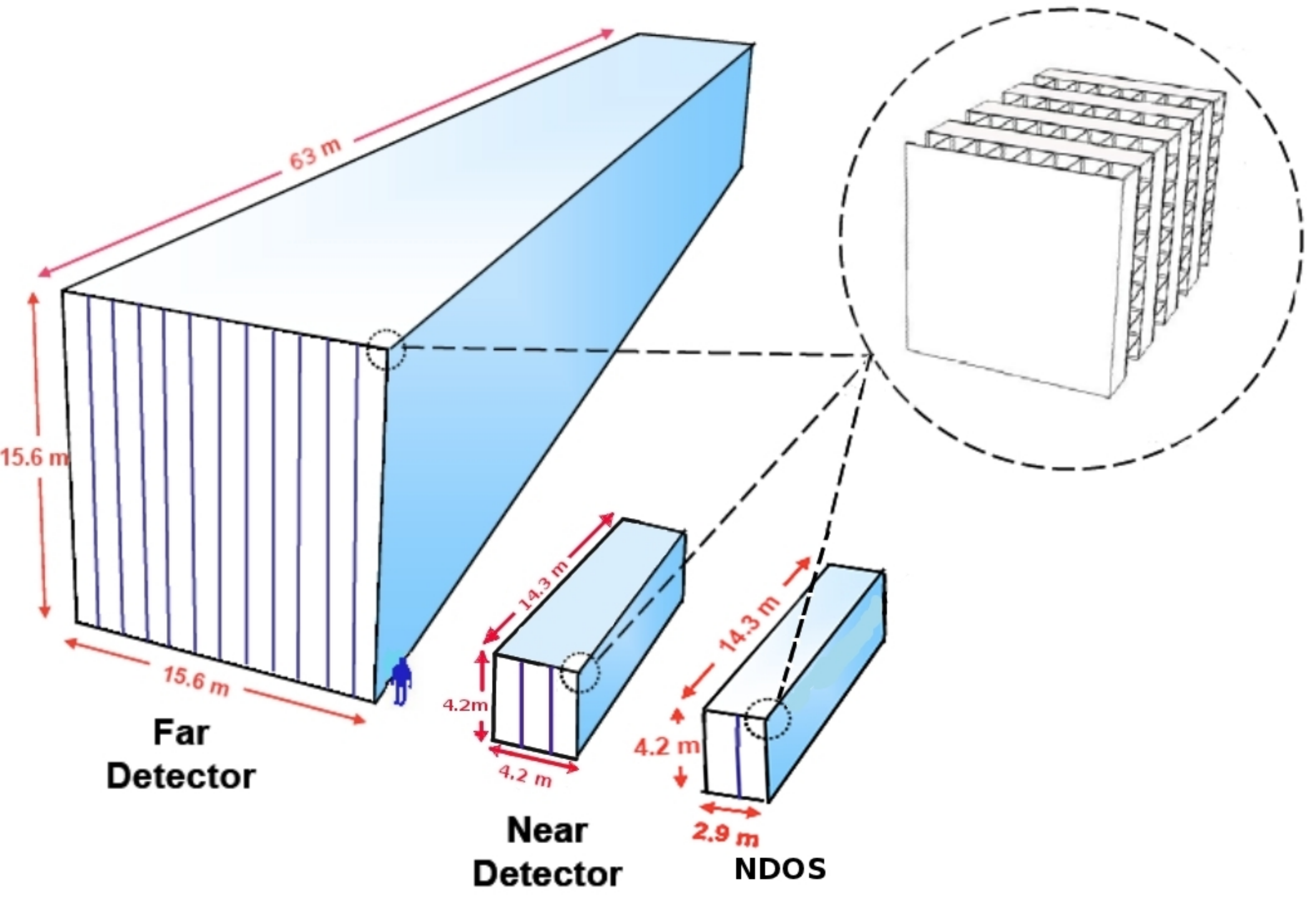}
\caption{Drawings of the \nova far and near detectors.  The human figure at 
the base of the far detector is for scale.}
\label{fig:novaDetectors}
\end{figure}

The \nova near detector will be identical to the far detector except that it will be 
smaller, 3 modules high by 3 modules wide, with 192 planes.  Behind the near 
detector proper will be a  muon ranger, a sandwich of 10 10-cm 
iron plates each followed by two planes of liquid scintillator detectors. 
\nova has also constructed a near detector prototype called the NDOS 
(Near Detector On the Surface) which has been running since November 2010 
on the surface at Fermilab, off axis to both the NuMI and Booster neutrino
beams.
Figure~\ref{fig:novaDetectors} contains a drawing of the \nova detectors.  

Light is extracted from each liquid scintillator cell by a U-shaped 0.7-mm
wavelength-shifting fiber, the ends of which terminate on a pixel of a 
32-pixel avalanche photodiode (APD), which is mounted on the module.
The APD is custom-made for the \nova experiment by the Hamamatsu
Corporation to optimize the match to the two fiber ends per pixel.
Light from the far end of the cell is preferentially attenuated at the
lower wavelengths, so that the peak of the spectrum is at about 540 nm.
The use of APDs is crucial for the experiment since they have a quantum
efficiency of approximately 85\% at this wavelength compared to 10\%
for a photomultiplier with a bialkali photocathode.
The system is designed to produce a minimum of 20 photoelectrons
from the far end of the cell for the passage of a minimum ionizing
particle at normal incidence. The APD is run at a gain of 100, so low
noise is required for efficient operation. The APD is cooled to 
$-$15$^\circ$ C by a thermoelectric cooler to reduce the thermal noise of 
the APD to an acceptable limit.

The \nova front-end electronics runs in continuous digitization mode at 2
MHz for the far detector and 8 MHz for the near detector. It delivers 
GPS time-stamped, pedestal subtracted, and zero-suppressed data to the
data acquisition system (DAQ).  At the far detector, the DAQ buffers the 
data for up to 20 seconds while awaiting a beam spill time message from
Fermilab via Internet.  All data within a 30 \musec window around the 
10 \musec beam spill will be recorded for offline analysis.



\section{Results on $\numu \rightarrow \nutau$: the dominant oscillation mode}
\label{sec:resultsDom}
The dominant oscillation mode for all long-baseline accelerator
experiments performed to date is $\numu \rightarrow \nutau$. This
channel was used by K2K~\cite{Ahn:2002up,Ahn:2006zza} and
MINOS~\cite{Michael:2006rx} to provide essential confirmation of the
neutrino oscillations observed by Super-Kamiokande in atmospheric
neutrinos~\cite{Fukuda:1998mi}. Accelerator experiments with their
fixed baselines, $L$, and high energy resolution detectors allow
precise measurement of $L/E$\@. In turn, this allows resolution of the
oscillatory quantum-mechanical interference pattern and precise
measurements of $|\dm|$ and $\sntheta$: these results are described
here in section~\ref{sec:resultsDm2Sin2}. The corresponding
measurements for muon antineutrinos are described in
section~\ref{sec:resultsDm2BarSin2Bar}.

Direct observation of tau appearance by OPERA will further confirm
$\numu \rightarrow \nutau$ as the dominant mode of oscillation and the
results from the first half of their data
set~\cite{Agafonova:2010dc,Agafonova:2012zz} are described in
section~\ref{sec:resultsTau}.


\subsection{Precision measurement of $|\dm|$ and $\sntheta$}
\label{sec:resultsDm2Sin2}
\begin{figure}
\centering
\includegraphics[trim = 0.1mm 0.1mm 0.1mm 0.1mm, clip, width=0.6\columnwidth]{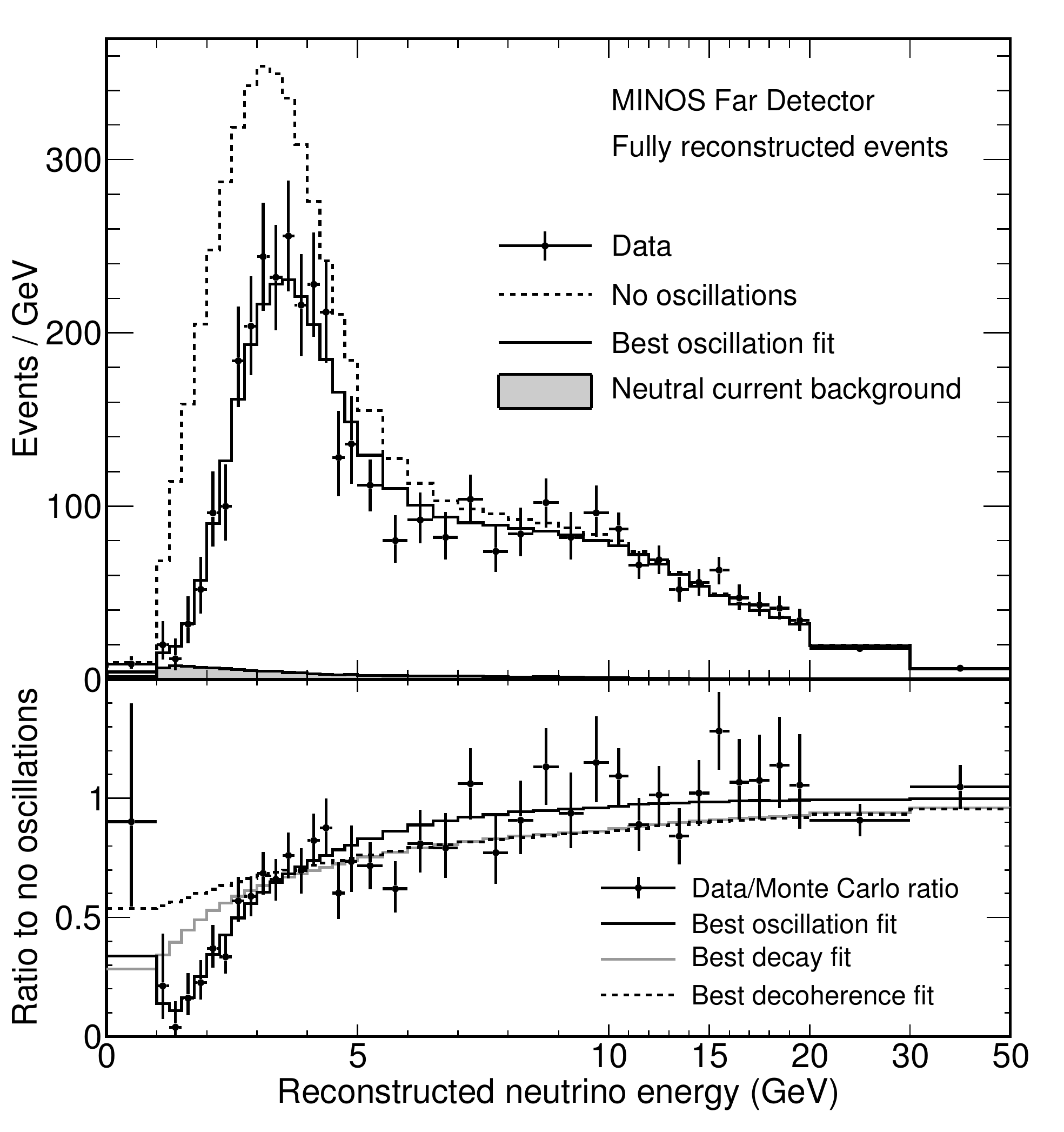}
\caption{The energy spectrum of fully reconstructed muon neutrino
  candidate events in the MINOS Far detector (top pane). Both the no
  oscillation hypothesis and the best oscillation fit are shown. The
  shaded region shows the expected neutral-current background. The
  ratio to no oscillations (bottom pane) displays the best fits to
  models of neutrino decay and neutrino decoherence, where they are
  seen to be disfavored at high significance (7\,$\sigma$ and
  9\,$\sigma$ respectively).}
\label{fig:enSpect} 
\end{figure}

In an accelerator experiment, measurement of $|\dm|$ and $\sntheta$ is
performed by observing the energy dependent disappearance of muon
neutrinos. The fixed baselines, $L$, are known to high precision and
so contribute a negligible uncertainty to measurement of $L/E$, which
is dominated by the energy resolution of the detectors. The energy at
which the maximum disappearance occurs is a measure of $|\dm|$ and the
disappearance probability at that point is given by
$\sntheta$. Figure~\ref{fig:enSpect} shows the energy spectrum of muon
neutrino candidate events in the MINOS far detector where the energy
dependent deficit can be clearly seen, with the maximum disappearance
occurring at around 1.4~GeV for the 735~km baseline.

A crucial ingredient to enabling precise measurements of the
oscillation parameters is event-by-event identification of whether the
observed interactions are neutral-current~(NC) or charged-current~(CC)
events. In the absence of sterile neutrinos, the spectrum of NC events
is unchanged due to oscillations and has to be separated from the muon
neutrino CC sample. For the experiments performed to date,
identification of the flavor of CC events has been of secondary
importance to the separation of NC events since the vast majority of
CC events are muon flavor. Given the tau production threshold at a
neutrino energy of around 3.5~GeV, this appearance mode is naturally
suppressed in K2K, MINOS, T2K and \nova due to their lower beam
energies and so relatively few $\nutau$~CC interactions occur. The
appearance of electron neutrinos is a sub-dominant effect (detailed in
section~\ref{sec:resultsSubdom}) that contributes, for example, only
around 1\% of the event rate in MINOS\@. The performance of the
different experiments in selecting a $\numu$~CC event sample is
discussed below.


\subsubsection{K2K $\numu$ Disappearance Results}
\begin{figure}
\centering
\includegraphics[trim = 0.1mm 0.1mm 0.1mm 0.1mm, clip, width=0.6\columnwidth]{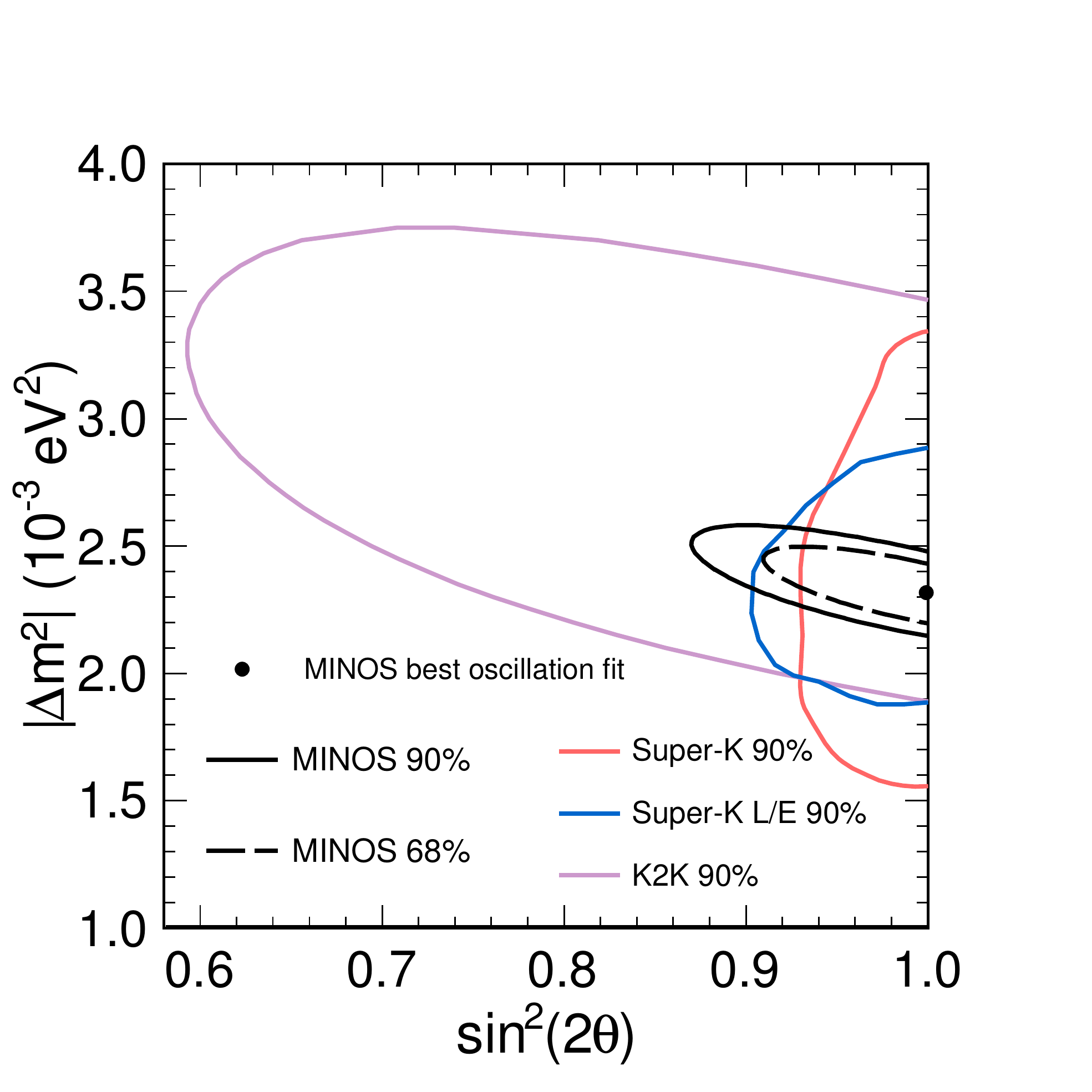}
\caption{The 90\% confidence regions for $|\dm|$ and
  $\sntheta$. Results shown are published contours from
  K2K~\cite{Ahn:2006zza}, MINOS~\cite{Adamson:2011ig} and
  Super-Kamiokande~\cite{Ashie:2004mr,Ashie:2005ik}. For the latest
  but still preliminary results see Figure~\ref{fig:sinVsdm2Prelim}.}
\label{fig:sinVsdm2} 
\end{figure}

\ \newline K2K was the first accelerator long-baseline experiment,
taking data from 1999--2004. The neutrino beam was produced and
measured at KEK in Japan and then observed 250~km away at the
Super-Kamiokande detector. K2K saw 112 beam-originated events in the
fiducial volume of Super-Kamiokande with an expectation of
158.1$^{+9.2}_{-8.6}$ without oscillation~\cite{Ahn:2006zza}. The
water \Cer detector allowed separation of 58 single-ring muon-like
events in which a distortion of the energy spectrum was seen. At the
K2K beam energy these muon-like events contained a high fraction of
quasi-elastic events and the incoming neutrino energy was
reconstructed using two-body kinematics. Combining information from
both the shape of the energy spectrum and the normalization, K2K
determined that the probability of obtaining their data in the case of
null-oscillations was 0.0015\%~(4.3\,$\sigma$) thus confirming the
Super-Kamiokande atmospheric neutrino results. The K2K
90\%~C.L. allowed region in the $|\dm|$\,--\,$\sntheta$ plane is shown
by the magenta line in~Figure~\ref{fig:sinVsdm2}.


\subsubsection{MINOS $\numu$ Disappearance Results}
\label{sec:minosNuMuDisap}
\ \newline MINOS started data taking in 2005 and ran for 7 years
through April 2012. Around 80\% of the data was taken with the beam
optimized to produce neutrinos and the remaining 20\% antineutrinos
(see section~\ref{sec:resultsDm2BarSin2Bar} for a description of the
$\numubar$ disappearance results). The first $\numu$ disappearance
results from MINOS are given in~\cite{Michael:2006rx} and detailed in
a longer paper~\cite{Adamson:2007gu}. Updated results are given
in~\cite{Adamson:2008zt} and those presented here are taken
from~\cite{Adamson:2011ig}. Additionally, the results from the
preliminary analysis using the full MINOS data
set~\cite{Nichol:Neu2012} are also summarised here.

The geometry of the MINOS detectors allows three dimensional
reconstruction of tracks and showers. Using the reconstructed vertex
information a fiducial volume cut was made that separated incomplete
and partially reconstructed events occurring at the edge of the
detector from those that were fully reconstructed. As mentioned above,
a crucial step in this analysis was the separation of $\numu$~CC events
from NC events. For the first results a particle identification
parameter was constructed using probability density functions for the
event length, the fraction of the energy contained in the track, and
the average pulse height per plane. The later results used an improved
technique based on a k-nearest-neighbor algorithm~(kNN). This kNN
technique used the energy deposition along a track and its fluctuation
to discriminate muons from spurious tracks reconstructed from hadronic
activity in NC interactions. For the most recent analysis an overall
efficiency for selecting $\numu$~CC events of 90\% was achieved. The
first results made a selection on the charge-sign of the muon but later
analyses have included the 7\% antineutrino component of the
neutrino-enhanced beam, which had a significantly higher average
energy~\cite{Adamson:2011ch}.

Near detector data was used to substantially reduce systematic effects
on this measurement that would otherwise arise from limited knowledge
of the neutrino flux and cross-sections. Both the Near and Far
Detectors measured a product of flux times cross-section and by doing a
relative measurement, the uncertainties on that product canceled to
first order. However, the flux iwas not the same at the Near and Far
detectors: one saw a line-source of neutrinos and the other saw what
was effectively a point source. The Far Detector flux was populated by
neutrinos from more forward decaying pions and so the spectrum was
somewhat harder than at the Near detector. The beamline simulation
incorporated and was used to estimate these largely geometrical
effects.

Due primarily to the flux and cross-section uncertainties, the Near
detector data differed from the simulation by up to 20\% as a function
of energy. An extrapolation procedure used the Near Detector
measurements to predict the Far Detector energy spectrum via a number
of steps as follows: subtracting the estimated background from the
Near Detector energy spectrum; deconvolving the effects of Near
detector energy resolution; using a transfer matrix to account for the
different flux at the Far Detector; weighting each energy bin
according to the oscillation probability; reintroducing the effect of
energy resolution at the Far Detector; and adding in the estimated Far
Detector background. With all these steps complete an oscillated Far
Detector prediction was obtained for comparison with the data.

Several sources of systematic uncertainty were accounted for in this
measurement. The three largest uncertainties on the measurement of
$|\dm|$ were on the absolute energy scale of hadronic showers, the
absolute energy scale of muons and the relative normalization of event
rates between Near and Far Detectors. Other uncertainties included NC
contamination, the relative hadronic energy scale, cross-sections and
beam flux. Overall, the statistical error on the MINOS measurement of
$|\dm|$ was still more significant than the systematic uncertainty.

The largest three systematic uncertainties on the measurement of
$\sntheta$ were on the NC contamination, cross-sections, and the
relative hadronic energy scale. However, the MINOS measurement of
$\sntheta$ was dominated by the statistical uncertainty, with the
systematic uncertainty being smaller by more than a factor of four.

Every NuMI beam event with a reconstructed muon was included in the
likelihood fit to extract the oscillation parameters. These events
were split into 7 event categories to extract the maximum
information. Partially reconstructed events, where the neutrino
interacted in the rock outside the detector or in the outer edges of
the detector, were a separate category and only their reconstructed
muon information was used (any shower energy was ignored due to its
limited use for this sample). Fully reconstructed $\numu$~CC candidate
events were separated by the charge-sign of the muon. Positively
charged events formed their own single sample but the negatively
charged events were divided into 5 categories using their estimated
energy resolution (for example, a highly-elastic CC event where most
of the neutrino energy was carried away by the muon was measured more
precisely than an inelastic event where shower energy fluctuations
smeared the measurement). The four dominant systematic uncertainties
were included as nuisance parameters and the mixing angle was
constrained by the physical boundary at $\sntheta=1$.

\begin{figure}
\centering
\includegraphics[trim = 0.1mm 0.1mm 0.1mm 0.1mm, clip, width=0.6\columnwidth]{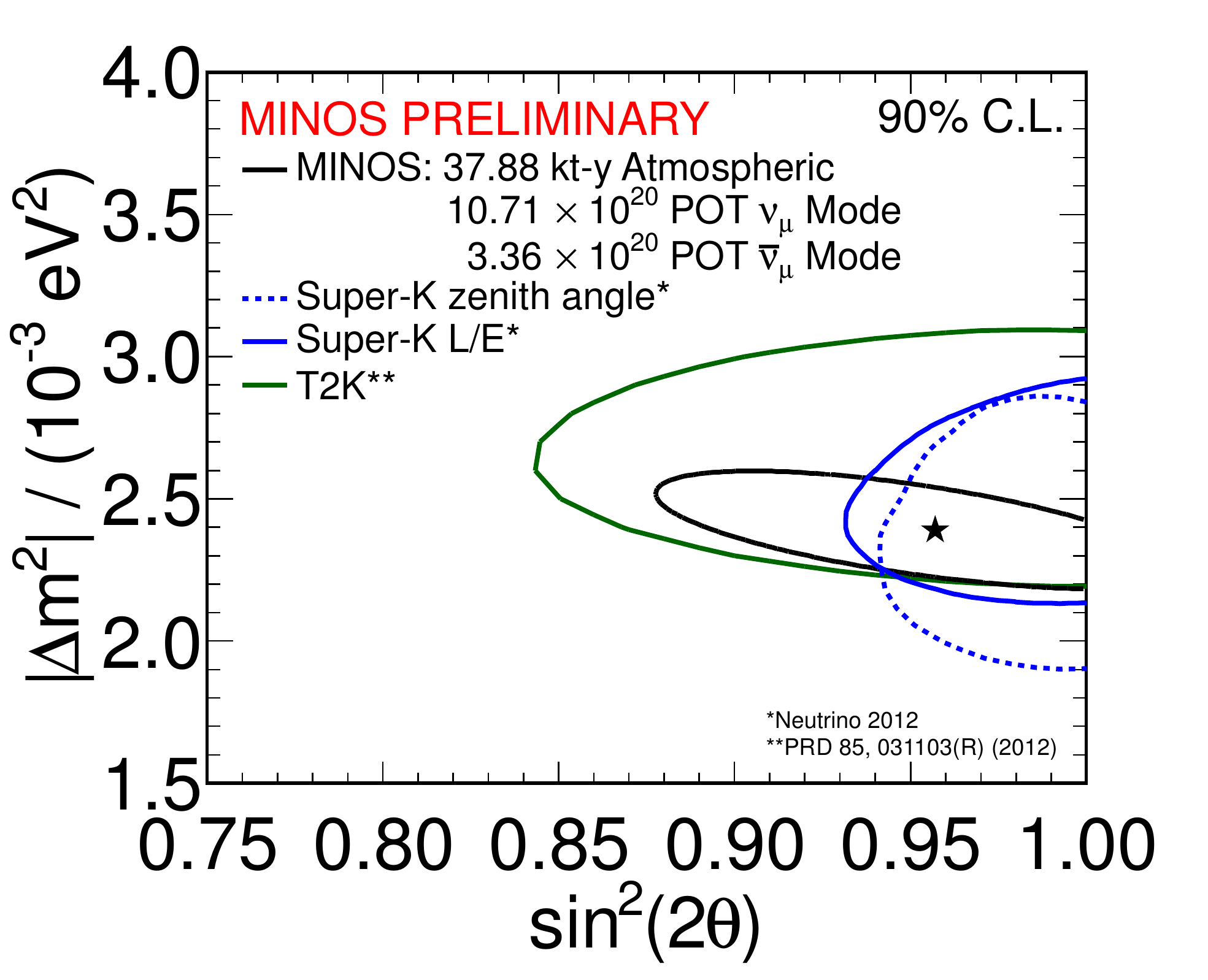}
\caption{Preliminary 90\% confidence regions for $|\dm|$ and
  $\sntheta$ (except for T2K, which is published). Results are shown
  for MINOS~\cite{Nichol:Neu2012}, T2K~\cite{Abe:2012gx} and
  Super-Kamiokande~\cite{Itow:Neu2012}. The MINOS results shown here
  are a combination of NuMI beam data and atmospheric neutrino data.}
\label{fig:sinVsdm2Prelim} 
\end{figure}

Thousands of beam neutrino interactions have been recorded at the
MINOS Far detector and used, as described above, to make the world's
most precise measurement of
$|\dm|=(2.32^{+0.12}_{-0.08})\times10^{-3}$~eV$^2$ while constraining
$\sntheta<0.90$ at
90\%~C.L.~\cite{Adamson:2011ig}. Figure~\ref{fig:enSpect} shows the
fully reconstructed events recorded by MINOS where the distortion of
the energy spectrum expected by oscillations can be seen and contrasts
with that expected from alternative models of neutrino disappearance
such as neutrino decay or decoherence (they are excluded at 7 and
9\,$\sigma$ respectively). The MINOS contours associated with this
published result are shown in Figure~\ref{fig:sinVsdm2} (updated but
preliminary results from MINOS are shown in
Figure~\ref{fig:sinVsdm2Prelim}).

Recently, preliminary MINOS results using the complete data set have
been released~\cite{Nichol:Neu2012}. The total neutrino-enhanced beam
exposure is $10.7\times10^{20}$~POT, 50\% more than the previous
result given above. Furthermore, two additional data sets are
included: firstly, the antineutrino-enhanced beam
data~($3.36\times10^{20}$~POT); and secondly, atmospheric neutrinos
and antineutrinos~(37.9~kiloton-years). While still well within the
previous 1\,$\sigma$ contours, the best fit point for this new
analysis has moved slightly away from maximal mixing to
$|\dm|=(2.39^{+0.09}_{-0.10})\times10^{-3}$~eV$^2$ and
$\sntheta=0.96^{+0.04}_{-0.04}$ (the shift in upwards in $|\dm|$ being
correlated with the shift downward in $\sntheta$, due to the required
overall normalization being similar to the previous result).

The MINOS preliminary 90\%~C.L. allowed region in the
$|\dm|$\,--\,$\sntheta$ plane is shown in
Figure~\ref{fig:sinVsdm2Prelim} by the solid black contour. The latest
results from Super-Kamiokande~\cite{Itow:Neu2012} (preliminary) and
T2K~\cite{Abe:2012gx} are shown alongside for comparison. All the
results presented here use the 2-flavor approximation.


\subsubsection{T2K $\numu$ Disappearance Results}
\begin{figure}
\centering
\includegraphics[trim = 0.1mm 0.1mm 0.1mm 0.1mm, clip, width=0.6\columnwidth]{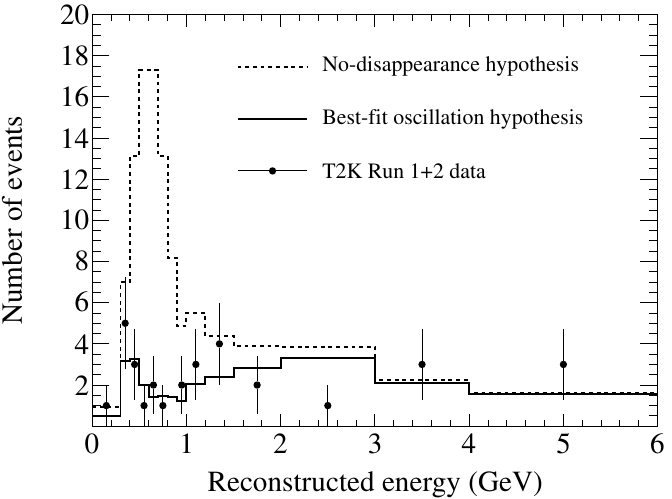}
\caption{Reconstructed neutrino energy spectra of T2K $\nu_\mu$
 disappearance analysis.}
\label{fig:T2K-numudisappspec}
\end{figure}

\ \newline T2K started taking data in 2010 and was the first
experiment to use an off-axis beam to observe muon neutrino
disappearance~\cite{Abe:2012gx}. The exposure for the first result was
$1.43\times10^{20}$~POT and is expected to increase substantially over
the next few years. In the Super-Kamiokande far detector, 31
fully-contained muon-like ring events were observed against an
expectation of $104\pm14$(syst) without neutrino oscillations. The
observed neutrino energy spectrum alongside the predicted spectra with
and without oscillation are shown in
Figure~\ref{fig:T2K-numudisappspec}.

The values of the oscillation parameters obtained are consistent with
both MINOS results and Super-Kamiokande atmospheric
neutrinos. Interestingly, the T2K constraints on $\sntheta$ already
approach the limit set by MINOS. This demonstrates the sensitivity of
T2K where the energy peak of the narrow band, off-axis, beam is
positioned close to the oscillation maximum and consequently a large
fraction of the $\numu$ flux disappears. The T2K contours are shown in
Figure~\ref{fig:sinVsdm2Prelim} alongside the latest MINOS results.


\subsection{Measurements of $|\dmbaratm|$ and $\snthetabar$}
\label{sec:resultsDm2BarSin2Bar}
MINOS accumulated 20\% of its total exposure with the NuMI beam
configured to enhance production of antineutrinos and made the first
direct measurement of muon antineutrino
disappearance~\cite{Adamson:2011fa}. The CPT theorem, that provides
the foundation of the standard model, predicts identical disappearance
of neutrinos and antineutrinos in vacuum and the measurements
described here allow precision tests of that hypothesis as well as
other models of new physics. The first antineutrino result from MINOS
reported tension with the neutrino results but with further data the
results are now consistent~\cite{Adamson:2012rm,Nichol:Neu2012}. In
addition to these results, the 7\% antineutrino component of the
neutrino-enhanced beam has also been analyzed~\cite{Adamson:2011ch};
these data provided a higher statistics sample of $\numubar$ events in
the 5--15~GeV range, allowing the oscillation probability to be
measured with greater precision in that region. The MINOS magnetized
detectors were essential to obtaining a high purity sample of
$\numubar$~CC events and making the measurements reviewed here.

The antineutrino-enhanced beam flavor composition, described in
section~\ref{sec:beams}, was 40\%~$\numubar$, 58\%~$\numu$,
2\%~$\nue+\nuebar$~\cite{HimmelThesis}. The reason for the large
number of neutrinos was two-fold: firstly, the antineutrino
cross-section is about 2--3 times lower than for neutrinos; and
secondly, the yield of negative pions from the beam target was lower
than for positive pions. However, the ratio of antineutrinos to
neutrinos in the NuMI beam varied strongly as a function of energy and
below 6~GeV about 80\% of the interactions were antineutrinos (and
that's where the oscillation effect was largest for
MINOS). Discrimination of muon neutrinos from antineutrinos was
performed on an event-by-event basis by analyzing the track curvature
in the detector's magnetic field. Efficiency and purity wass estimated
from the MC simulation at 91.6\% and 99.0\% respectively for the Far
Detector.

With the magnetized detectors able to cleanly separate positive and
negative muons, the rejection of NC events was an important requirement
for this analysis. The k-nearest-neighbor multivariate technique used
for the neutrino analysis (see section~\ref{sec:resultsDm2Sin2}) was
used to separate $\numubar$~CC events from NC. The procedure for
extrapolating Near Detector antineutrino data to make a Far Detector
prediction was essentially the same as for the neutrino analysis. The
detector and beamline simulations were reperformed for antineutrinos
to calculate, for example, the required detector resolution
deconvolution matrix and flux transfer matrix for $\numubar$. A slight
modification to the oscillation step of the extrapolation was required
to allow neutrinos and antineutrinos to oscillate differently in the
simulation.

Systematic uncertainties on the measurement of the antineutrino
oscillation parameters were similar to those described for neutrinos in
section~\ref{sec:minosNuMuDisap} above. An additional uncertainty was
included on the level of neutrino contamination and the knowledge of
the neutrino oscillation parameters. The MINOS measurement of neutrino
parameters is not yet systematically limited and given both the factor
of 3 lower exposure recorded for antineutrinos
($3.36\times10^{20}$~POT) and the reduced number of $\numubar$ per
POT, the antineutrino measurement is dominated by statistical
uncertainties.

\begin{figure}
\centering
\includegraphics[trim = 0.1mm 0.1mm 0.1mm 0.1mm, clip, width=0.6\columnwidth]{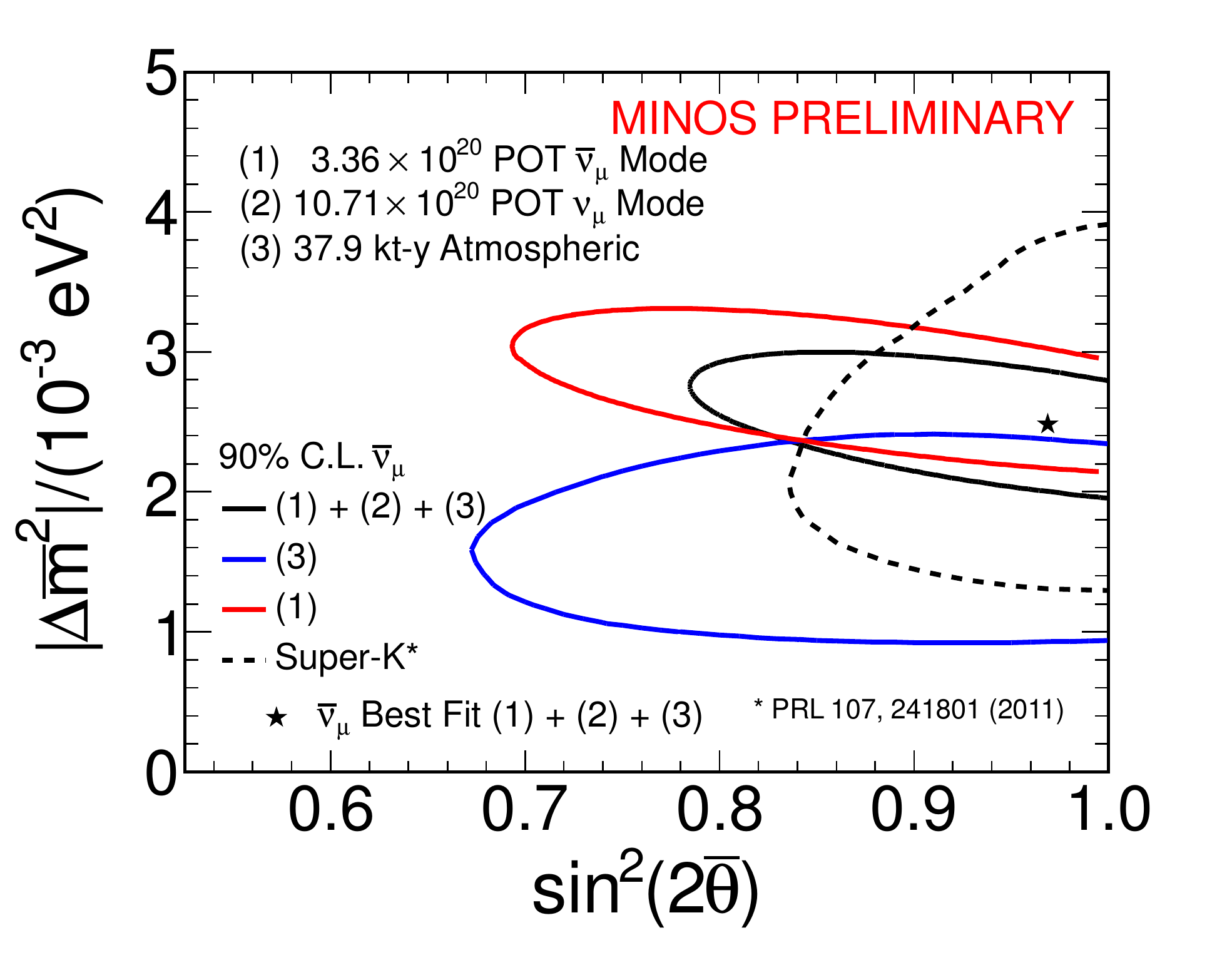}
\caption{The 90\% confidence regions for antineutrino parameters
  $|\dmbaratm|$ and $\snthetabar$. Antineutrino results are shown from
  Super-Kamiokande~\cite{Abe:2011ph} (dashed black) alongside the
  latest preliminary results from MINOS~\cite{Nichol:Neu2012} (solid
  black). The MINOS results used three data sets: (1) atmospheric
  antineutrinos; (2) antineutrinos from the NuMI beam operating in
  antineutrino-enhanced mode; and (3) antineutrinos from the
  neutrino-enhanced beam. The red contour shows the result from just
  the NuMI beam data and the blue contour from just the atmospheric
  antineutrino data.}
\label{fig:sinbarVsdm2bar} 
\end{figure}

Recently, a preliminary version of the MINOS measurements of
antineutrino oscillation parameters using the full data set have been
released~\cite{Nichol:Neu2012}. This analysis incorporates three
distinct data sets: the antineutrino-enhanced NuMI beam data
($3.36\times10^{20}$~POT); the antineutrinos in the neutrino-enhanced
beam ($10.7\times10^{20}$~POT); and atmospheric antineutrino data
(37.9~kiloton-years). The antineutrino mass splitting was measured to
be $|\dmbaratm|=(2.48^{+0.22}_{-0.27})\times10^{-3}$~eV$^2$ and the
mixing angle $\snthetabar=0.97^{+0.03}_{-0.08}$ with
$\snthetabar>0.83$ at 90\% C.L\@. The antineutrino contour from MINOS
is shown in Figure~\ref{fig:sinbarVsdm2bar} by the solid black
line. Also shown for comparison is the result from the
Super-Kamiokande measurement (dashed black) of the combined flux of
atmospheric muon neutrinos and antineutrinos~\cite{Abe:2011ph}. The
red contour shows the result from just the NuMI beam data and the blue
contour from just the MINOS atmospheric antineutrino data. The MINOS
measurements provide the highest precision on the antineutrino mass
squared splitting while Super-Kamiokande measures the antineutrino
mixing angle most precisely.

The uncertainty on the difference in the atmospheric mass squared
splittings of neutrinos and antineutrinos is currently dominated by
the statistical precision on the antineutrino measurements, by about a
factor of 2--3. In the future, \nova will improve measurement of all
the disappearance related parameters for neutrinos and
antineutrinos. Importantly for future precision tests of CPT symmetry,
several systematic errors on the difference between $|\dm|$ and
$|\dmbaratm|$ will be significantly smaller than the systematic
uncertainty on the two absolute measurements taken separately.


\subsection{Searches for $\nutau$ Appearance}
\label{sec:resultsTau}
The observation of $\nutau$ appearance with a $\numu$ source would
directly confirm the hypothesis of $\numu \rightarrow \nutau$
oscillations as the cause of the disappearance affect observed by
atmospheric and accelerator experiments.  This is the goal of the OPERA
experiment~\cite{Acquafredda:2009zz}. Furthermore, there is currently
no observation at the 5-sigma level of the appearance of neutrino
flavors due to oscillations, only disappearance. The next few years
should see the conclusive observation of both $\nutau$ appearance with
OPERA and $\nue$ appearance with T2K and \nova, demonstrating key
aspects of the 3-flavour neutrino oscillation model.

The kinematic threshold for $\tau$ production from $\nutau$
interactions is around 3.5~GeV and at that energy the first maximum of
the oscillation probability occurs at a baseline of approximately
2500~km. For a fixed baseline, matching the energy of the beam with
the peak of the product of $\nutau$ cross-section times oscillation
probability maximizes the number of $\nutau$ interactions in the
detector for a given integrated flux: this is largely what the OPERA
experiment has done with the CNGS beam. As described in
section~\ref{sec:beams}, the experiments using the CNGS and NuMI beams
have very similar baselines, 730~km vs.\ 735~km respectively, but
differ substantially in their average neutrino energies of 17~GeV and
3~GeV respectively due to the different physics goals of the
experiments.

The OPERA experiment at LNGS started taking data in 2008 with the CNGS
beam~\cite{Acquistapace:1998rv,Meddahi:1051376} and in 2010 they
published the observation of their first $\nutau$ candidate
event~\cite{Agafonova:2010dc}. As described in
section~\ref{sec:detectors}, the OPERA detector consists of
lead-emulsion bricks with electronic detectors to pinpoint the bricks
in which neutrino interactions occurred.

\begin{figure}
\centering
  \includegraphics[width=0.4\columnwidth]{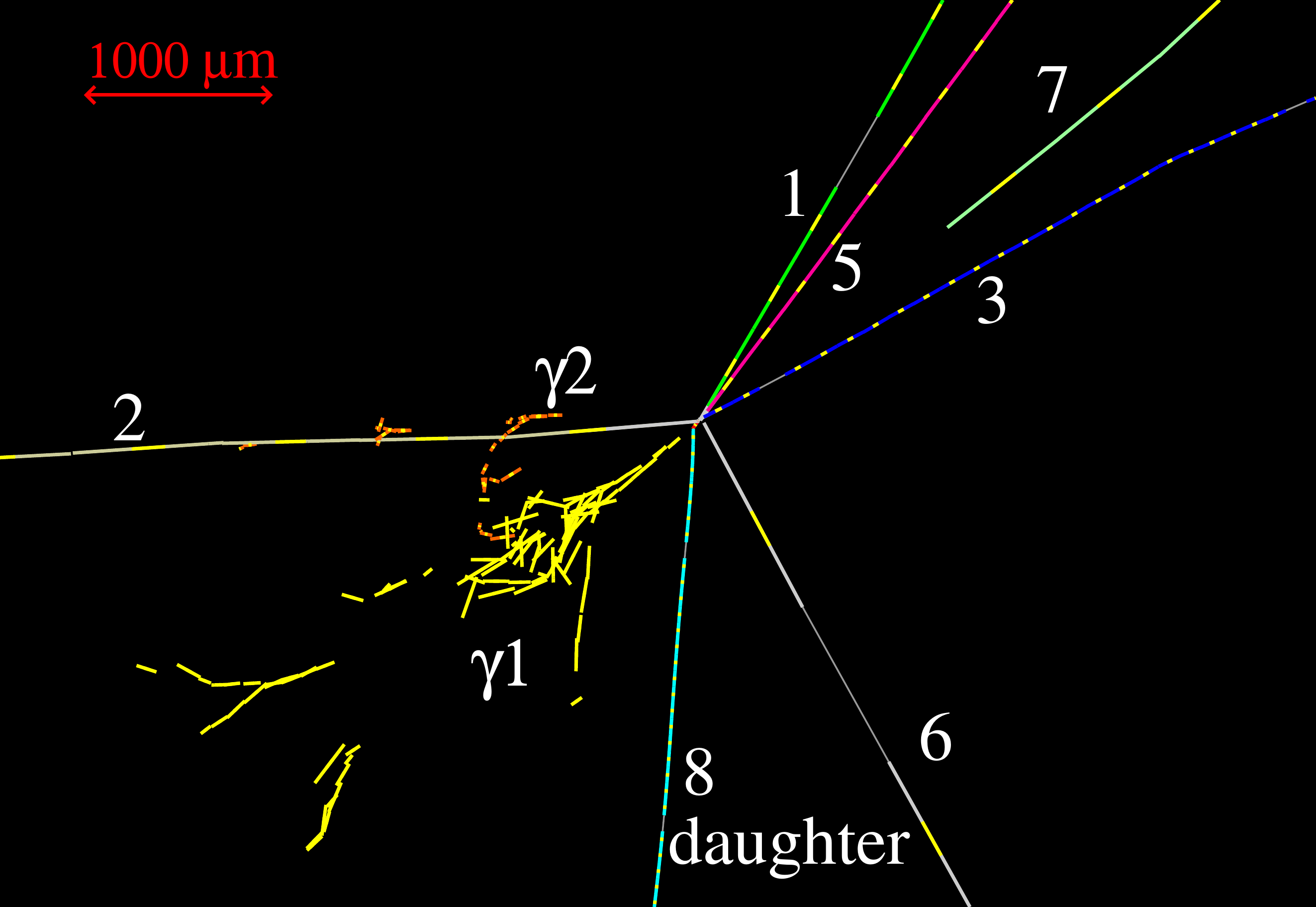}
    \includegraphics[width=0.4\columnwidth]{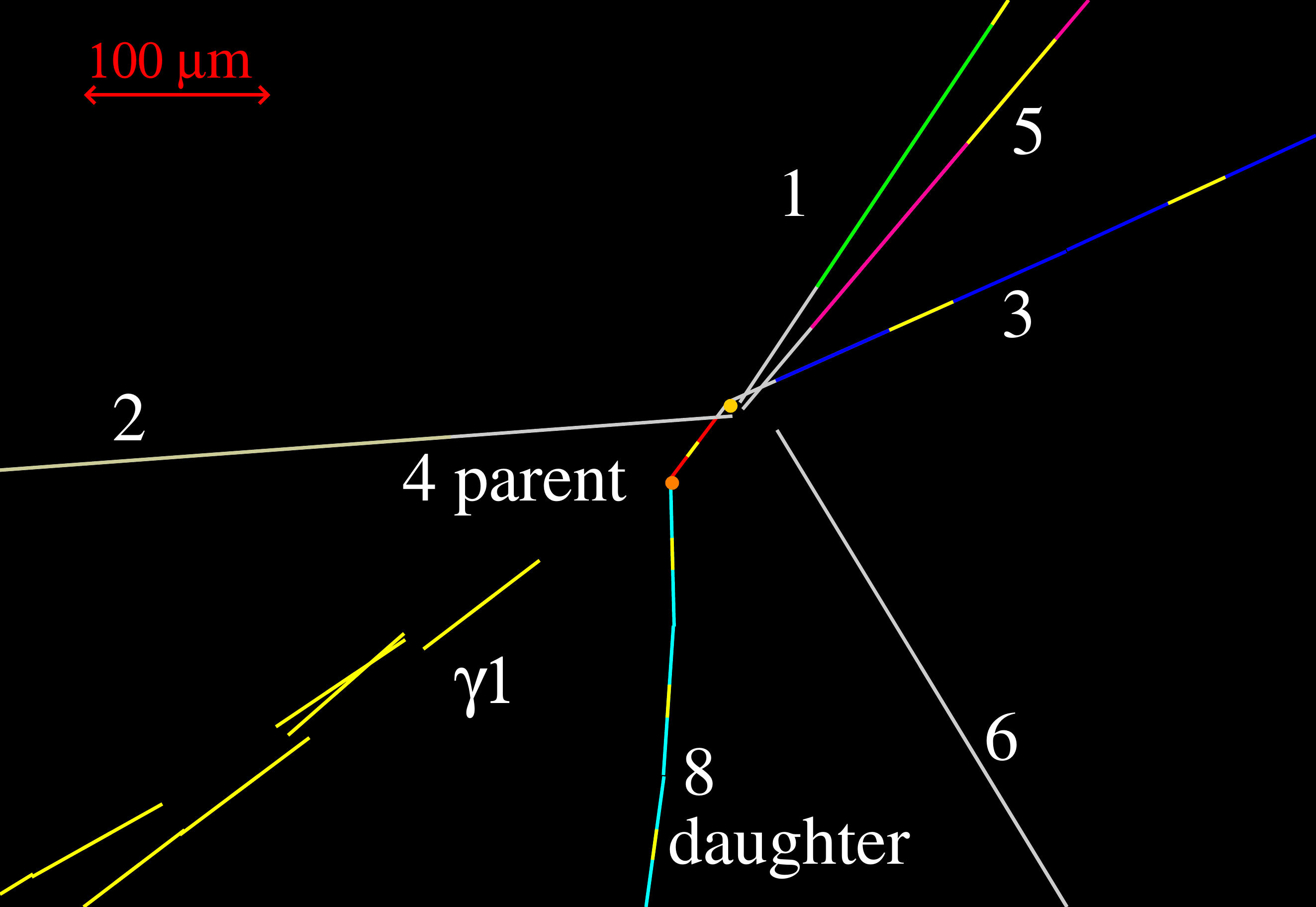}\\
\includegraphics[trim = 0.1mm 0.1mm 0.1mm 0.1mm, clip, width=0.8\columnwidth]{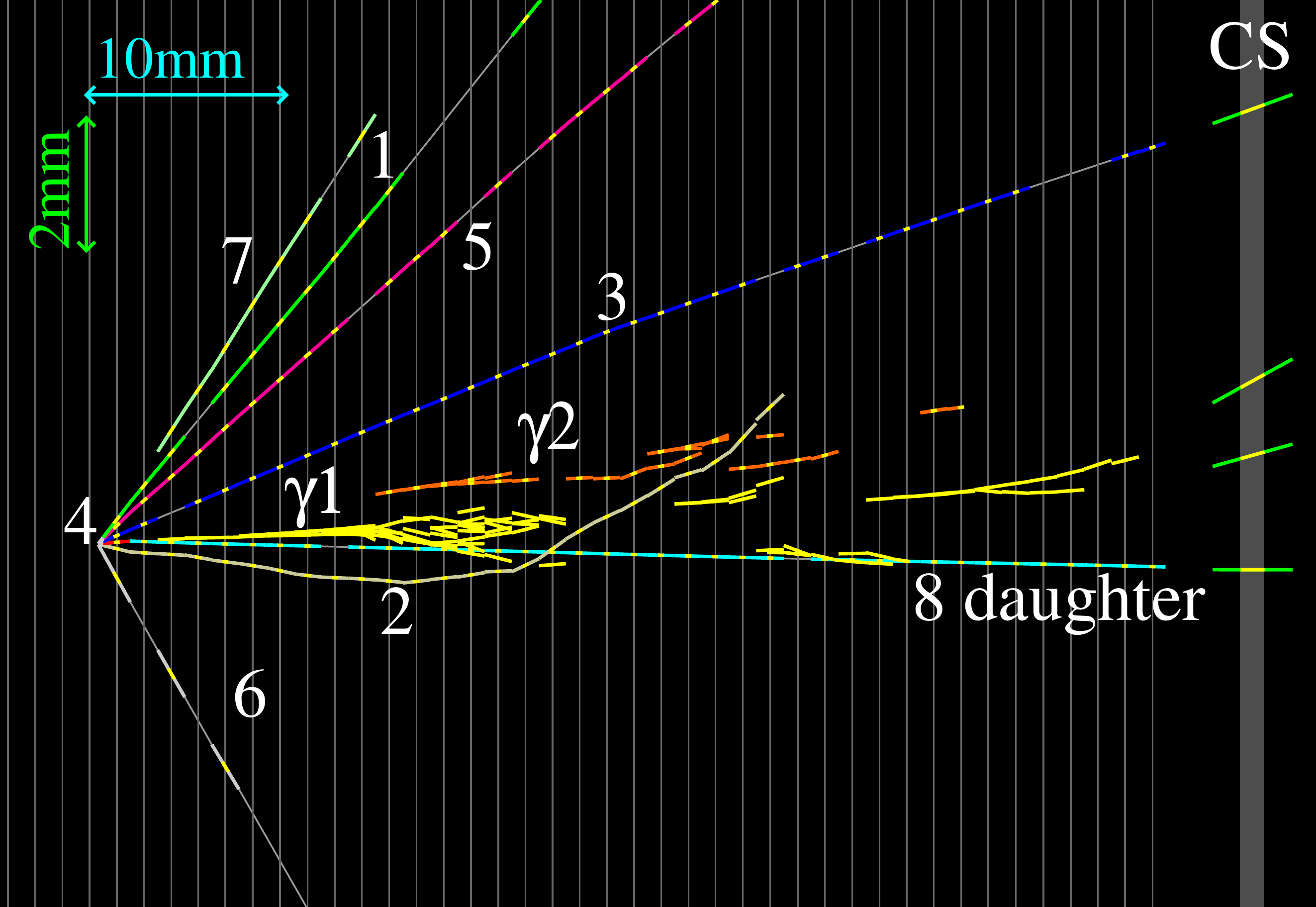}
\caption{The first candidate $\nutau$ event observed by the OPERA
  experiment. The top plots show the transverse view with the right
  plot being a zoom of the left. The bottom plot shows the
  longitudinal view. The short red track (labeled as ``4 parent'') is
  identified as being due to the $\tau$ lepton and the track of what
  is thought to be the tau-daughter is shown in turquoise (labeled as
  ``8 daughter''). A kink is clearly seen, particularly in the zoomed
  transverse view (top right), and demarked by the change in color
  from red to turquoise along the track. A detailed description of the
  likely candidates for each of the numbered tracks is given in the
  OPERA paper~\cite{Agafonova:2010dc}.}
\label{fig:tau} 
\end{figure}

The first candidate $\nutau$ event observed by OPERA is shown in
Figure~\ref{fig:tau}. A detailed description of the likely candidates
for each of the numbered tracks is given
in~\cite{Agafonova:2010dc}. This event is compatible with the decay
$\tau^- \rightarrow \rho^- \nu_\tau$ with the $\rho(770)$ decaying to
a $\pi^0$ and $\pi^-$.

A preliminary analysis of further data has recently been released and
a second $\nutau$ candidate has been
observed~\cite{Nakamura:Neu2012}. This event was seen in the 2010-11
data set and it satisfies the selection criteria for $\nutau
\rightarrow 3$~hadrons. In the data set analyzed to date, the
preliminary background estimate was 0.2 events and 2.1 signal events
were expected. The Poisson probability of observing 2 or more events
given a background expectation of 0.2 is 1.75\%.

Atmospheric neutrino experiments have a relatively large number of
$\nutau$ events in their data samples, given the broad range of
available energies and the Earth's 13,000~km
diameter. Super-Kamiokande has published 2.4\,$\sigma$ evidence for
$\nutau$ appearance~\cite{Abe:2006fu} using candidate events selected
for the expected shape of $\nutau$ interactions and characteristics of
$\tau$ leptons. This statistical separation is a complementary
approach to OPERA's goal of directly observing individual $\nutau$
events. At the time of writing a new SK result was published on the
arXiv that provides evidence for $\nutau$ appearance at the
3.8\,$\sigma$ confidence level~\cite{Abe:2012jj}. In the future,
MINOS+ will also have a relatively large number of $\nutau$ events
(around 90/year with the $\numu$-enhanced beam) and with sufficient
rejection of backgrounds will have sensitivity to this oscillation
channel~\cite{Tzanankos:2011zz}.


\section{Results on $\numu \rightarrow \nue$: the sub-dominant oscillation mode}
\label{sec:resultsSubdom}
With the baselines and neutrino energies (the $L/E$) used by the
experiments described in this review, $\numu \rightarrow \nue$ is a
sub-dominant oscillation mode (although at an $L/E$ 25 times larger,
the solar mass splitting would have a significant effect and $\nue$'s
would then make up the majority of the flux).

Measurements of the sub-dominant $\numu \rightarrow \nue$ oscillation
mode are of great importance for a number of reasons: firstly, its
discovery will demonstrate the full 3-flavor neutrino oscillation
model; secondly, with a non-zero value of $\thetaonethree$ a door is
opened to discovering CP violation in the lepton sector; and thirdly,
by exploiting the neutrino--matter interaction that the neutrinos and
antineutrinos experience as they propagate through the Earth, the
neutrino mass hierarchy (the sign of $\Delta m^2_{32}$) can also be
determined.

Measurements of the sub-dominant mode made using accelerator neutrino
beams are highly complementary to those made using nuclear
reactors. The reactor neutrino experiments Double
Chooz~\cite{Abe:2011fz}, Daya Bay~\cite{Ahn:2012nd} and
RENO~\cite{Ahn:2012nd} have recently observed sub-dominant neutrino
oscillations via the disappearance of $\nuebar$ over a distance of
around 1.5~km. This channel is only sensitive to $\thetaonethree$ and
so a direct measurement can be made. In contrast, the
accelerator experiments are sensitive to $\thetaonethree$, the CP
phase, the mass hierarchy and the octant of $\thetatwothree$,
enabling a rich set of measurements to be made using a combination of
different baselines and energies with neutrinos and/or antineutrinos.

In this section the electron neutrino appearance results from K2K,
MINOS and T2K are presented. A key feature of these experiments is
their ability to distinguish the rare occurrence of electron flavor
neutrino interactions from among the many more $\numu$~CC events and
NC events from all neutrino flavors. For example, electron neutrino
events in MINOS contribute only around 1\% of the event rate. The
significant majority of $\numu$~CC events are relatively easy to
reject due to the presence of the muon. However, in highly inelastic
$\numu$~CC events the muon can escape detection and should the
hadronic shower have a significant electromagnetic component (from,
for example, $\pi^0\rightarrow \gamma\gamma$) then it can be
misidentified as an electron neutrino event.


\subsection{K2K $\nue$ Appearance Results}
\label{sec:resultsK2KNueApp}
The first long-baseline accelerator neutrino experiment to search for
electron neutrino appearance was
K2K~\cite{Ahn:2004te,Yamamoto:2006ty}. This measurement exploited the
ability of the Super-Kamiokande detector to distinguish muons and
electrons, which had been well established for the earlier atmospheric
neutrino results. As such, the primary background for K2K was events
containing a $\pi^0$ from a NC interaction. This background occurs
when one of the two gammas from the $\pi^0$ decay is not
reconstructed, due to highly asymmetric energies or a small opening
angle between the two gammas. Beam $\nue$ events are around 1\% at the
KEK site and the background from such electron neutrinos intrinsic to
the beam was estimated to be only 13\% of the total background.

At the limit set by the CHOOZ experiment~\cite{Apollonio:2002gd} and
with an exposure of $9.2\times10^{19}$ protons on target, K2K expected
to see only a few events and so it was critical that the background
was reduced to a very low level. The basic selection of electron
neutrino events is as follows: the first step is to require electron
\Cer-ring candidates; secondly, any events with electron-equivalent
energy below 100~MeV are removed to reject charged pions and electrons
from muon decay; and thirdly, no candidate may have a muon decay
within a 30~$\mu$s time window. To improve the rejection of the
$\pi^0$ background a dedicated algorithm to calculate the invariant
mass under the assumption that there were two rings was also used. The
total background expectation with the above cuts was
1.7$^{+0.6}_{-0.4}$~events (in the case of no oscillation). The
overall efficiency for selection of $\nue$ signal events in the
simulation is around~50\%.

The fraction of the background coming from NC interactions that
produce a single $\pi^0$ (NC\,1$\pi^0$) was 70\% so constraining the
associated systematic uncertainty was crucial. To do this a 1~kiloton
water \Cer Near detector was used to measure the NC\,1$\pi^0$/CC
interaction ratio and the uncertainty was constrained to the 12\%
level. Many other sources of systematic uncertainty were considered and
the largest individual one concerned the $\pi^0$ mass cut and that
uncertainty was constrained using atmospheric neutrino data. The other
systematics also included the detector efficiency, water properties,
neutrino flux at SK, and several neutrino interaction model
uncertainties. In total the background uncertainty was between
24--39\% depending on the run period.

K2K observed 1 event that passed their selection criteria, consistent
with the background expectation. These data allowed a 90\%~C.L. limit
to be set on the maximum electron neutrino appearance probability of
0.13, at the oscillation parameters measured by K2K via $\numu$
disappearance (see section~\ref{sec:resultsDm2Sin2}). Such an
appearance probability corresponds to an approximate limit of
$\snthetaonethree<0.26$.


\subsection{MINOS $\nue$ Appearance Results}
\label{sec:resultsMINOSNueApp}
The first MINOS $\nue$ appearance result was released in
2009~\cite{Adamson:2009yc} and two further results with more data and
analysis improvements have since been
published~\cite{Adamson:2010uj,Adamson:2011qu}. The MINOS detectors
were optimized for measuring muon neutrino interactions at the few-GeV
scale. The steel planes are 1.4 radiation lengths thick and the strip
width is 4.1~cm (compared to the Moli\`{e}re radius of 3.7~cm) giving
a relatively coarse view of an electron shower. Absolutely crucial for
controlling the systematic uncertainties on these measurements is the
functionally identical design of the Near and Far detectors. As with
K2K, the dominant background is from NC interactions. Although,
$\numu$~CC events also contribute significantly to the background
along with intrinsic $\nue$ events in the beam and $\nutau$ events
that have oscillated from $\numu$.

Determining the composition of the background is important for this
analysis since at the Far detector a fraction of the $\numu$ events
have oscillated away and therefore the background from $\numu$~CC
events is reduced. The other effect of oscillations is to introduce a
background from $\nutau$ in the Far detector that does not exist in
the Near detector. In contrast, the NC events do not oscillate away
and to first order that background component is the same in the Near
and Far detectors. MINOS took a data-driven approach to determining
the background composition by comparing the data with the simulation
for a number of data sets taken with the NuMI beam in special
configurations. For example, with the magnetic horns turned off the
peak in the energy spectrum disappears, which drastically changes the
CC/NC ratio as a function of energy. Similarly, data taken with the
beam configured to produce higher energy neutrinos has an enhanced NC
fraction at low energies. A fit to the ND data and MC across all these
special data sets was used to estimate the background composition and
determine the uncertainties on each component.

The selection of electron neutrino candidate events starts out with
fiducial volume cuts and ensuring the event is in time with the
low-duty-cycle NuMI beam. Electron showers penetrate only a few
(typically 6--12) planes and are transversely compact so any events
with tracks longer than 24~planes or with a track extending more than
15~planes beyond the end of a reconstructed shower are rejected. A
requirement is also made that events contain at least 5~contiguous
planes with an energy deposition at least half that of a minimum
ionizing particle. Any events with an energy less than 1~GeV or
greater than 8~GeV are also removed. After these pre-selection cuts
77\% of the signal, 39\% of NC events and 8.5\% of $\numu$~CC events
remain.

Further reduction of backgrounds is achieved by a more sophisticated
analysis of the energy deposition patterns in preselected events. The
first two MINOS results used an artificial neural network with 11
variables characterizing the transverse and longitudinal profile of
events. For the most recent MINOS analyses, a nearest-neighbor
``library event matching'' (LEM) technique is used. Each data event is
compared, one-by-one, to a large library of tens of millions of
simulated events. Since the detector is homogeneous, events occurring
throughout the volume are translated to a fixed reference location and
then compared at the level of individual strips. This approach is
computationally intensive and is made more manageable in two notable
ways: firstly, fluctuations in the energy deposition of individual
strips are allowed for; and secondly, library events are shifted by
$\pm$1~plane in search of a better match. The final LEM discriminant
is formed using a neural network that takes as its inputs the event
energy along with three variables derived from the 50~best-matched
events. A cut of ${\rm LEM}>0.7$ selects $(40.4\pm2.8)$\% of signal
events.

The predictions for the Far detector signal and backgrounds as a
function of energy and LEM uses the Near detector data as the starting
point. The simulated ratio of Near and Far detector rates for each
background type is used as the conversion factor to translate the Near
detector data into a Far detector prediction. 

Two data samples provide sidebands that allow many of the procedures
developed for this analysis to be tested and the accuracy of the
simulation to be probed. Firstly, $\numu$~CC events with cleanly
identified muons provide a sample of known hadronic showers once the
muon hits are removed. These muon-removed events are a lot like NC
interactions and the predicted and observed events at the Far detector
agree well. The second sideband is the ${\rm LEM}<0.5$ region that
contains almost no $\nue$ appearance events. The Far detector
prediction for this ${\rm LEM}<0.5$ region is obtained in the same way
as for the signal region and so all stages of the analysis up to the
final signal extraction are exercised: for example, determining the
background composition and extrapolating the Near detector data is
done in the same way.

A fit to the data, binned as a function of the LEM discriminant and
reconstructed energy, was performed using the full 3-flavor
oscillation framework including matter effects. The influence of the
already measured oscillation parameters was included when constructing
the contours.

Updated MINOS results were released this summer for neutrinos, along
with the first appearance results for
antineutrinos~\cite{Nichol:Neu2012}. With an exposure of
$10.6\times10^{20}$~POT in the neutrino-enhanced beam and assuming
$\snthetaonethreeNOBrackets=0$ ($\snthetaonethreeNOBrackets=0.1$,
$\delta=0$, normal mass hierarchy) MINOS expected to see 128.6 (161.1)
events in the Far detector; 152 events were observed.

With an exposure of $3.3\times10^{20}$~POT in the
antineutrino-enhanced beam and assuming $\snthetaonethreeNOBrackets=0$
($\snthetaonethreeNOBrackets=0.1$, $\delta=0$, normal mass hierarchy)
MINOS expected to see 17.5 (21.2) events in the Far detector; 20
events were observed.

\begin{figure}
\centering
\includegraphics[trim = 0.1mm 0.1mm 0.1mm 0.1mm, clip, width=0.6\columnwidth]{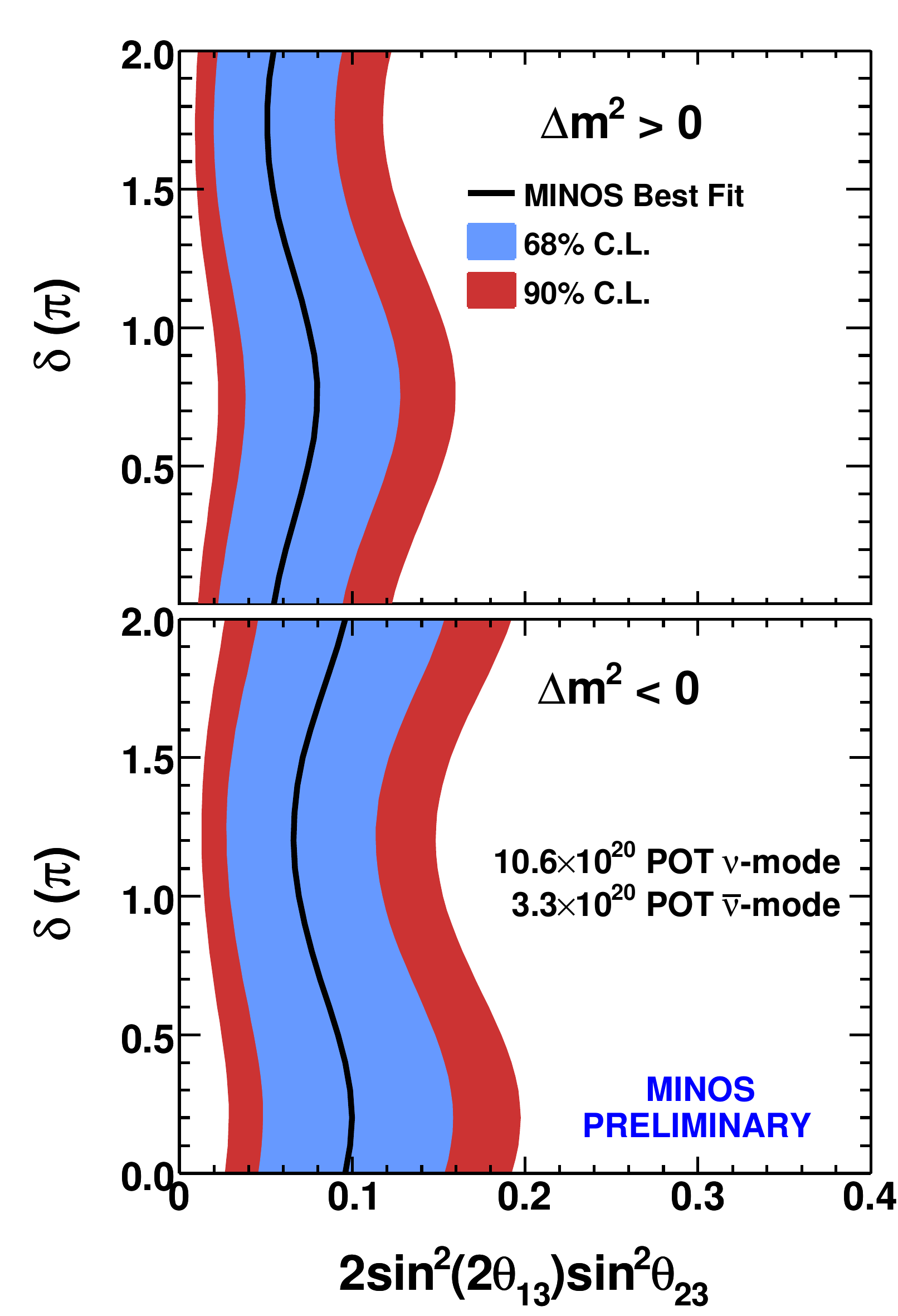}
\caption{MINOS allowed regions for the CP violating phase and
  $2\snthetaonethree\snthetatwothree$, obtained using the full data
  set of both neutrinos and antineutrinos. The top (bottom) plot
  assumes the normal (inverted) mass hierarchy. All values of the CP
  violating phase are consistent with the data and so the best fit
  parameters are shown by the black line. The blue (red) band shows
  the regions allowed at 68\% (90\%) confidence level. The
  $\thetaonethree=0$ hypothesis is disfavored at the 96\% confidence
  level. These results were preliminary at the time of
  writing~\cite{Nichol:Neu2012}. }
\label{fig:nueMINOS_joint} 
\end{figure}

The allowed regions as a function of the CP violating phase, $\delta$,
and $2\snthetaonethree\snthetatwothree$ are shown in
Figure~\ref{fig:nueMINOS_joint}. For $\delta=0$ and the normal
(inverted) mass hierarchy a best fit of
$2\snthetaonethree\snthetatwothree = 0.053~(0.094)$ is obtained; the
90\%~C.L. allowed range is
$0.01<2\snthetaonethree\snthetatwothree<0.12$
($0.03<2\snthetaonethree\snthetatwothree<0.19$) and the
$\thetaonethree=0$ hypothesis is disfavored at the 96\% confidence
level. These results are consistent with both the T2K result described
below in section~\ref{sec:resultsNueT2K} and with the reactor neutrino
experiments.

\begin{figure}
\centering
\includegraphics[trim = 0.1mm 0.1mm 0.1mm 0.1mm, clip, width=0.6\columnwidth]{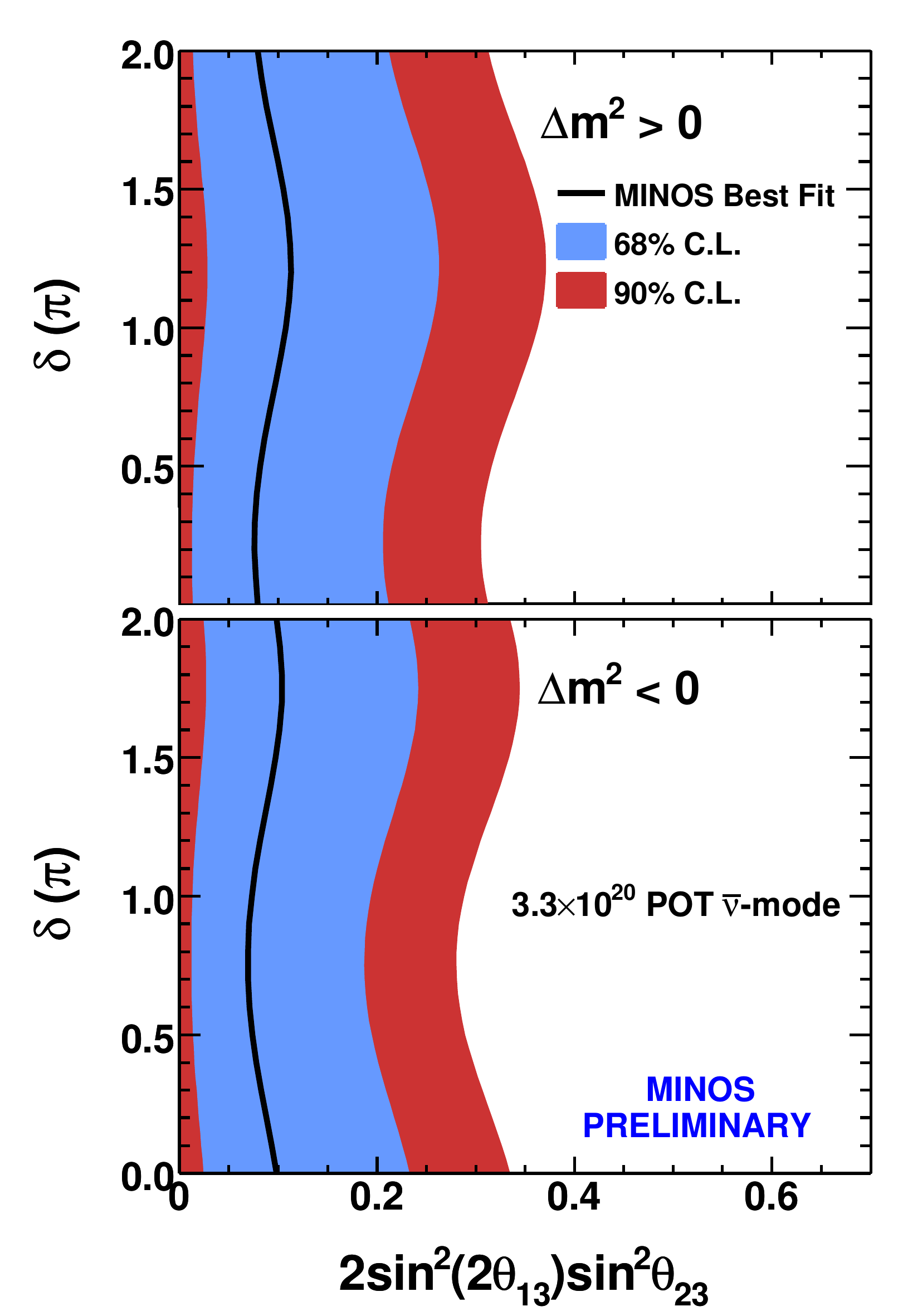}
\caption{MINOS allowed regions for antineutrinos as a function of the
  CP violating phase and $2\snthetaonethree\snthetatwothree$. See
  Figure~\ref{fig:nueMINOS_joint} for the full description. These
  results were preliminary at the time of
  writing~\cite{Nichol:Neu2012}.}
\label{fig:nueMINOS_rhc} 
\end{figure}

Figure~\ref{fig:nueMINOS_rhc} shows the results from the first
measurement of electron antineutrino appearance. The data set used for
this measurement was obtained with the NuMI beam set to enhance
production of antineutrinos. The limits on
$2\snthetaonethree\snthetatwothree$ are consistent with those from
neutrinos. Although, the smaller exposure and lower antineutrino
cross-section means that the limits are not as strong as for
neutrinos. Significant improvement in measurement of electron
antineutrino appearance is not expected until \nova takes data using
the NuMI beam configured for enhanced $\numubar$ production (see
section~\ref{sec:future}).


\subsection{T2K $\nue$ Appearance Results}
\label{sec:resultsNueT2K}
The primary goal of the T2K experiment is to discover electron
neutrino appearance and precisely measure the oscillation probability if
it exists. The experimental setup is optimized for this purpose.

T2K reported the first evidence of electron neutrino appearance
(2.5\,$\sigma$ significance, p-${\rm value}=0.7$\%) in June 2011 based
on $1.43\times 10^{20}$~POT data taken before the Great East Japan
Earthquake on 11th March 2011~\cite{Abe:2011sj}.

The goal of the analysis is to select $\nue$~CC interactions at high
efficiency and with the background contamination as low as
possible. At the peak of the T2K neutrino energy spectrum, around
600~MeV, the interaction of neutrinos is dominated by CC quasi-elastic
interaction (CCQE), $\nu_e + n \rightarrow e^- + p$, and that was
chosen as the target signal interaction. The benefit of CCQE
interaction is that with just a measurement of the momentum of the
final lepton, the parent neutrino energy can be reconstructed with a
good energy resolution of around 80~MeV.

The signature for signal events in the Super-Kamiokande detector is a
single showering (electron-like) ring in the expected energy region.
The two major sources of background events are the intrinsic electron
neutrino contamination in the beam mainly produced by muon decay in
the decay volume, and inelastic NC interaction of all flavors that
contain a $\pi^0$ in the final state. The $\gamma$s from $\pi^0$s are
detected in SK by the \Cer light from their electromagnetic showers,
which can be indistinguishable from the \Cer light distribution
produced by an electron. For example, if one of the two $\gamma$s from
the $\pi^0$ decay is missed, the event topology in SK becomes very
similar to that of the signal, i.e.\ a single electron-like ring.

Selection criteria for the signal event are as follows. The ``fully
contained in fiducial volume'' (FCFV) events are selected by
requiring: no event activity in either the outer detector or in the
100~$\mu$s before the event trigger time; at least 30~MeV
electron-equivalent energy deposited in the inner detector (defined as
visible energy $E_{vis}$); and the reconstructed vertex to be in the
fiducial volume of 22.5~kilotonnes. The event timing is required to be
within the range from -2~$\mu$s to 10~$\mu$s around the beam trigger
time.

Further selection cuts require events with the number of rings equal
to 1 and a PID consistent with being electron-like. The visible energy
is required to be $E_{vis} > 100$~MeV to reduce NC
elastic-interactions and decay electron backgrounds. It is also
required to have no associated delayed electron signal to reduce the
background from invisible $\pi \rightarrow\mu$ decay.  To suppress
misidentified $\pi^{0}$, a second electron-like ring is forced to be
reconstructed and a cut on the two-ring invariant mass $M_{inv} <
105$~MeV/c$^2$ is imposed. Finally, the neutrino energy $E_\nu^{rec}$,
computed using the reconstructed momentum and direction of the ring
assuming CCQE kinematics and neglecting Fermi motion, is required to
be $E_\nu^{rec} < 1250$~MeV.

The $\nue$ appearance signal efficiency is estimated with MC to be
66\% while rejection for $\numu + \numubar$~CC, beam $\nue$~CC, and NC
are $>99$\%, 77\%, and 99\%, respectively.

\begin{figure}
\centering
\includegraphics[width=0.6\columnwidth]{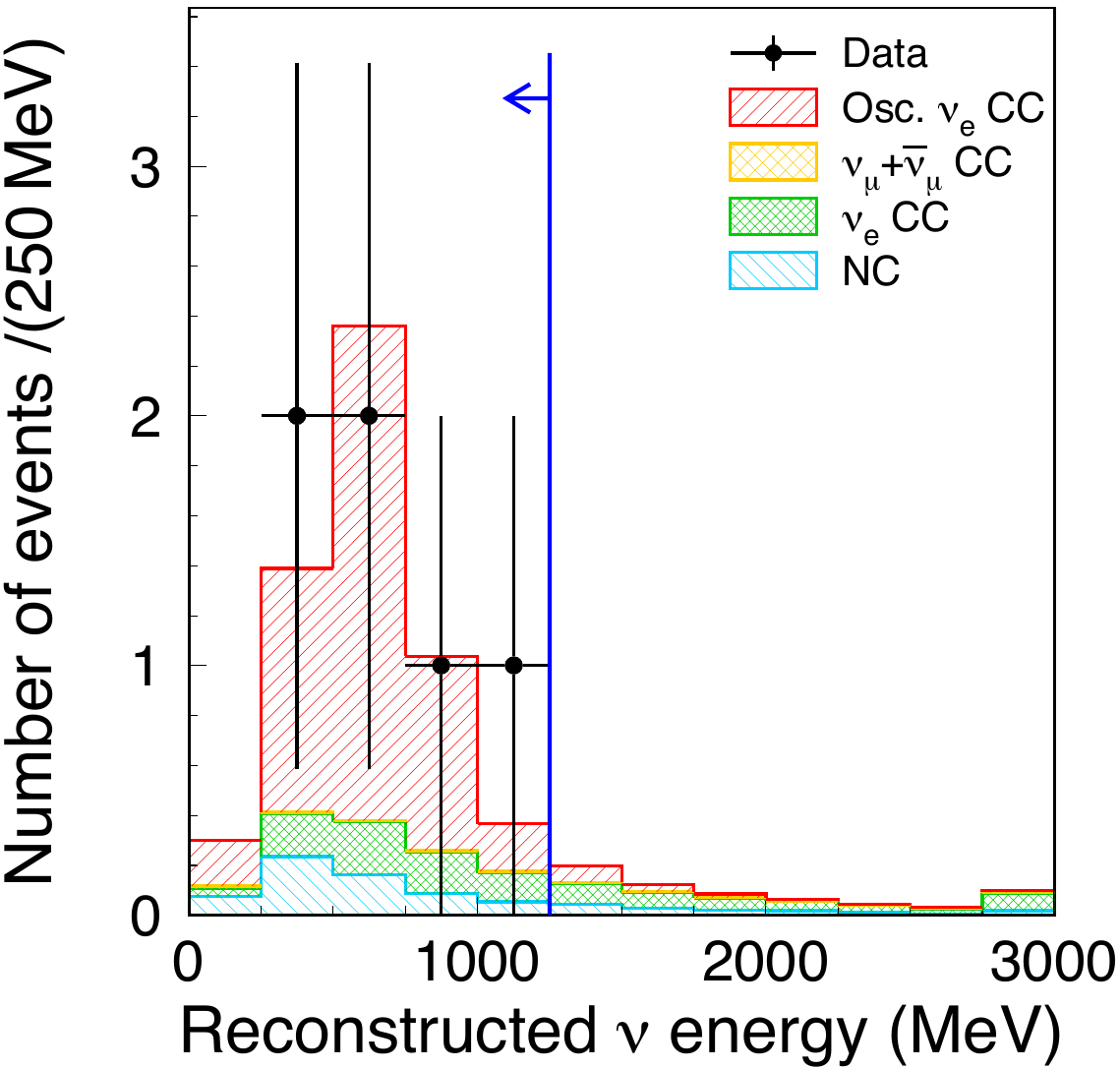}
\caption{Reconstructed neutrino energy $E_\nu^{rec}$ spectra for T2K
  $\nue$ appearance search. The black points show the 6 candidate
  events observed in SK using $1.43\times 10^{20}$~POT data. Using
  $\snthetaonethree=0.1$ the red histogram is the predicted appearance
  signal, the expected background shown in yellow is for muon
  neutrinos, green is for the electron neutrinos intrinsic to the beam
  and blue is for the NC events.}
\label{fig:T2KnueEnu} 
\end{figure}

The selection is applied to the data and 6 events in SK are selected
as signal candidates from all data before the earthquake, corresponding
to $1.43\times 10^{20}$~POT. The $E_\nu^{rec}$ distribution of the
observed events together with the signal and background expectations are
shown in Figure~\ref{fig:T2KnueEnu}.

The expected signal and background events are estimated using the far
detector MC simulation with the constraints and inputs from
measurements of near detector $\numu$~CC events and external
data. These external data include hadron production measurements made
by the NA61 experiment~\cite{Abgrall:2011ae,Abgrall:2011ts} using
30~GeV protons impinging on the neutrino production target and also
neutrino interaction cross-sections measured by previous experiments
such as MiniBooNE\@.

The off-axis near detector measures the number of inclusive $\numu$~CC
events by selecting events with a single negative muon.  The ratio of
the observed number of events to that from the MC simulation is $1.036
\pm 0.028({\rm stat})^{+0.044}_{-0.037} ({\rm det. syst}) \pm 0.038
({\rm phys. syst})$. This near detector ratio is multiplied by the
number of events from the far detector simulation to give the
predicted number of events in the far detector data. This method
provides partial cancellation of uncertainties in the absolute flux
and cross-sections at the far detector.

The number of background events thus obtained when
$\snthetaonethree=0$ is estimated to be $1.5\pm 0.3$~(syst). The major
contributions to the background systematic error come from the beam
flux (8.5\%), cross-section (14\%) and far detector systematic error
(15\%). The probability that the observed number of events becomes 6
or larger if $\snthetaonethree=0$ is calculated to be 0.7\%, which
corresponds to a 2.5\,$\sigma$ excess.

\begin{figure}
\centering
\includegraphics[width=0.6\columnwidth]{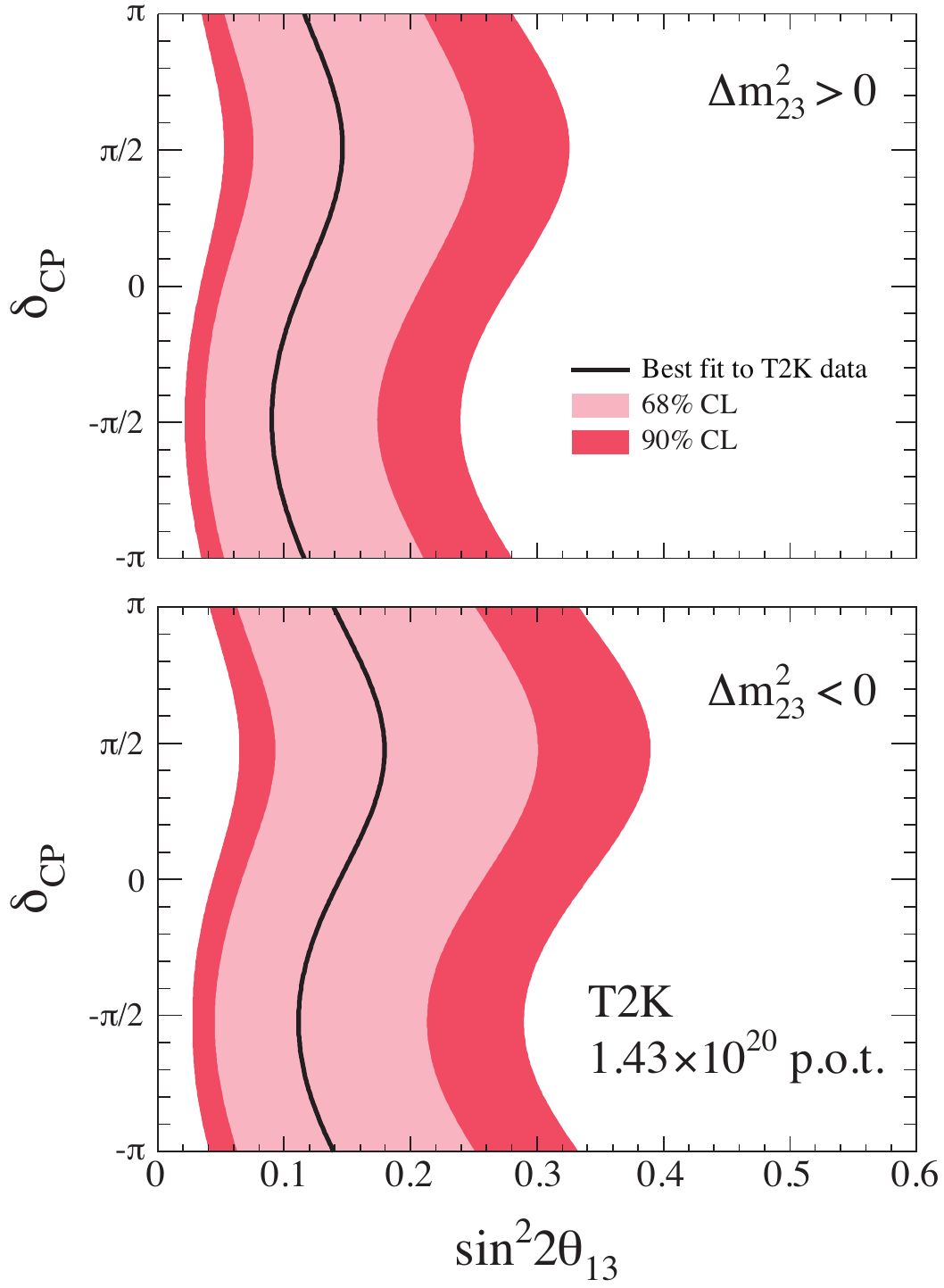}
\caption{Allowed regions in the $\snthetaonethreeNOBrackets$--$\delta$
  plane from the T2K $\nue$ appearance measurement. Light (dark) red
  areas are 68\%~C.L. and 90\%~C.L. regions. Solid black curves are
  best fit relations.}
\label{fig:T2Knuecont} 
\end{figure}

The constraints on the oscillation parameters are evaluated also by
using only the number of events. The confidence intervals are $0.03
(0.04) < \sin^22\theta_{13} < 0.28 (0.34)$ at 90\%~C.L. and the best
fit parameters are $\snthetaonethree = 0.11 (0.14)$ for the normal
(inverted) hierarchy assuming $\sintttwotwothree = 1$, $\Delta
m_{32}^2 = 2.4\times 10^{-3}$~eV$^2$ and
$\delta=0$. Figure~\ref{fig:T2Knuecont} shows the T2K allowed regions
of parameters in the $\snthetaonethreeNOBrackets$--$\delta$ plane.

To summarize the T2K $\nue$ appearance search, 6 signal candidate
events are detected while the expected number of background events at
$\snthetaonethree=0$ is $1.5\pm 0.3$. The probability to observe 6 or
more events without $\nue$ appearance is 0.7\%, which corresponds to
2.5\,$\sigma$ significance\footnote{In Summer 2012, T2K updated the
  results with $3.01\times 10^{20}$ POT of
  data~\cite{Nakaya:Neu2012}. The observed number of events is~11
  while the expected background is $3.22\pm 0.43$ at
  $\snthetaonethree=0$, which corresponds to 3.2\,$\sigma$
  significance and provides further firm evidence of $\nue$
  appearance.}.  Constraints on the
$\snthetaonethreeNOBrackets$--$\delta$ space are given for both the
normal and inverted mass hierarchy.


\section{Results on New Physics Searches}
\label{sec:resultsNewPhysics}
The provision of intense and relatively well understood neutrino beams
along with large detectors has opened up whole new avenues to look for
new physics. Here we focus on three main areas:
section~\ref{sec:resultsSterile} describes the searches for sterile
neutrinos; section~\ref{sec:tof} briefly summarizes neutrino velocity
measurements; and section~\ref{sec:lorentz} describes searches for
Lorentz symmetry violation.


\subsection{Searches for sterile neutrinos}
\label{sec:resultsSterile}
While the conventional picture of oscillations between three active
neutrino flavors is well established, the possibility of mixing with
one or more unseen sterile neutrinos is not excluded. Neutral-current
(NC) interaction cross-sections are identical for the three active
flavors and so no change in the NC event rate would be observed as a
function of $L/E$ in the standard neutrino model. MINOS provided the
first limits on the fraction of mixing to sterile neutrinos allowed at
the atmospheric mass splitting in~\cite{Adamson:2008jh}, with details
given in a longer paper~\cite{Adamson:2010wi}. Earlier, in 2000,
Super-Kamiokande had excluded the possibility of maximal
$\numu\rightarrow\nust$ oscillations at 99\%~C.L.~\cite{Fukuda:2000np}
by exploiting the effect such oscillations would have on both the NC
event rate and the number of $\numu$ and $\nutau$ candidate events
(the difference in the neutrino-matter interaction of $\numu$ and
$\nutau$ compared to $\nust$ is significant for atmospheric neutrinos
of the energy measured by SK). More recent observations of $\nutau$
appearance~\cite{Abe:2006fu,Abe:2012jj} also constrain oscillations to
sterile neutrinos, although limits are not directly given in those
papers. The current best limits on the fraction of mixing to sterile
neutrinos are from MINOS and given in~\cite{Adamson:2011ku}.

Selection of NC events in the MINOS detectors requires careful study
since the visible energy is relatively low and there is no distinct
feature to the events (for example, missing transverse momentum is not
easily observed in the MINOS detectors). NC candidate events can have
signal in as few as 4 scintillator strips. The high rate environment
of the ND, where there are around 16 events per 10~$\mu$s beam spill,
requires additional selections on timing and topology: events must be
separated by at least 40~ns and events that occur within 120~ns of
each other must be separated in the beam direction by at least 1~m. To
select an NC-candidate event sample the length of the event has to be
less than 60~planes and any track in the event must not extend beyond
the end of a shower by more than 5~consecutive planes.

\begin{figure}
\centering
\includegraphics[trim = 0.1mm 0.1mm 0.1mm 0.1mm, clip, width=0.6\columnwidth]{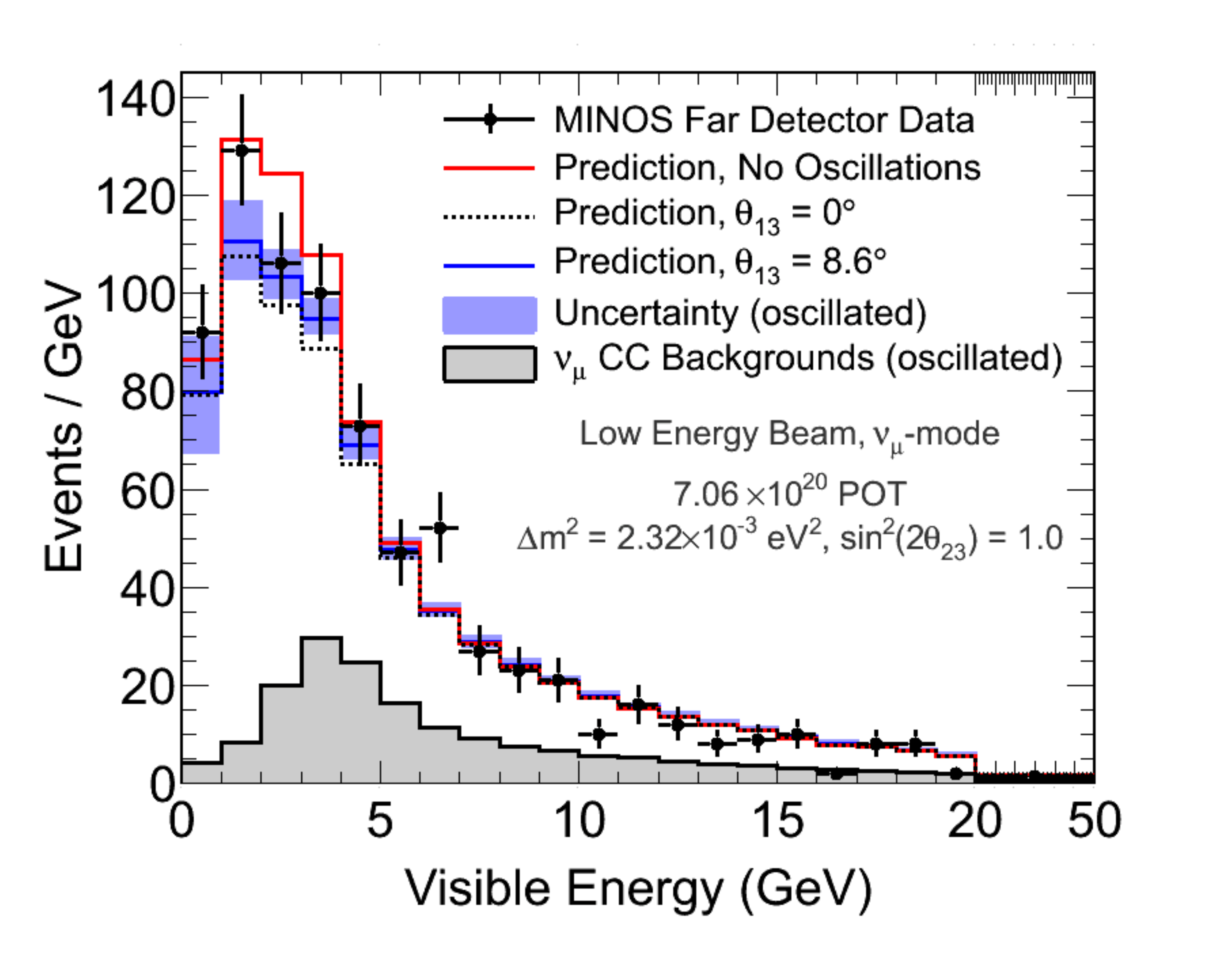}
\caption{Visible energy spectrum of MINOS Far detector neutral-current
  event candidates~\cite{Nichol:Neu2012}. The data are shown by the
  black points. The prediction obtained from the Near detector data is
  shown for three cases: no oscillations (red); oscillations with
  atmospheric parameters and $\thetaonethree=8.6^{\circ}$ (dashed
  black); and oscillations with atmospheric parameters and
  $\thetaonethree=0^{\circ}$ (dashed black). The contamination of the
  NC spectrum from $\numu$~CC events is shown by the gray histogram.}
\label{fig:sterileNCSpect} 
\end{figure}

An extrapolation procedure similar to that used in the $\nue$
appearance analysis (see section~\ref{sec:resultsMINOSNueApp}) is used
to form the Far detector prediction for the NC
spectrum. Figure~\ref{fig:sterileNCSpect} shows the visible energy
spectrum of Far detector candidate NC events. The data can be seen to
be consistent with no oscillation to sterile neutrinos.

Many sources of systematic uncertainty on the MINOS NC results are
similar to the $\numu$ disappearance and $\nue$ appearance
measurements (see sections~\ref{sec:minosNuMuDisap}
and~\ref{sec:resultsMINOSNueApp} respectively), for example the
absolute and relative energy scale of hadronic showers, and the
relative event rate normalization. Uncertainties specific to the NC
measurement are in the Near and Far detector selection, and in the
CC~background. The latest results, given below, are approaching the
systematic limit for how much further these measurements can be
improved by MINOS.

A straightforward phenomenological approach to presenting the limits
on the allowed level of sterile neutrino mixing is to consider the
fraction, $f_s$, of the disappearing $\numu$ flux that could oscillate
to $\nust$. MINOS finds $f_s < 0.22~(0.40)$ at 90\%~C.L., where the
number in brackets is the limit assuming maximal $\nue$ appearance at
the CHOOZ limit. The alternative approach to presenting the limits is
in the context of a specific model. MINOS has considered two models:
firstly, one where the fourth mass eigenstate $m_4=m_1$; and secondly
where $m_4>>m_3$. The 90\%~C.L. limits obtained from MINOS data are
$\theta_{24}<7^\circ(8^\circ)$ and $\theta_{34}<26^\circ(37^\circ)$ in
the $m_4>>m_3$ model, and $\theta_{34}<26^\circ(37^\circ)$ in the
$m_4=m_1$ model. In the future, the MINOS+ experiment will extend the
sensitivity to sterile neutrinos, in particular through also
constraining the disappearance of $\numu$ and $\numubar$ (see
section~\ref{sec:futSensNewPhysics}).


\subsection{Neutrino Velocity}
\label{sec:tof}
In 2007 MINOS made the first measurement of neutrino velocity in a
long-baseline experiment~\cite{Adamson:2007zzb}. The time of flight
between the Near and Far detectors separated by $734\,298.6\pm0.7$~m
was measured to be $-126\pm32({\rm stat})\pm64({\rm syst})$~ns
w.r.t.\ the calculated time for light to travel the same distance,
which corresponds to $(v-{\rm c})/{\rm
  c}=(5.1\pm2.9)\times10^{-5}$. This result was systematically limited
by uncertainties in the timing system and its overall sensitivity
comparable with previous neutrino velocity measurements from
short-baseline experiments~\cite{PhysRevLett.43.1361}.

Dedicated upgrades to the OPERA experiment's timing system along with
high statistics neutrino event samples gave substantially improved
sensitivity to the neutrino velocity. In September 2011 they released
their result $(v-{\rm c})/{\rm c}=[2.37\pm0.32({\rm
    stat})^{+0.34}_{-0.24}({\rm syst})]
\times10^{-5}$~\cite{Adam:2011zb}, which generated huge world wide
media interest. However, in February 2012 the OPERA collaboration
released a statement, available on their website, saying that two
errors in the timing system had been found that could potentially
bring the neutrino velocity back into line with expectations from
special relativity. This was followed by a measurement from the ICARUS
experiment~\cite{Antonello:2012hg}, also located in the LNGS
laboratory, that was of similar sensitivity to OPERA but consistent
with expectations. Around the time of writing OPERA released an
updated result $(v-{\rm c})/{\rm c}=[0.27 \pm 0.31({\rm
    stat})^{+0.34}_{-0.33}({\rm syst})]
\times10^{-5}$~\cite{Adam:2011zb}, confirming that they had understood
the anomaly in their first result. Results from
Borexino~\cite{AlvarezSanchez:2012wg} and LVD~\cite{Agafonova:2012rh}
are also consistent with OPERA and ICARUS. These results from four of
the experiments located at Gran Sasso are the world's most precise
measurements of the neutrino velocity and they are approaching their
ultimate systematic limit. Future measurements that use different
beamlines and hence have a lower number of correlated systematic
uncertainties will be important. MINOS, and in future MINOS+, will
exploit recent investments in their timing systems with the aim of
reducing the systematic uncertainties further~\cite{Adamson:Neu2012}.


\subsection{Searches for Lorentz Symmetry Violation}
\label{sec:lorentz}
MINOS has investigated whether neutrinos have a preferred
direction in space and hence violate Lorentz symmetry and consequently
also CPT symmetry. This search was performed in the context of the
Standard Model Extension
theory~\cite{Colladay:1996iz,Colladay:1998fq,PhysRevD.69.105009} that
provides a model-independent framework with coefficients to quantify
the various ways Lorentz symmetry could be violated. The experimental
observable for these searches is a sidereal variation in the rate of
neutrino interactions. MINOS has results for $\numu$ and $\numubar$ in
the Near detector as well as $\numu$ in the Far
detector~\cite{Adamson:2008ij,Adamson:2012hp,Adamson:2010rn}.

The rotation of the Earth rotates the neutrino beam in the
sun-centered inertial reference frame with the sidereal frequency of
$2\pi / 23^{\rm h}56^{\rm m}04.090\,53^{\rm s}$. The offset of the
sidereal frequency from the Earth's rotational frequency of
$2\pi/24^{\rm h}$ is experimentally advantageous since diurnal effects
can potentially average out over the course of a year. The MINOS
analysis was performed by examining the data as a function of local
sidereal phase (LSP), which is simply the local sidereal time divided
by the length of a sidereal day. Each neutrino event was placed in an
LSP histogram and the protons on target for each beam spill used in
the analysis were placed in a second LSP histogram. The ratio of the
two histograms gave the normalized number of neutrino events observed
as a function of LSP\@. Fast Fourier transforms to determine the power
associated with sinusoidal functions at the sidereal frequency and its
second harmonic were performed. To date, no sidereal variation of the
neutrino event rate has been detected.

In addition to long-baseline accelerator experiments, searches for
Lorentz symmetry violation have been performed by several other
neutrino experiments. This has allowed many of the coefficients in the
SME to be constrained over a wide range of directions, baselines and
neutrino energies. A comprehensive summary of experimental limits is
given in~\cite{Kostelecky:2008ts}.


\section{Future Sensitivities}
\label{sec:future}

The expected future physics sensitivities of experiments currently
running, or about to start taking data, are outlined
here. Section~\ref{sec:futSensOsc} describes the prospects for
measurements of the standard 3-flavor neutrino oscillation parameters
and section~\ref{sec:futSensNewPhysics} focuses on models of new
physics.


\subsection{Oscillation physics}
\label{sec:futSensOsc}
As of 2012, all three mixing angles are known to be nonzero and have
been measured to reasonably good accuracy. However, there is no
significant information on the mass ordering, the $\theta_{23}$ octant
or CP violation yet.  The main goals of long-baseline experiments in
the next decade will be to determine or obtain indications of the
present unknowns by improving the precision of the measurements as
much as possible. Since the CP violation term in the $\nue$ appearance
probability depends on all the mixing angles in some way, it is
important to improve the precision of $\theta_{23}$ through $\numu$
disappearance measurements as well as $\nu_e$ appearance. Further, if
\sintttwotwothree is not unity, then the determination of the
$\theta_{23}$ octant will tell us whether $\nuthree$ couples more
strongly to $\numu$ or $\nutau$.

T2K plans to accumulate up to $750~{\rm kW}\times 5\times10^7$~seconds
equivalent POT, which is about $8\times 10^{21}$~POT and 26~times the
exposure so far. The \nova sensitivities discussed below all assume
that \nova will run for three years in neutrino mode and three years
in antineutrino mode, for a total of $36 \times 10^{20}$ POT.  These
predicted sensitivities are largely based on analysis techniques that
were used by the MINOS experiment.  \nova expects to be able to
achieve somewhat better sensitivities as it incorporates additional
techniques allowed by \nova's finer segmentation and greater active
fraction.


\subsubsection{\numu Disappearance}
\begin{figure}
\centering
\includegraphics[width=0.8\columnwidth]{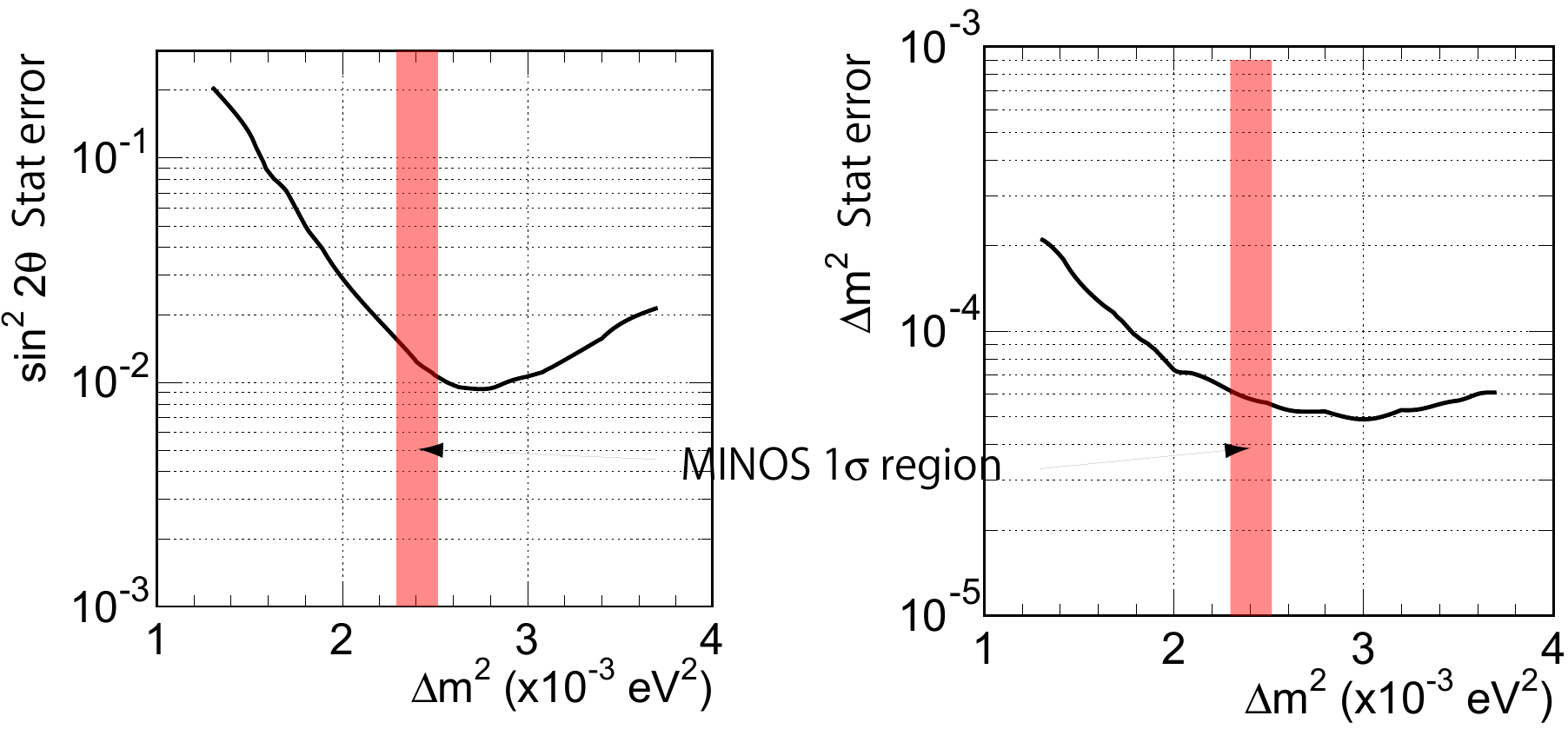}
\caption{T2K expected statistical precision on the oscillation
  parameters $\sintttwotwothree$ and $\dmsqtwo$ assuming an exposure
  of $750~{\rm kW}\times 5\times 10^7$~seconds as a function of true
  $\dmsqtwo$ \cite{T2Kproposal}. The 1\,$\sigma$ confidence intervals
  for $\dmsqtwo$ from MINOS are indicated by red hatches.}
\label{fig:T2Knmdisappstat} 
\end{figure}

\ \newline The disappearance of \numu charged current events measures
\sintttwotwothree and $\dmsqtwo$. The expected statistical precision
of the T2K $\nu_\mu$ disappearance measurements at $750~{\rm kW}\times
5\times 10^7$~seconds are plotted in
Figure~\ref{fig:T2Knmdisappstat}~\cite{T2Kproposal}.
The statistical precision reaches $\delta(\sin^22\theta_{23})\sim 1$\%
and $\delta(\dmsqtwo) \sim 0.05\times 10^{-3}$~eV$^2$. The goal for
the systematic uncertainties is to reach the same level as for the
statistical errors for both of the parameters.

\begin{SCfigure}[50.0][!htb]
\centering
\includegraphics[width=0.6\textwidth]{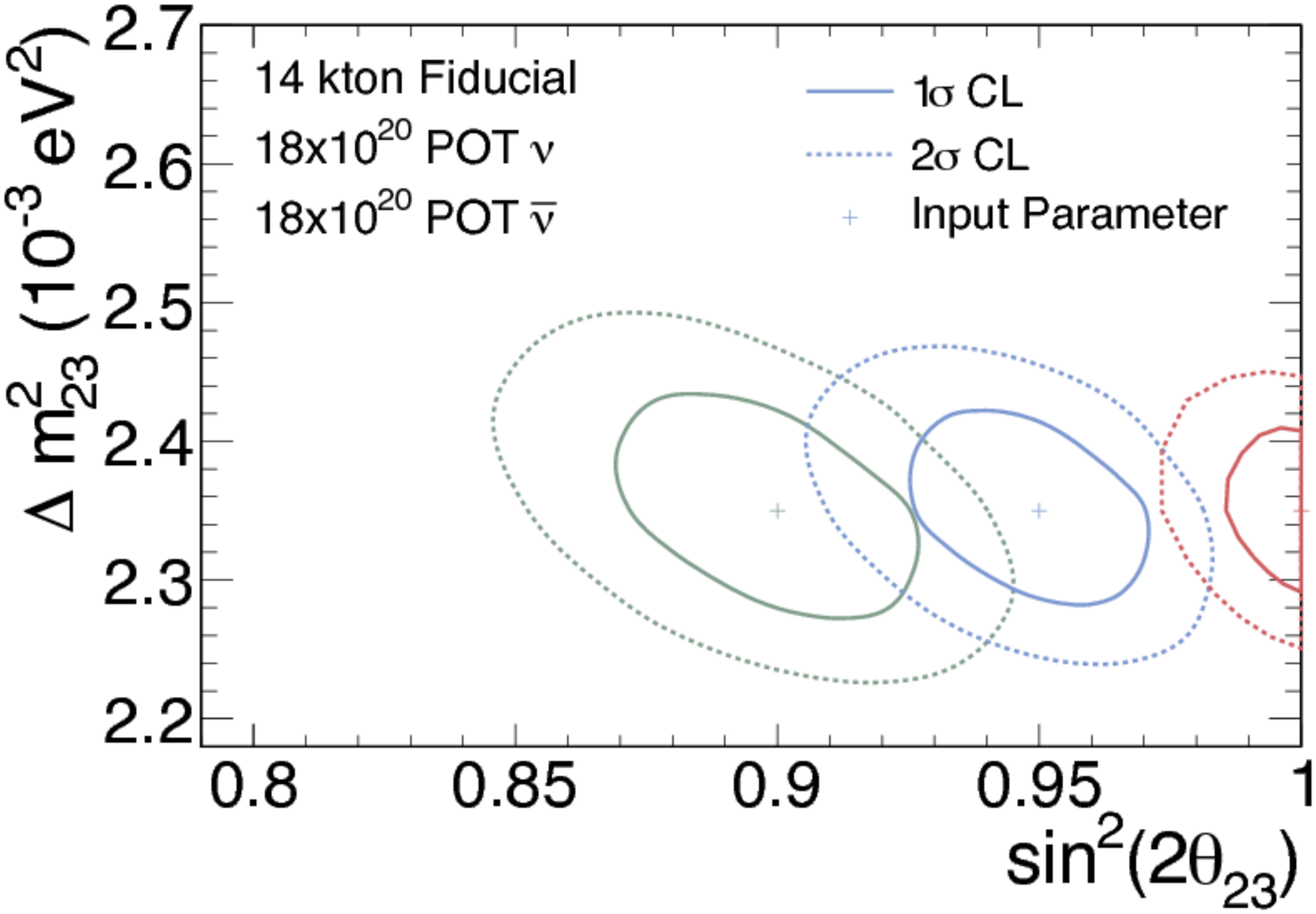}
\caption{One and two standard deviation \nova sensitivity contours for
  a joint measurement of $\dmsqtwo$ and \sintttwotwothree for three
  possible values of these parameters indicated by the plus signs.
  The single parameter measurement of \sintttwotwothree will be
  somewhat more sensitive than the extreme limits of the displayed
  contours.\vspace{0.75in}}
\label{fig:s22th23}
\end{SCfigure}

The latest MINOS measurement of \sintttwotwothree is $0.96 \pm
0.04$~\cite{Nichol:Neu2012}.  For the reasons cited above, \nova
should be able to make a measurement that is about a factor of two to
three more sensitive.  Figure~\ref{fig:s22th23} shows the \nova
sensitivity for three possible values of \sintttwotwothree.  \nova
will gain further information about $\thetatwothree$ from $\numutonue$
oscillations, as discussed below.


\subsubsection {$\numutonue$ Oscillations}
\begin{figure}[!htb]
\begin{minipage}[t]{3.0in}
\centering
\includegraphics[width=3.0in]{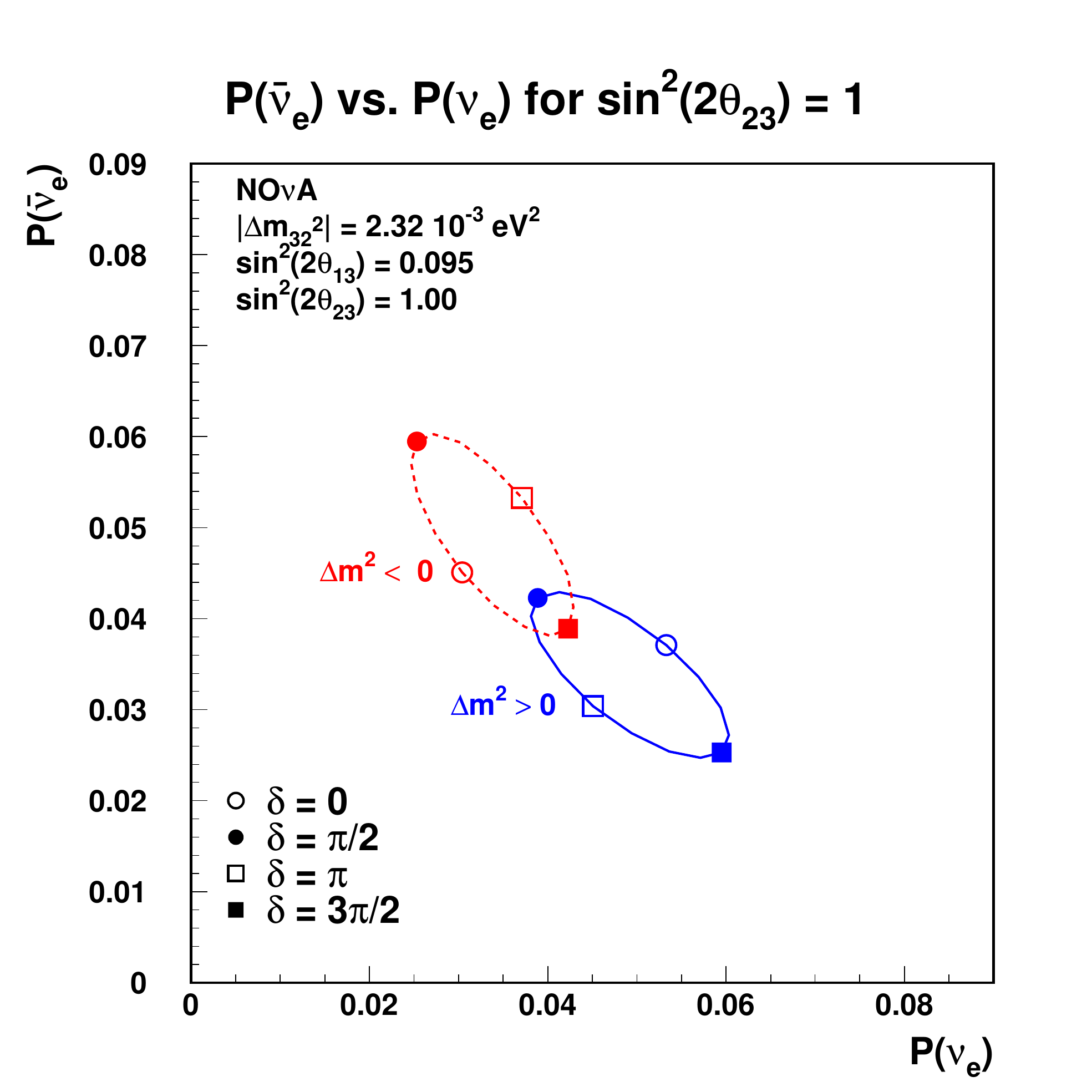}
\caption{Bi-probability plot for $\sintttwotwothree=1.00$.  See text
  for explanation.}
\label{fig:finder100}
\end{minipage}
\hfill
\begin{minipage}[t]{3.0in}
\centering
\includegraphics[width=3.0in]{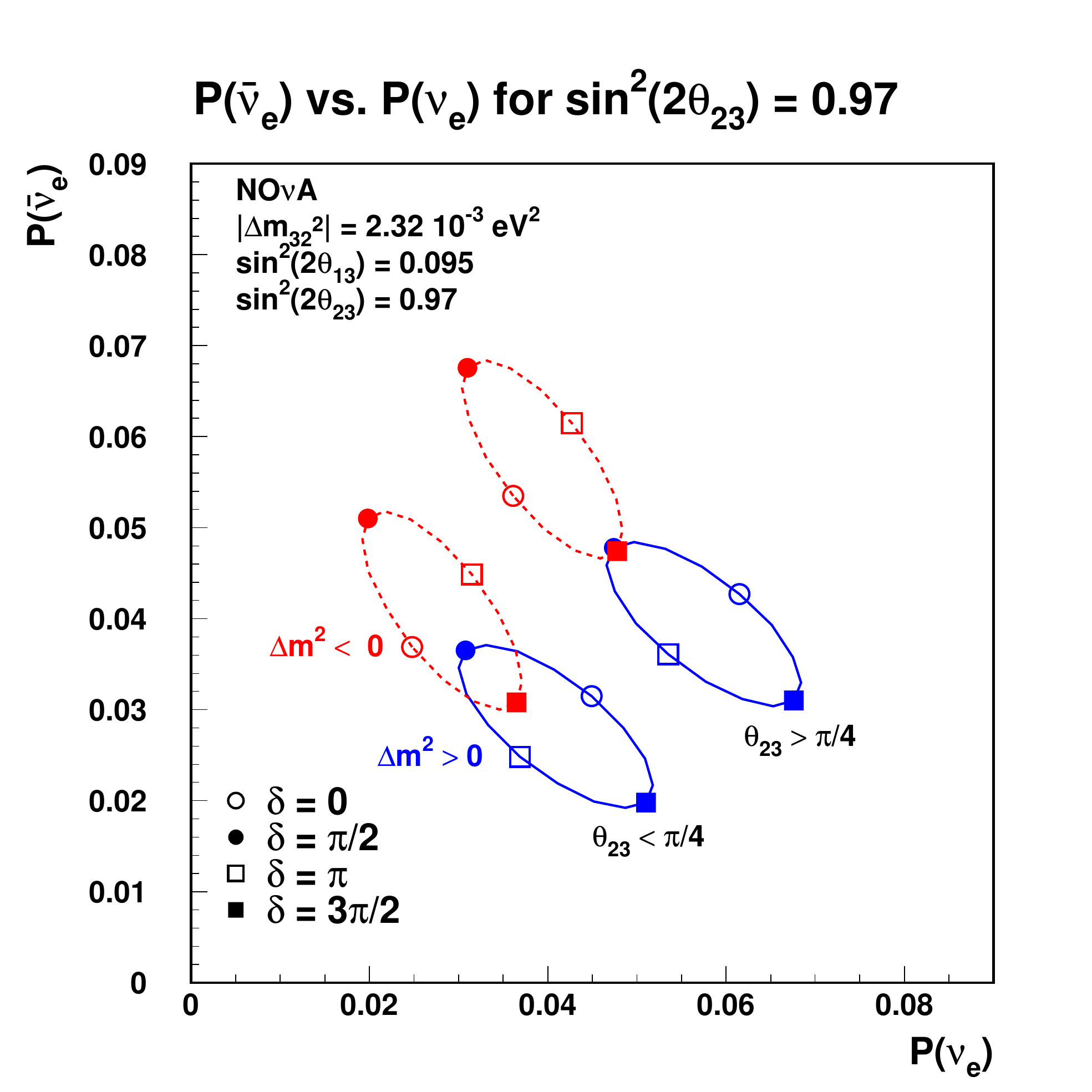}
\caption{Bi-probability plot for $\sintttwotwothree=0.97$.  See text
  for explanation.}
\label{fig:finder97}
\end{minipage}
\end{figure}

\ \newline The parameters for \numutonue oscillations are considerably
more complex than for $\numu$ disappearance.  This process is largely
proportional to both\ \sintttwoonethree and \sintttwothree, with large
perturbations caused by the mass ordering (through the matter effect)
and by CP violation.  A convenient way to see the dependences is
through bi-probability plots.  These plots show the loci of possible
\nova measurements of \numutonue and \numubartonuebar oscillation
probabilities, given a set of parameters.  These parameters include
\sintttwoonethree, which is fixed at 0.095, a value consistent with
the recent reactor
measurements\cite{Abe:2012tg,Dwyer:Neu2012,Ahn:2012nd}, and
\sintttwotwothree.  Figures~\ref{fig:finder100} and \ref{fig:finder97}
show bi-probability plots for \sintttwotwothree = 1.00 and 0.97,
respectively.  The CP-violating phase $\delta$ traces out the ovals
and the multiplicity of ovals represents the two possible mass
orderings and, for Figure~\ref{fig:finder97}, the ambiguity of whether
$\thetatwothree$ is larger or smaller than $\pi/4$.

\begin{figure}[!htb]
\begin{minipage}[t]{3.0in}
\centering
\includegraphics[width=3.0in]{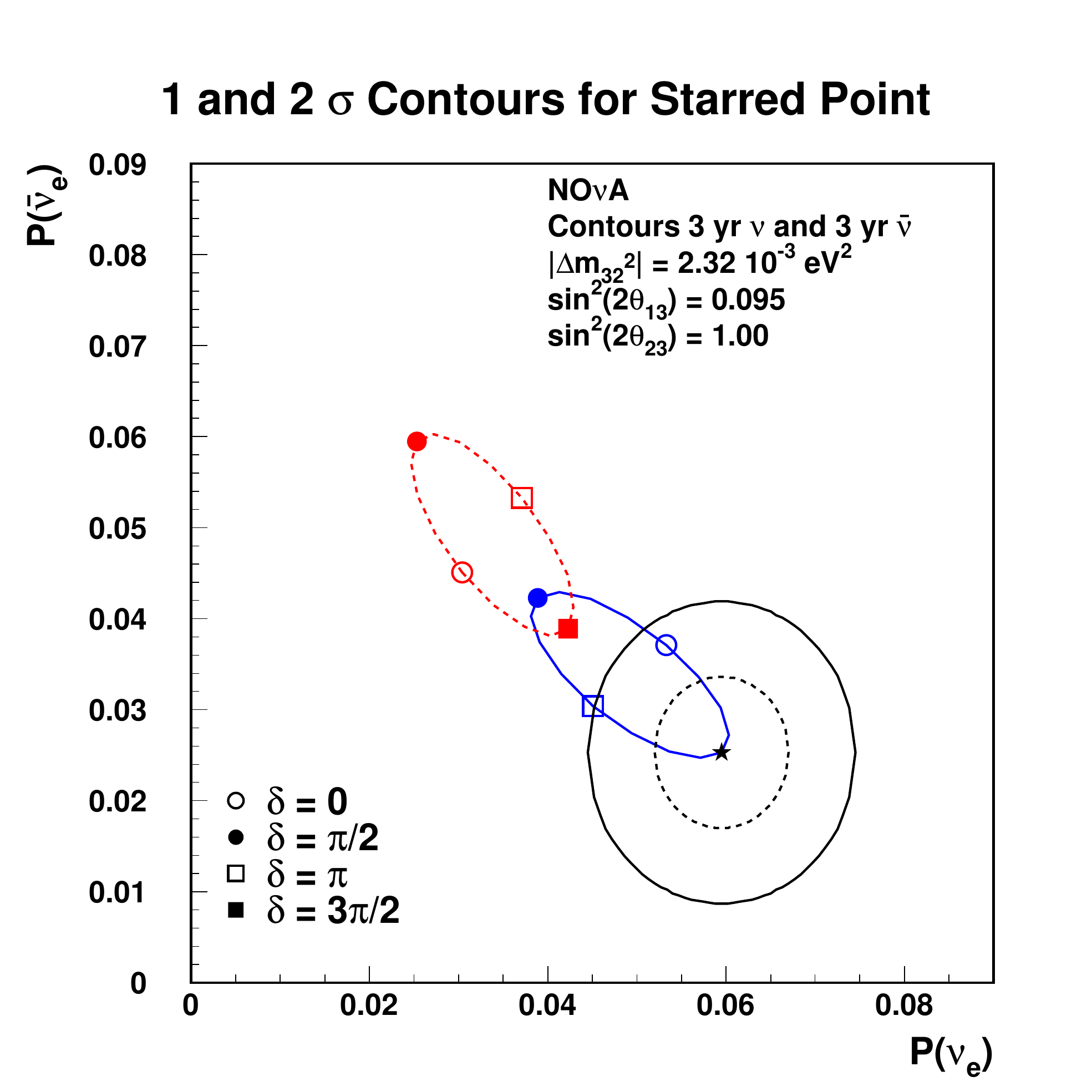}
\caption{Bi-probability plot for $\sintttwotwothree=1.00$ with \nova
  expected 1 and 2 standard deviation contours superimposed on the
  starred point.}
\label{fig:disp100n333}
\end{minipage}
\hfill
\begin{minipage}[t]{3.0in}
\centering
\includegraphics[width=3.0in]{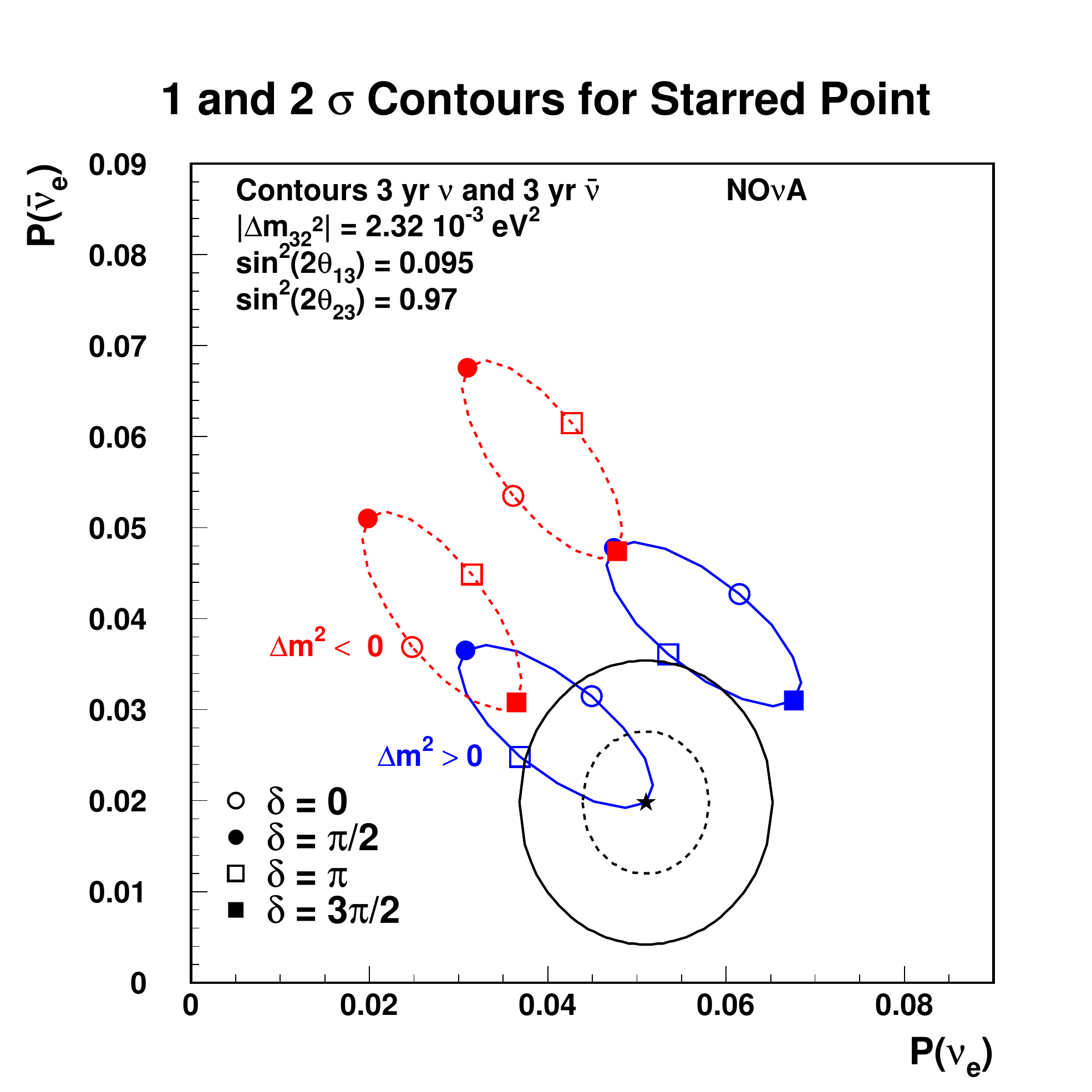}
\caption{Bi-probability plot for $\sintttwotwothree=0.97$ with \nova
  expected 1 and 2 standard deviation contours superimposed on the
  starred point.}
\label{fig:disp97ln333}
\end{minipage}
\end{figure}

A useful way to visualize what \nova will be able to do is to
superimpose one and two standard deviation contours on the
bi-probability plots.  For example, Figures~\ref{fig:disp100n333} and
\ref{fig:disp97ln333} show these contours for a favorable set of
parameters, normal mass ordering and $\delta=3\pi/2$.  The mass
ordering is resolved to more than two standard deviations, the
$\theta_{23}$ ambiguity is resolved to two standard deviations, and CP
violation is established to almost two standard deviations.  This
occurred because the matter effect and the CP-violating effect went in
the same direction, so there was no ambiguity.

\begin{SCfigure}[10.][!htb]
\centering
\includegraphics[width=3.0in]{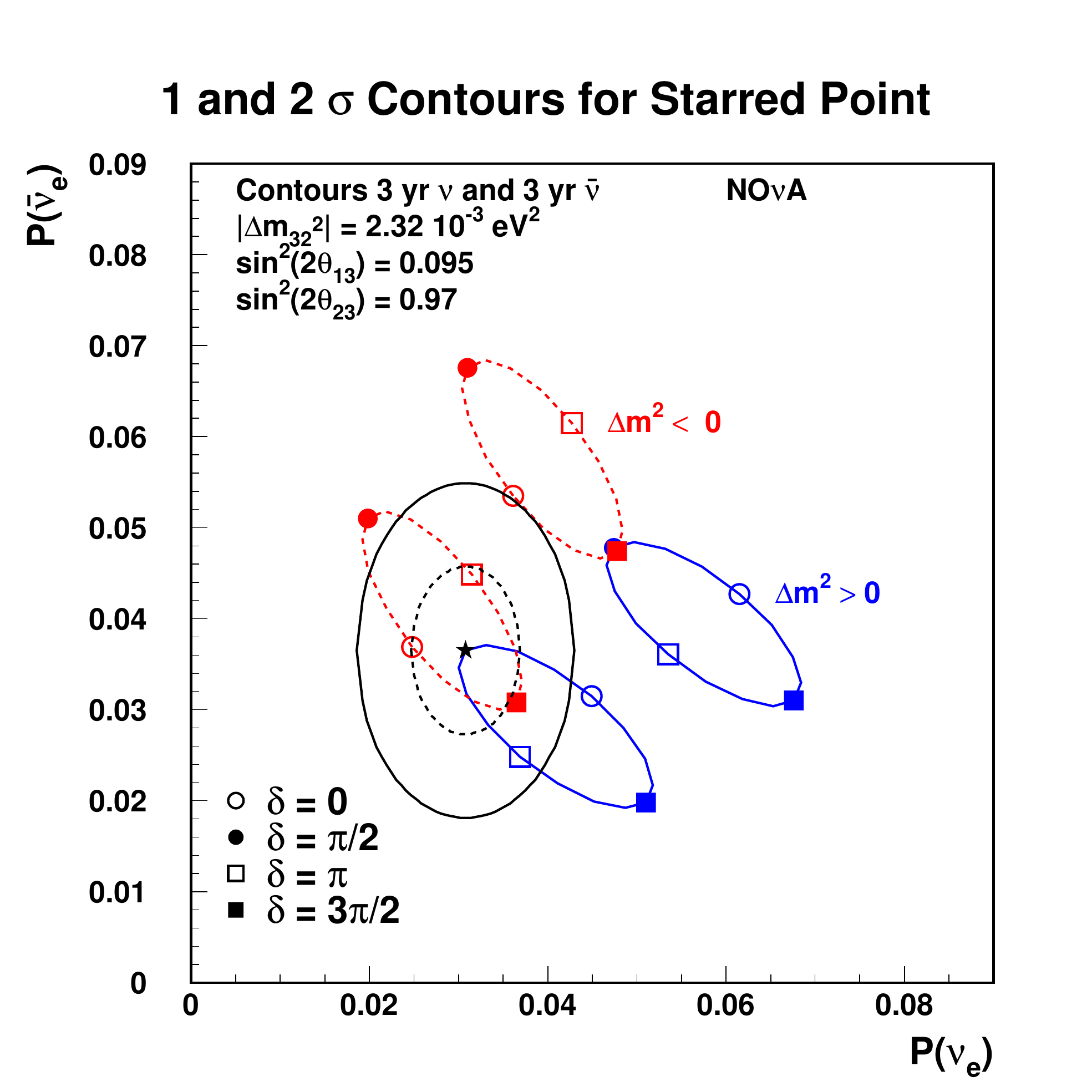}\hspace{0.5in}
\caption{Bi-probability plot for $\sintttwotwothree=0.97$ with \nova
  expected 1 and 2 standard deviation contours superimposed on the
  starred point.
\vspace{1.5in}}
\label{fig:disp97ln133}
\end{SCfigure}

An unfavorable set of parameters would be one in which the matter
effect and the CP-violating effect go in opposite directions so that
there is an ambiguity as to which direction each one went.  An example
of that is shown in Figure~\ref{fig:disp97ln133}.  The $\theta_{23}$
ambiguity is resolved, but the mass ordering is not, and therefore
there is little information on the CP-violating phase.  If nature
gives us this situation, then the only way to resolve the mass
ordering in the short term is to compare \nova measurements of
\numutonue\ oscillations with those from an experiment with a
different baseline.  The only experiment that meets that requirement
is T2K, which has a 295~km baseline.

\begin{figure}[!htb]
\begin{minipage}[t]{3.0in}
\centering
\includegraphics[width=3.0in]{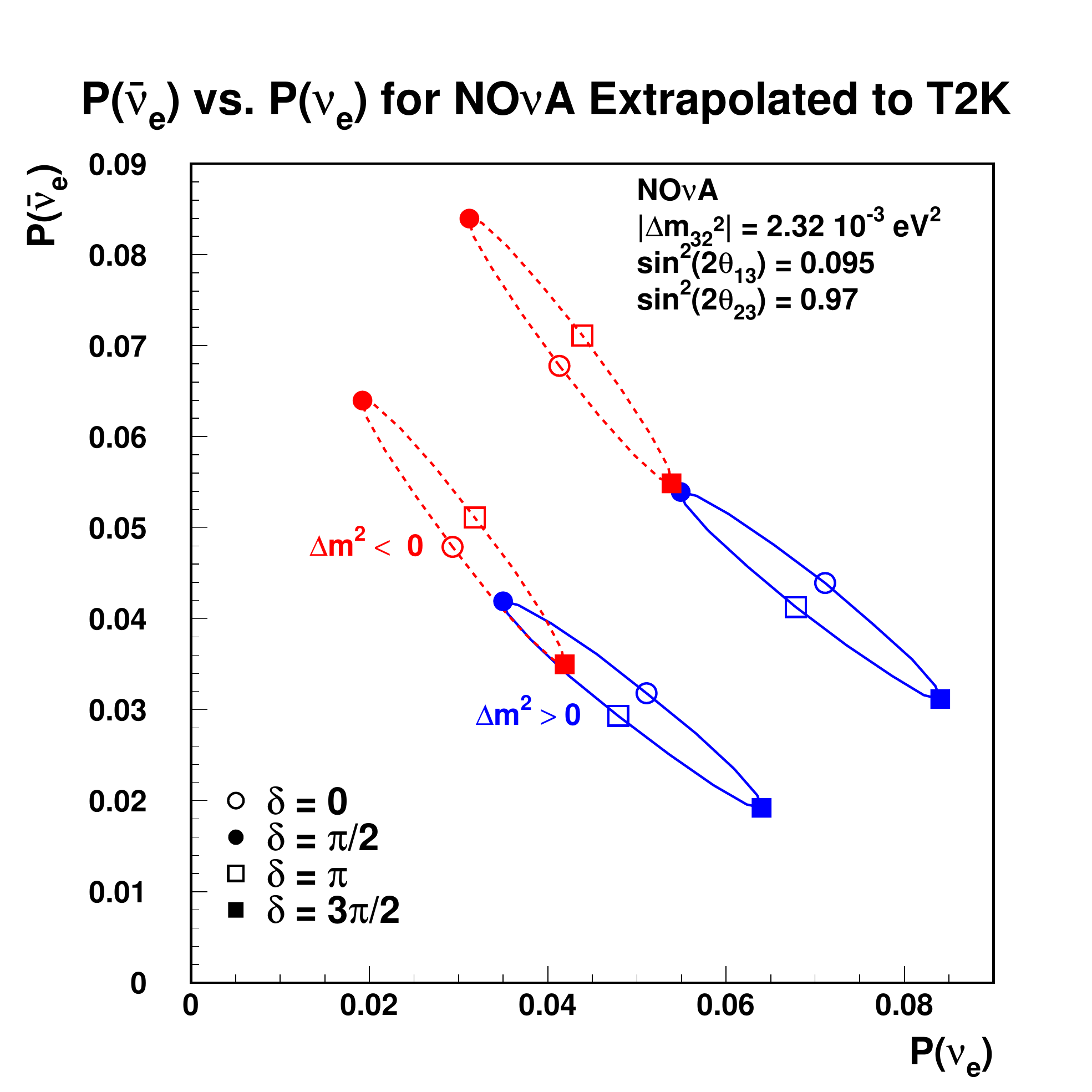}
\caption{Bi-probability plot for $\sintttwotwothree=0.97$ with \nova
  extrapolated to the average oscillation phase of T2K}
\label{fig:finder97ex}
\end{minipage}
\hfill
\begin{minipage}[t]{3.0in}
\centering
\includegraphics[width=3.0in]{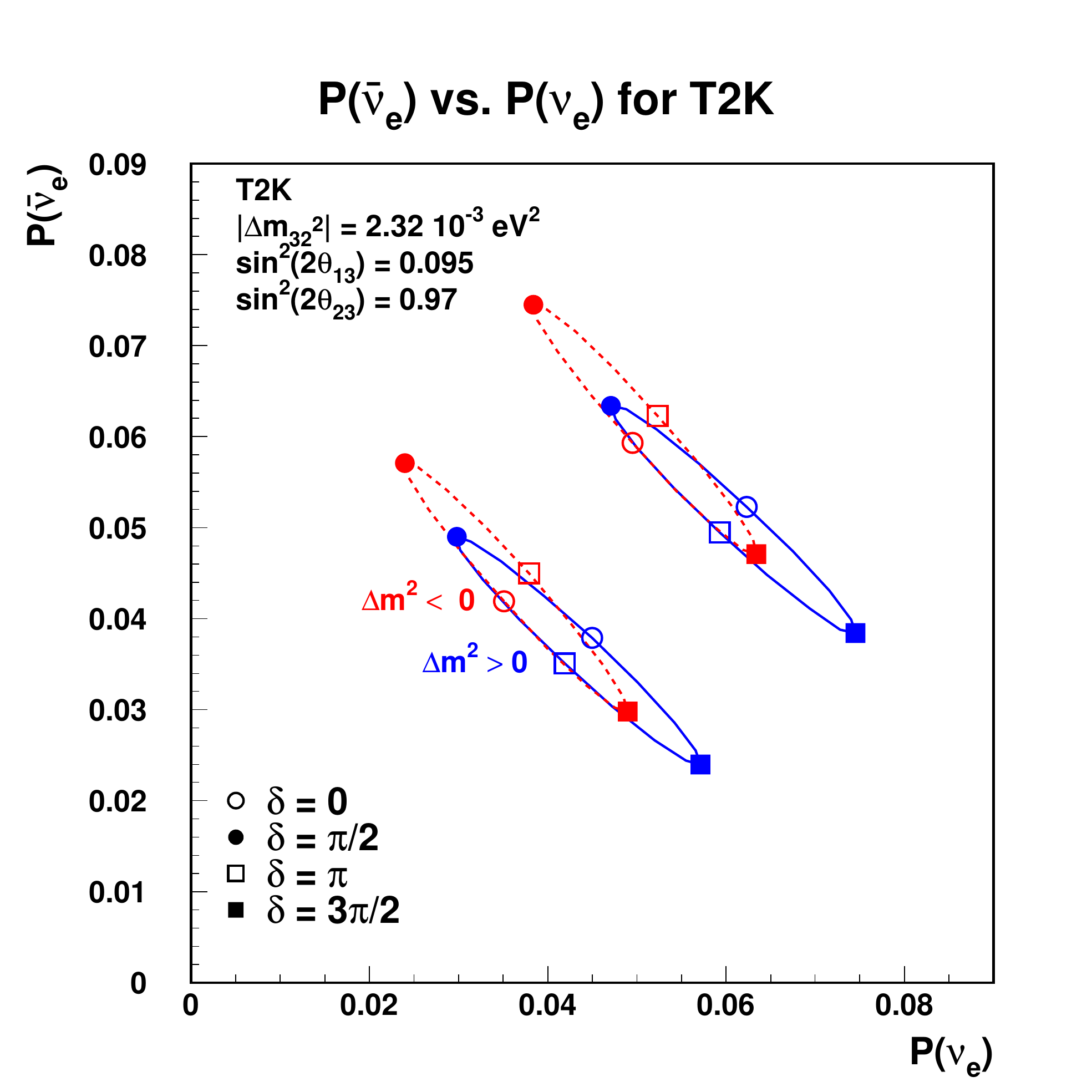}
\caption{Bi-probability plot for $\sintttwotwothree=0.97$ for T2K.}
\label{fig:finderT2K97}
\end{minipage}
\end{figure}

The algorithm for resolving the mass ordering is quite simple.  If
\nova measures a higher probability of \numutonue\ oscillations than
T2K, then the mass ordering is normal; if it is the opposite, it is
inverted.  That is because \nova and T2K will see the identical
CP-violation, but T2K will see a much smaller matter effect due to its
shorter baseline.  The only catch in this algorithm is that the
comparison must be done at the same point in the oscillation phase,
and the two experiments run at different average oscillation phases.
Figures~\ref{fig:finder97ex} and \ref{fig:finderT2K97} show the
bi-probability plots in which the \nova measurements have been
extrapolated to the same oscillation phase as the T2K measurements.  A
comparison of the two plots shows that the algorithm works for all
values of $\delta$.

\begin{figure}[!htb]
\begin{minipage}[t]{3.0in}
\centering
\includegraphics[width=3.0in]{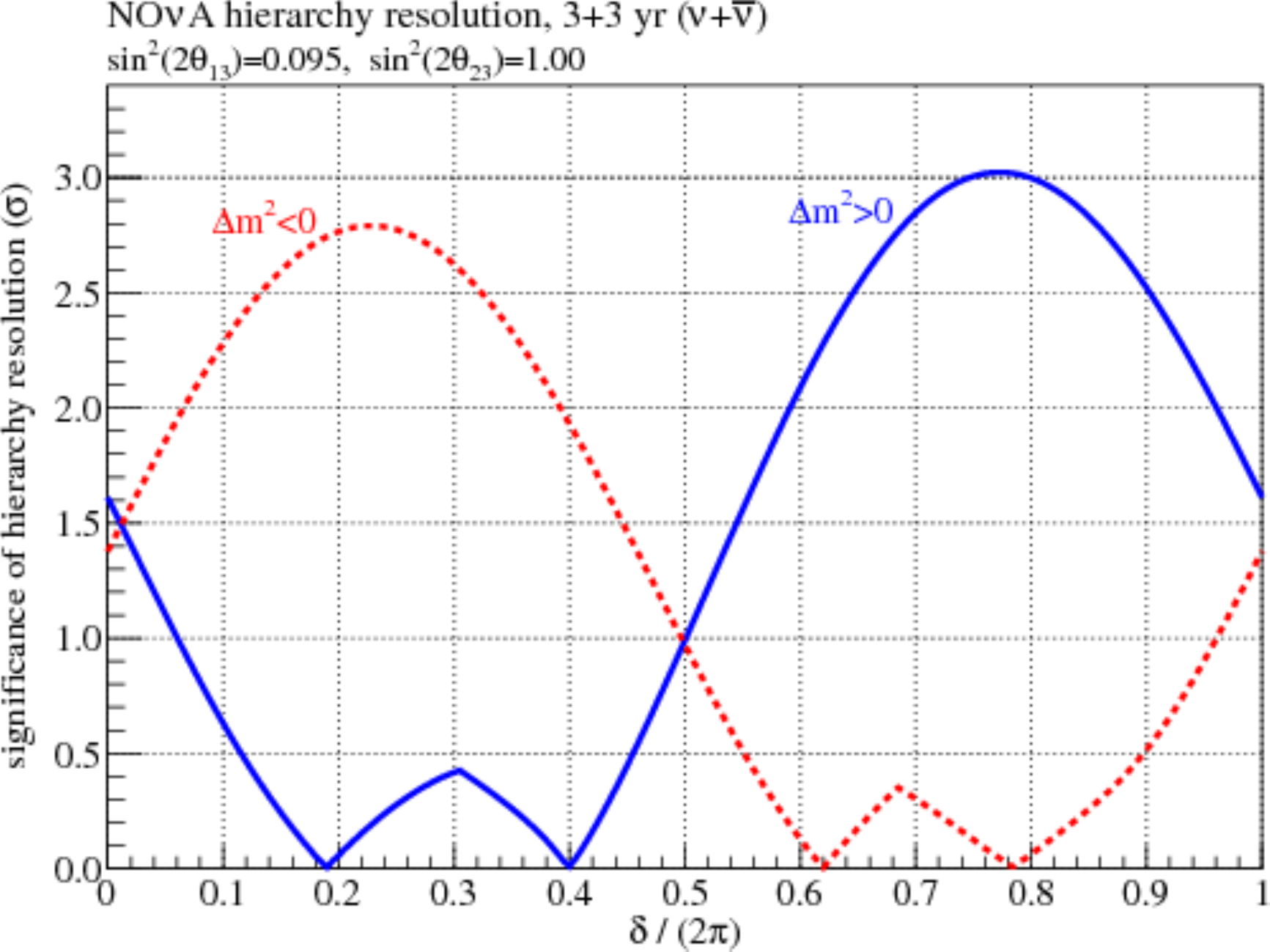}
\caption{Significance of the resolution of the mass ordering as a
  function of $\delta$ in standard deviations.  These sensitivities
  are for \nova alone for the two possible orderings and
  \sintttwotwothree = 1.0.  The zeros correspond to the crossing of
  the ovals in Figure~\ref{fig:finder100}.}
\label{fig:novaMassOrdering}
\end{minipage}
\hfill
\begin{minipage}[t]{3.0in}
\centering
\includegraphics[width=3.0in]{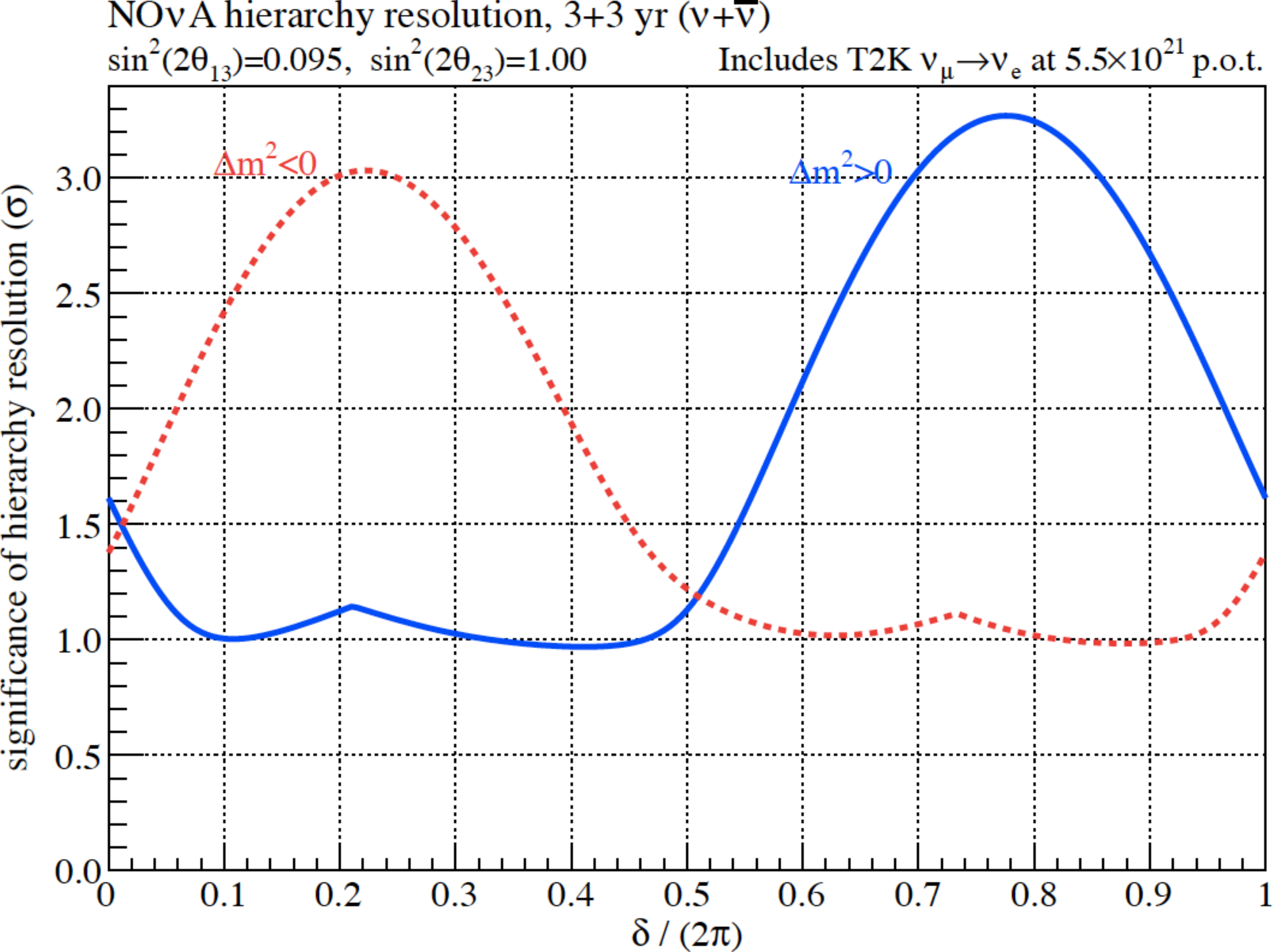}
\caption{Same as the figure to the left except that information from
  the T2K experiment has been included.}
\label{fig:novaT2KmassOrdering}
\end{minipage}
\end{figure}

\begin{figure}[!htb]
\begin{minipage}[t]{3.0in}
\centering
\includegraphics[width=3.0in]{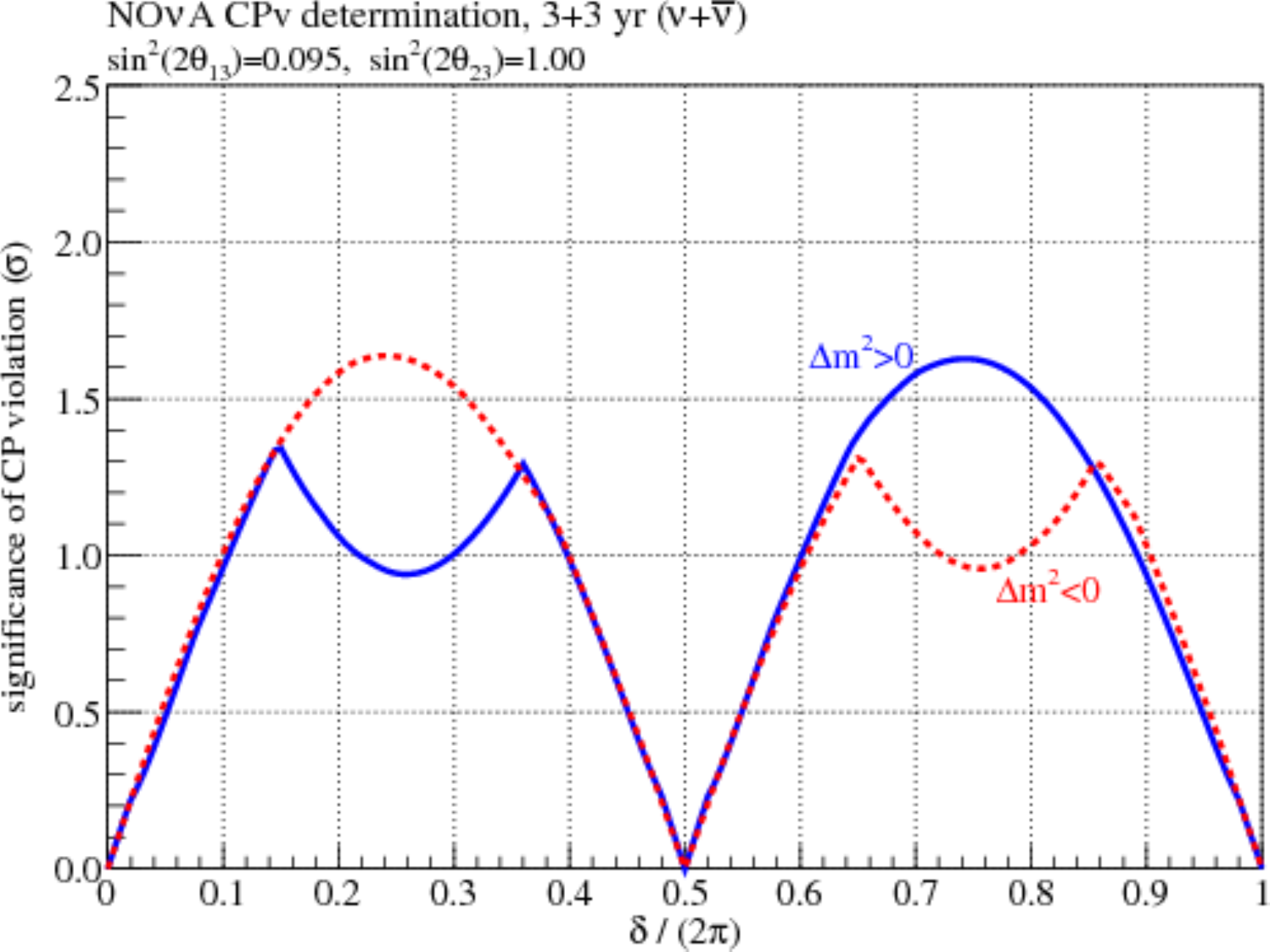}
\caption{Significance of the determination that CP violation occurs in
  neutrino oscillations as a function of $\delta$ in standard
  deviations. These sensitivities are for \nova alone for the two
  possible orderings and $\sintttwotwothree=1.0$.  The significance
  goes to zero at $\delta=0$ and $\delta=\pi$ since there is no CP
  violation at those points.  The dips in the peaks occur because the
  mass ordering has not been resolved for the ordering containing the
  dips.}
\label{fig:CPv}
\end{minipage}
\hfill
\begin{minipage}[t]{3.0in}
\centering
\includegraphics[width=3.0in]{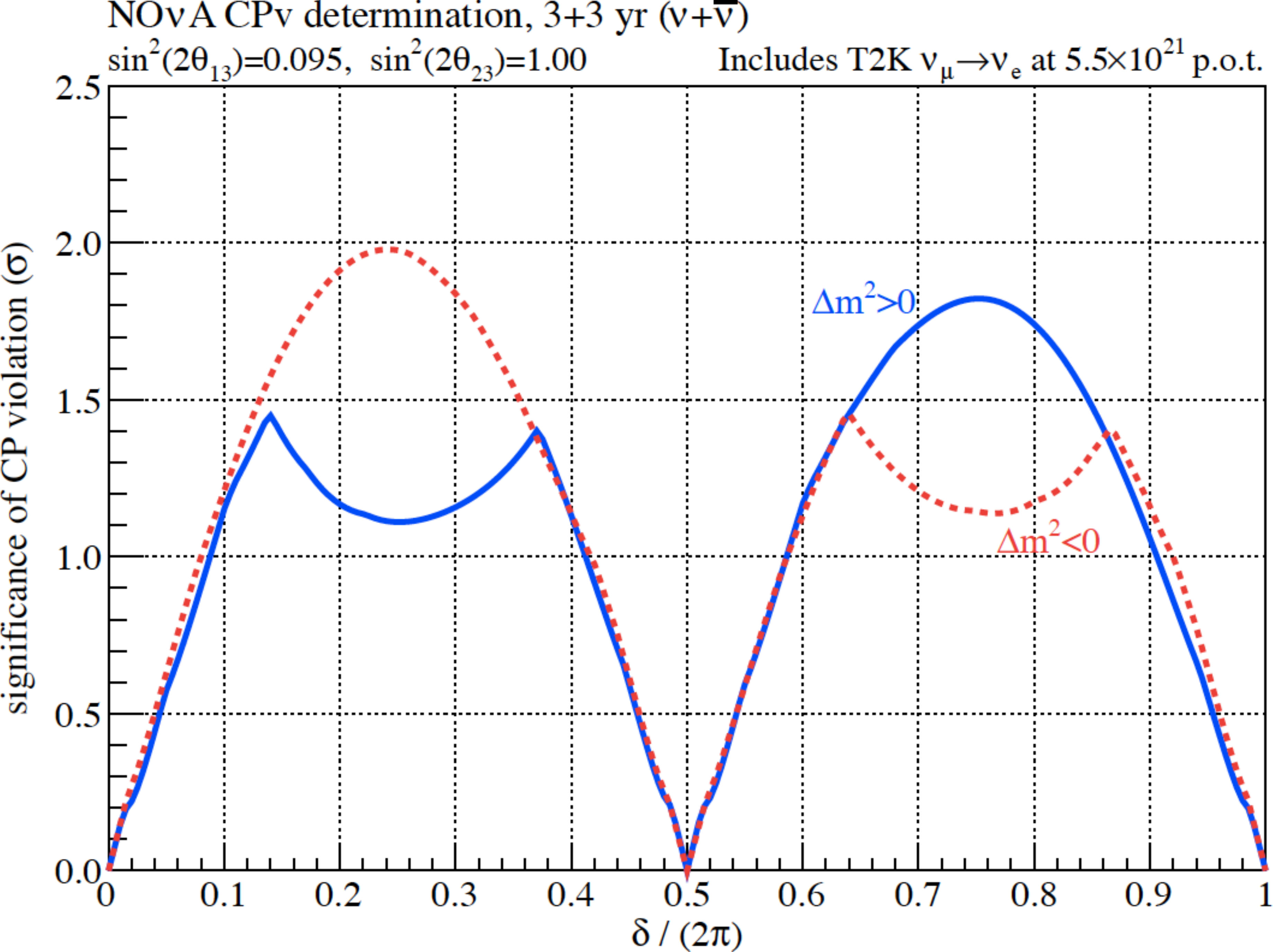}
\caption{Same as the figure to the left except that information from
  the T2K experiment has been included.}
\label{fig:novaT2KCPv}
\end{minipage}
\end{figure}

Unfortunately, the combined statistical power of \nova and T2K at the
end of the nominal six-year \nova run will be insufficient to resolve
the mass ordering at the two standard deviation level using this
strategy.  However, it is unlikely that either the American or the
Japanese neutrino program will end at that time.  With anticipated
improvements in both programs, in the worst case, the mass ordering
should be resolved in the next decade.
Figures~\ref{fig:novaMassOrdering} and \ref{fig:CPv} summarize the
\nova sensitivities for resolving the mass ordering and determining
that there is CP violation in the leptonic sector, respectively.
These figures are for \nova alone and use only the total measured
oscillation rate.  There will be some gain in sensitivity in using the
measured energy dependence and, as mentioned previously, improvements
in the analysis.  Figures~\ref{fig:novaT2KmassOrdering} and
\ref{fig:novaT2KCPv} show the same information, but include the
information from T2K that is expected to be available at the end of
the nominal six-year \nova run.
 

\subsection{Searches for new physics}
\label{sec:futSensNewPhysics}
Future data to be accumulated by long-baseline experiments offer novel
avenues to search for new physics in several
ways. MINOS+~\cite{Tzanankos:2011zz} will run with the NuMI beam
providing a flux that is least a factor of two higher in energy and
power than for MINOS\@. This wide band beam will yield thousands of
interactions a year in the Far detector with well measured $L/E$\@. In
combination with a precise prediction for the spectrum of interactions
from the Near detector, precision probes of new physics will be
performed. \nova and T2K experiments will exploit their narrow band
beams that have well defined energies. The \nova detectors with their
fine granular sampling of events (1 plane is 0.15 radiation lengths,
see section~\ref{sec:detectorsNOvA}) will provide enhanced ability to
distinguish the different neutrino interaction types.

\begin{figure}
\centering \includegraphics[trim = 0.1mm 0.1mm 0.1mm 0.1mm, clip,
  width=0.47\columnwidth]{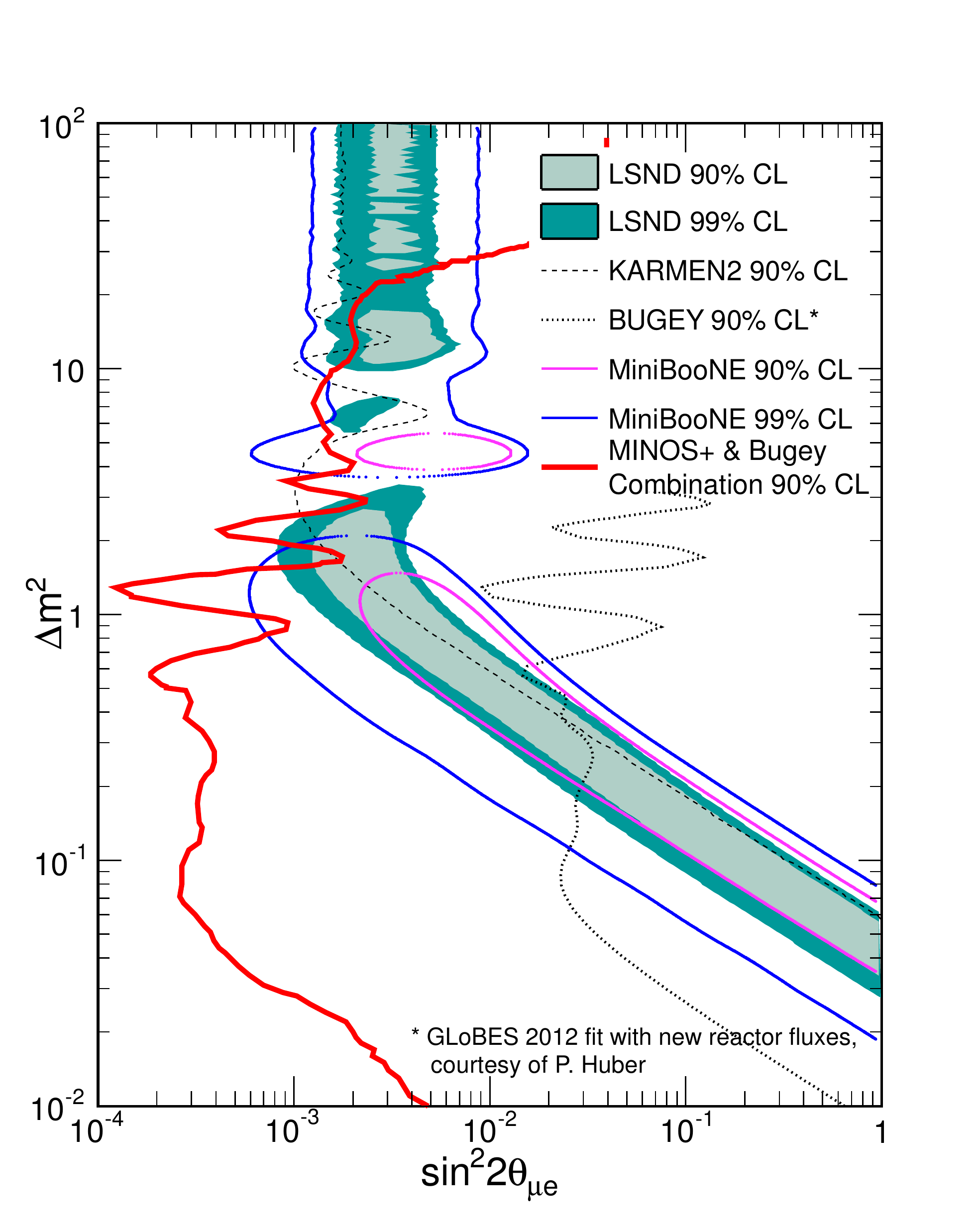}
\includegraphics[trim = 0.1mm 0.1mm 0.1mm 0.1mm, clip,
  width=0.47\columnwidth]{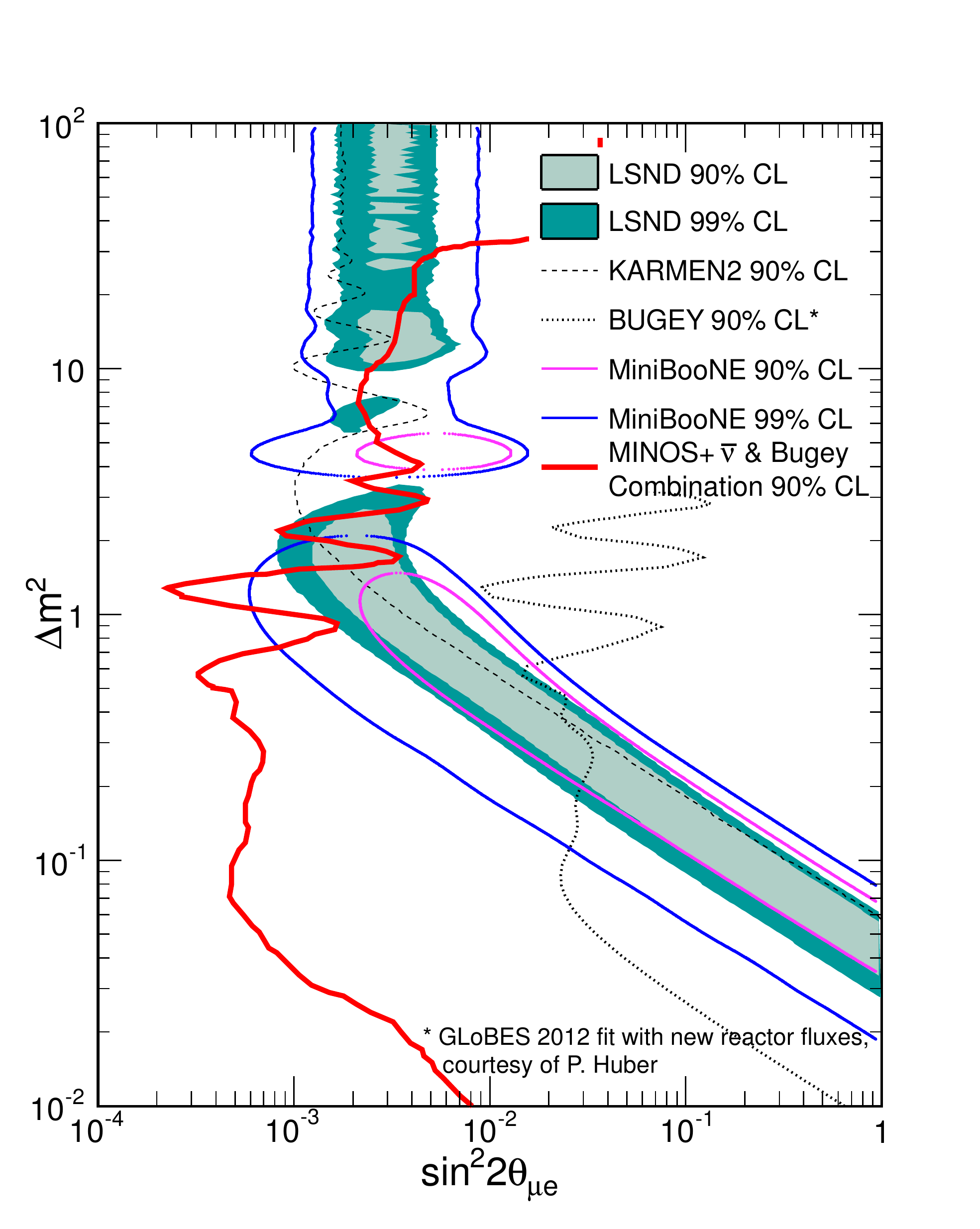}
\caption{Expected sensitivities for MINOS+ combined with Bugey data to
  $\sin^2(2\theta_{\mu e})$ as relevant for sterile neutrino
  searches. 90\%~C.L. contours are shown for exposures of
  $1.2\times10^{21}$~POT in both neutrino-enhanced (left) and
  antineutrino-enhanced (right) NuMI beam
  configurations~\cite{Nichol:Neu2012}. The regions of parameter space
  allowed by MiniBooNE and LSND experiments along with the limits from
  KARMEN~\cite{Armbruster:2002mp} are also shown.}
\label{fig:sterileContours} 
\end{figure}

Sterile neutrinos are one of the major areas of interest that will be
probed by upcoming experiments. \nova will improve on the MINOS
searches for a deficit in the rate of NC interactions in the Far
detector (see section~\ref{sec:resultsNewPhysics}), with significantly
better rejection of the dominant background coming from $\numu$~CC
events. In addition to studies of NC events, MINOS+ will use the
complementary approach to looking for sterile neutrinos that involves
constraining the disappearance of $\numu$ and hence, via unitarity,
will constrain the appearance of $\nue$ (that short-baseline
experiments are directly sensitive
to). Figure~\ref{fig:sterileContours} shows what MINOS+ expects to add
to the world's constraints on muon-electron mixing at mass squared
splittings between $10^{-2}$~eV$^2$ and 10~eV$^2$ (i.e. larger than
the atmospheric and solar mass splittings). The red curve in
Figure~\ref{fig:sterileContours} is the expected combined sensitivity
of MINOS+ and the Bugey reactor experiment~\cite{Declais:1994su}:
Bugey constrains the $\theta_{14}$ mixing angle with its $\nuebar$
disappearance measurements while MINOS+ aims to constrain
$\theta_{24}$ via the $\numu$ disappearance mode. Predicted
90\%~C.L. sensitivities for MINOS+ combined with Bugey data are shown
for exposures of $1.2\times10^{21}$~POT in both neutrino-enhanced
(left) and antineutrino-enhanced (right) NuMI beam configurations:
these contours show that MINOS+ has the sensitivity to exclude
substantial regions of parameter space allowed by
MiniBooNE~\cite{AguilarArevalo:2010wv} and LSND~\cite{Aguilar:2001ty}
results.

In addition to searching for sterile neutrinos, MINOS+ will have a
rich physics program that includes more precise measurements of
$|\dmatm|$ and $|\dmbaratm|$, a search for tau neutrinos, non-standard
interactions, extra-dimensions, measurements of neutrino
time-of-flight and atmospheric neutrinos.


\section{Conclusion}
\label{sec:conclusion}
Accelerator long-baseline experiments have made many measurements of
neutrino oscillations, extracting fundamental neutrino mixing
parameters and mass squared differences. The quantum mechanical
interference pattern expected from neutrino oscillations has been
observed with high statistics. 

The most precise measurements to-date of $|\dmatm|$ for both neutrinos
and antineutrinos were made by a long-baseline neutrino oscillation
experiment. Measurement of the largest neutrino mixing angle,
$\thetatwothree$, has reached the level of precision obtained using
atmospheric neutrinos and second generation long-baseline experiments
will soon improve the precision considerably further. Evidence for
electron neutrino appearance in a beam of muon neutrinos has recently
been obtained and is consistent with new results that demonstrate the
disappearance of reactor electron antineutrinos due to
$\thetaonethree$.

Using a dedicated accelerator long-baseline experiment, candidate tau
neutrino events have been directly observed in a beam of muon
neutrinos and analysis of the complete data set is expected to reveal
several more $\nutau$ candidates. Searches for oscillations into
sterile neutrinos have set stringent limits on various models and
these will improve further in the future. Long-baseline experiments
have also been exploited in searches for Lorentz violation and to make
world-leading measurements of the neutrino velocity.

The second generation long-baseline experiments currently taking data,
or soon to start, will exploit the relatively large value of
$\thetaonethree$ with the aim of measuring the mass hierarchy,
determining the octant of $\thetatwothree$, searching for CP violation
and exploring models of new physics. Over the next decade, these
experiments promise a rich program of research with the sensitivity to
make fundamental discoveries.

\section{References}
\addcontentsline{toc}{chapter}{Bibliography}
\bibliographystyle{styleFiles/jeffStyle.bst}
\bibliography{biblio}

\end{document}